\documentclass[10pt,aps,showpacs,nofootinbib,prd,aps,epsf,floats,floatfix,
amsmath,amssymb,amsfonts,axodraw,floatfix,graphicx,twocolumn]{revtex4-1}
\usepackage{amsmath, amssymb}
\usepackage{paralist}
\newcommand{\mathsym}[1]{{}}

\usepackage{graphicx}
\usepackage{psfrag}
\usepackage{amsmath}
\usepackage{amssymb}
\usepackage{bm}
\usepackage{stackrel}
\usepackage[normalem]{ulem}
\usepackage{hepunits}
\newsavebox{\PSLASH}
\sbox{\PSLASH}{$p$\hspace{-1.8mm}/}

\newcounter{saveeqn}

\newsavebox{\notrightarrow}
\sbox{\notrightarrow}{$\to$\hspace{-4mm}/}

\newsavebox{\PARTIALSLASH}
\sbox{\PARTIALSLASH}{$\partial$\hspace{-1.6mm}/}

\newsavebox{\ASLASH}
\sbox{\ASLASH}{$A$\hspace{-2.1mm}/}

\newsavebox{\KSLASH}
\sbox{\KSLASH}{$k$\hspace{-1.8mm}/}

\newsavebox{\LSLASH}
\sbox{\LSLASH}{$\ell$\hspace{-1.8mm}/}

\newsavebox{\QSLASH}
\sbox{\QSLASH}{$q$\hspace{-1.8mm}/}

\newsavebox{\DSLASH}
\sbox{\DSLASH}{$D$\hspace{-2.2mm}/}

\newsavebox{\DbfSLASH}
\sbox{\DbfSLASH}{${\mathbf D}$\hspace{-2.8mm}/}

\newsavebox{\DELVECRIGHT}
\sbox{\DELVECRIGHT}{$\stackrel{\rightarrow}{\partial}$}

\newcommand{\blue}{\IfColor{\textCadetBlue}{}}
\newcommand{\black}{\IfColor{\textBlack}{}}
\newcommand{\red}{\IfColor{\textRed}{}}
\newcommand{\green}{\IfColor{\textOliveGreen}{}}
\newcommand{\lila}{\IfColor{\textRedViolet}{}}

\newcommand{\fm}{\mathop{\rm fm}\nolimits}
\newcommand{\fmc}{\mathop{\rm fm/c}\nolimits}

\newcommand{\rd}[1]{\mathop{{d}#1}}

\newcommand{\deriv}[2]{\frac{{d}#1}{{d}#2}}
\newcommand{\derivv}[1]{\frac{\mathop{{d}}}{\rd#1}}
\newcommand{\pderiv}[2]{\frac{\partial#1}{\partial#2}}
\newcommand{\pderivv}[1]{\frac{\partial}{\partial#1}}

\newcommand{\at}[2][]{#1|_{#2}}

\newcommand{\inv}[1]{\frac{1}{#1}}
\newcommand{\avec}[1]{\left|\boldsymbol{#1}\right|}

\newcommand{\tder}[1]{\mathfrak{D}#1}

\newcommand{\cd}[2]{\mathrm{\nabla}_{#1}{#2}}
\newcommand{\lieder}[2]{\mathcal{L}_{#1}{#2}}
\newcommand{\christoffel}[3]{\Gamma^{#1}_{#2#3}}

\DeclareMathOperator{\diag}{diag}

\begin{document}
\title{
Evolution of magnetic fields from the $\boldsymbol{3+1}$ dimensional self-similar and\\ Gubser flows in ideal relativistic magnetohydrodynamics
}
\author{M. Shokri}\email{m\_shokri@physics.sharif.ir}
\author{N. Sadooghi}\email{sadooghi@physics.sharif.ir}
\affiliation{Department of Physics, Sharif University of Technology,
		P.O. Box 11155-9161, Tehran, Iran}
\begin{abstract}
Motivated by the recently found realization of the $1+1$ dimensional Bjorken flow in  ideal and nonideal relativistic magnetohydrodynamics (MHD), we use appropriate symmetry arguments, and determine the evolution of magnetic fields arising from the $3+1$ dimensional self-similar and Gubser flows in an infinitely conductive relativistic fluid (ideal MHD). In the case of the $3+1$ dimensional self-similar flow, we arrive at a family of solutions, that are related through a differential equation arising from the corresponding Euler equation. To find the magnetic field evolution from the Gubser flow, we solve the MHD equations of a stationary fluid in a conformally flat $dS^{3}\times E^{1}$ spacetime. The results are then Weyl transformed back into the Minkowski spacetime. In this case, the temporal evolution of the resulting magnetic field is shown to exhibit a transition between an early time $1/t$ decay to a $1/t^{3}$ decay at a late time. Here, $t$ is the time coordinate. Transverse and longitudinal components of the magnetic fields arising from these flows are also found. The latter turns out to be sensitive to the transverse size of the fluid. In contrast to the result arising from the Gubser flow, the radial domain of validity of the magnetic field arising from the self-similar flow is highly restricted. A comparison of the results suggests that the (conformal) Gubser MHD may give a more appropriate qualitative picture of the magnetic field decay in the plasma of quarks and gluons created in heavy ion collisions.
\end{abstract}
\maketitle
\section{Introduction}\label{Introduction}\label{sec1}
\par
In heavy ion collisions (HICs), large electromagnetic fields are generated by the electric current produced by the accelerated motion of positively charged spectators, i.e. nucleons that do not participate in the collision (see \cite{rajagopal2018,Huang-review} for recent reviews). Concerning the evolution of electromagnetic fields, one may distinguish between the collision, early (pre-equilibrium) and quark-gluon plasma (QGP) stages  \cite{Huang-review}.  Quite a large number of attempts are devoted to estimate the strength and the spacetime evolution of electromagnetic fields in these stages. Depending on the energies and the impact parameters of the collisions, they are found to be of the order $eB\sim 1$-$10 m_{\pi}^2$ in the early stage \cite{warringa2007,Huang-review,skokov2009, zakharov2014}.\footnote{Here, $e$ is the electric charge and $m_{\pi}\sim 140$ MeV the pion mass.} Moreover, they are believed to be aligned in the transverse direction with respect to the reaction plane. In very short timescales of about $0.065\fmc$ ($0.005\fmc$) at the Relativistic Heavy Ion Collider (RHIC) and $0.005\fmc$ at the Large Hadron Collider (LHC), the spectators leave the scene, and a medium including highly excited partons, mostly dominated by gluons, remains \cite{hatsuda-book}. At this stage, this medium is far from equilibrium and is, because of this gluon dominance, almost a perfect insulator. Electromagnetic fields are thus believed to quickly decay in this stage. The decay is roughly of a $t^{-3}$ nature near the center of the collision \cite{Huang-review}. In a timescale of roughly $0.5\fmc$, the medium is deexcited into a plasma of quarks and gluons, and a local thermal equilibrium is approximately achieved.
The spacetime history of the QGP in this stage is well understood using the relativistic hydrodynamics (RHD) (see \cite{romatschke2017} for a  recent review). Inspired by the successes of RHD, it seems therefore to be natural to consider the relativistic magnetohydrodynamics (MHD) to provide a reliable effective picture of the interplay between electromagnetic fields and the QGP.
\par
In this framework, one may ask two types of questions: (1) How do electromagnetic fields evolve within the ultrarelativistic fluid, and (2) how do the fluid degrees of freedom, e.g. the fluid's velocity and temperature, are affected by electromagnetic fields. To the best of our knowledge, the first analytical realization of electromagnetic fields in MHD was presented in \cite{rischke2015}. Here, the evolution of magnetic fields in an infinitely conductive fluid was found in the presence of the $1+1$ dimensional Bjorken flow \cite{bjorken1983}. In this setup the decay of the magnetic field turns out to be of a $t^{-1}$ nature at the center of the collision. This is significantly slower than its early time decay. In a previous work \cite{shokri2017}, we relaxed  the assumption of infinite conductivity made in \cite{rischke2015}, and found the evolution of magnetic and electric fields in the presence of the Bjorken flow. We also studied the effects of electromagnetic fields on the evolution of QGP temperature.
Other attempts to study the effects of magnetic fields on the properties of the QGP created at the RHIC and LHC, and, in particular, to determine their lifetime are made in \cite{magnetic-rest}.
\par
A well-known poverty of the Bjorken flow, that prevents it from giving a qualitative picture of certain observables of HICs, is its lack of a transverse expansion \cite{hatsuda-book}. In particular, the spectra of final hadrons' transverse momentum signal the existence of a significant radial expansion of the QGP \cite{kolb2003,rajagopal2011}. This fact motivated several attempts on the generalization of the Bjorken flow to solutions including an appropriate transverse expansion. It is the main purpose of the present paper to focus on the $3+1$ dimensional self-similar flow from \cite{csorgo2002,csorgo2002-2} and the Gubser flow from \cite{gubser2010-1,gubser2010-2}, and to determine the magnetic field evolution arising from these flows. To do this, we present a realization of these flows in an ideal MHD using, in particular, similar symmetry arguments as in \cite{beckenstein1978}.
\par
A $3+1$ dimensional self-similar flow can be regarded as a combination of three Bjorken flows in three spatial directions. Although this flow is a simple spherical Hubble expansion, more symmetries can be introduced by similarity variables \cite{csorgo2002,csorgo2002-2,rezolla-book}.
Other attempts are made, e.g., in \cite{pechanski2009} to introduce more realistic elliptically-shaped solutions of hydrodynamic equations. On the other hand, the crucial observation that leads to the Gubser flow is that the Bjorken flow is based on the assumption of a translational invariance in the transverse plane, which, as aforementioned, prohibits an expansion of the fluid in transverse directions. Similarly, the transverse MHD setup introduced in \cite{rischke2015} is, as a realization of the same Bjorken flow, also based on the same symmetry. Gubser argued that such a symmetry is indeed a poor approximation for a small system such as the QGP created in HICs, and replaced it with a certain conformal symmetry \cite{gubser2010-1}. Similar techniques are used in \cite{hatta2016} to introduce other nonboost invariant flows as well as a generalization of the Gubser flow to the case of noncentral collisions \cite{hatta2014}.
\par 	
The organization of this paper is as follows: In Sec. \ref{sec2}, we briefly review the general equations of MHD in the ideal limit. In Sec.  \ref{sec3}, we first present a generalized form of the Bekenstein and Oron's treatment of symmetries in MHD from \cite{beckenstein1978}. To set a benchmark for this analysis, we then study the transverse MHD, previously considered in \cite{rischke2015,shokri2017}, and arrive at the same results, as expected. We close this section with remarks on a cylindrically symmetric flow with a longitudinal boost invariance. In Sec. \ref{sec4}, the method developed in Sec. \ref{sec3} is applied to the case of a $3+1$ dimensional self-similar flow \cite{csorgo2002} with a cylindrical similarity variable. We first find that a family of solutions exists, and that the exact form of the magnetic field evolution is thus ambiguous. We then present a number of possible solutions to this problem. Among others, we consider the stationary case, where the corresponding electric current vanishes. This solution is referred to as the zero current self-similar solution (ZCSSF).  In Sec. \ref{sec5}, we use the method of \cite{gubser2010-1}, and present a realization of the Gubser flow in MHD. In Sec. \ref{sec5A}, we first start with a brief review of the Gubser flow from a slightly different point of view than originally introduced in \cite{gubser2010-1}. We then show, in Sec. \ref{sec5B}, that the magnetic field arising from the implementation of the Gubser flow into ideal MHD has only one nonvanishing component in the longitudinal beam direction, and that the surviving longitudinal component is sensitive to the finite transverse size of the fluid. These results, however, turn out to be in contrast to what is generally believed to be the case in HICs \cite{warringa2007,Huang-review,skokov2009, zakharov2014}.
To overcome this problem, we apply, in Sec. \ref{sec6}, the technique of Weyl transformations from \cite{gubser2010-2} to ideal MHD. However, instead of using the $SO(3)$ symmetry group as in \cite{gubser2010-2}, we introduce a proper similarity variable to fix the four-velocity. We then determine the spacetime dependence of the magnetic field by a Weyl transformation from a combination of a three dimensional anti de Sitter spacetime and a one dimensional Euclidean space, denoted by $dS^{3}\times \boldsymbol{E}^1$, into a $3+1$ dimensional Miknowski spacetime, denoted by $M^{3,1}$. This turns out to be a cure for the aforementioned problem with the Gubser MHD. This novel solution is referred to as the conformal MHD solution (CMHD). We then numerically compare the ZCSSF and CMHD solutions in Sec. \ref{sec7}. In particular, we introduce a number of parameters to emphasize the role played by nonvanishing longitudinal components of these solutions. Section \ref{sec8} contains our conclusions and final remarks.
\par
In this paper, we take $\hbar=c=k_B=1$, and assume the mostly plus metric $\diag(-+++)$. The four-velocity is thus normalized as $u^\mu u_\mu = -1$.  We also use the total antisymmetric tensor  $\epsilon_{0123}=-1/\epsilon^{0123}=-\sqrt{-g}$ and the transverse projector  $\Delta^{\mu\nu}\equiv g^{\mu\nu}+u^\mu u^\nu$.
The covariant, covariant proper time and Lie derivatives are denoted by $\cd{\mu}{},\tder{}\equiv u^\mu\cd{\mu}{}$, and $\lieder{\xi}{}$, respectively (see Appendix \ref{appA} for more definitions).
\section{Equations of MHD}\label{sec2}
\setcounter{equation}{0}
Relativistic MHD is an extension of the RHD that includes electromagnetic degrees of freedom.
The corresponding constitutive equations consist of the energy-momentum conservation equation,
\begin{equation}\label{M1}
\cd{\mu}{T^{\mu\nu}}=0,
\end{equation}
as well as homogeneous and inhomogeneous Maxwell equations,
\begin{eqnarray}\label{M2}
\partial_\alpha F_{\beta\gamma}+\partial_\beta F_{\gamma\alpha}+\partial_\gamma F_{\alpha\beta}=0,
\end{eqnarray}
and
\begin{eqnarray}\label{M3}
\cd{\nu}{F^{\mu\nu}}=J^\mu.
\end{eqnarray}
The latter implies the electric current conservation,
\begin{equation}\label{M4}
\cd{\mu}{J^\mu}=0.
\end{equation}
Other conserved currents, such as baryon number or entropy density currents, may also be present in the theory. In \eqref{M1}, $T_{\mu\nu}$ is the total energy-momentum tensor, consisting of fluid and Maxwell energy-momentum tensors, $T^{\mu\nu}_{\tiny\mbox{F}}$ and $T^{\mu\nu}_{\tiny\mbox{EM}}$,
\begin{equation}\label{M5}
T^{\mu\nu} = T^{\mu\nu}_{\tiny\mbox{F}}+T^{\mu\nu}_{\tiny\mbox{EM}}.
\end{equation}
Neglecting the magnetization and electric polarization, and assuming the fluid to be nondissipative, $T^{\mu\nu}_{\tiny\mbox{F}}$ is given by
\begin{equation}\label{M6}
T^{\mu\nu}_{\tiny\mbox{F}}= \epsilon u^\mu u^\nu + p\Delta^{\mu\nu}.
\end{equation}
Here, $u^\mu$, $\epsilon$ and $p$ are the fluid velocity, energy density and pressure of the fluid, respectively. Moreover, the Maxwell tensor $T^{\mu\nu}_{\tiny\mbox{EM}}$ reads \cite{generalrefs}
\begin{equation}\label{M7}
T^{\mu\nu}_{\mbox{\tiny{EM}}} = F^\mu_{~\alpha} F^{\nu\alpha} - \inv{4}g^{\mu\nu}F_{\alpha\beta}F^{\alpha\beta},
\end{equation}
where $F_{\mu\nu}$, similar to any other antisymmetric rank two tensor, can be decomposed as \cite{rezolla-book}
\begin{equation}\label{M8}
F^{\mu\nu}=u^\mu E^\nu-u^\nu E^\mu+\epsilon^{\mu\nu\rho\sigma}B_\rho u_\sigma.
\end{equation}
Here,
\begin{equation}\label{M9}
E^\mu = F^{\mu\nu}u_\nu,
\end{equation}
and
\begin{equation}\label{M10}
B^\mu=\inv{2}\epsilon^{\mu\nu\rho\sigma}u_\rho F_{\rho\sigma}.
\end{equation}
In the local rest frame (LRF) of the fluid, \eqref{M9} and \eqref{M10} are given by
\begin{equation}\label{M11}
 E^\mu_{\mbox{\tiny{LRF}}} = (0,\boldsymbol{E}),\qquad B^\mu_{\mbox{\tiny{LRF}}} = (0,\boldsymbol{B}).
\end{equation}
For the electromagnetic field strength tensor $F_{\mu\nu}$, we identify $\boldsymbol{E}$  and $\boldsymbol{B}$ with the electric and magnetic three-vectors, as measured in the LRF of the fluid. Because of this identification, $E^{\mu}$ and $B^{\mu}$ from \eqref{M9} and \eqref{M10} are referred to as electric and magnetic four-vectors.\footnote{One should bear in mind that, in an arbitrary frame $E^\mu$ and $B^\mu$ are not purely electric and magnetic.}
Using  \eqref{M11}, the magnitudes of local electric and magnetic fields are thus given by
\begin{equation}\label{M12}
B \equiv \avec{B}=\sqrt{B^\mu B_\mu},\quad E\equiv\avec{E}=\sqrt{E^\mu E_\mu}.
\end{equation}
In its simplest form, the electric current, appearing in \eqref{M3}, is given by
\begin{equation}\label{M13}
J^\mu = \rho_e u^\mu + \sigma_{e} E^\mu,
\end{equation}
where $\rho_e$ is the proper charge density, and $\sigma_{e}$ is the fluid conductivity. For an infinitely conductive fluid, in order to keep the current finite, $E^\mu$ must tend to zero. This is the so called ideal MHD limit \cite{beckenstein1978,rischke2015,shokri2017}. Using \eqref{M1} and \eqref{M5}, the energy-momentum conservation can also be written as \cite{hernandez2017,beckenstein1978}
\begin{equation}\label{M14}
\cd{\mu}{T^{\mu\nu}_{\tiny\mbox{F}}}=F^{\nu\alpha}J_\alpha.
\end{equation}
Contracting both sides of \eqref{M14} with $u_\nu$, we arrive first at the energy equation,
\begin{equation}\label{M15}
\tder{\epsilon}+(\epsilon+p)\cd{\mu}{u^\mu}=0.
\end{equation}
This relation shows that in the ideal limit, the electromagnetic part is completely decoupled from the energy equation. To solve \eqref{M15}, one needs to provide the equation of state (EOS). We assume the EOS to be \cite{csorgo2002}
\begin{eqnarray}\label{M16}
\epsilon=\kappa p,
\end{eqnarray}
with $\kappa\equiv 1/c_s^2$, and $c_s$ being the sound velocity in the fluid. In what follows, we assume $c_{s}$ to be constant.
\par
Projecting, at this stage, \eqref{M14} into the transverse direction, i.e. the direction perpendicular to $u_{\mu}$, we arrive at the Euler equation, that, in the case of ideal MHD reads \cite{beckenstein1978}
\begin{equation}\label{M17}
\left(\epsilon+p+B^2\right)a^\mu = -\Delta^{\mu\nu}\left[\partial_\nu\left(p+\frac{B^2}{2}\right)-\cd{\rho}{\left(B_\nu B^\rho\right)}\right],
\end{equation}
where $a^{\mu}\equiv \tder u^{\mu}$ is the acceleration of the fluid.
In contrast to \eqref{M15}, the Euler equation is different from its pure hydrodynamical counterpart,
\begin{equation}\label{M18}
\left(\epsilon+p\right)a^\mu = -\Delta^{\mu\nu}\partial_\nu p.
\end{equation}
One should bear in mind that, in the ideal MHD limit, the electric current is ambiguous. Hence, only the homogeneous Maxwell equation from \eqref{M2} should be used to solve the energy equation (\ref{M15}) \cite{romero2005}. In what follows, the energy-momentum tensor and the inhomogeneous Maxwell equations \eqref{M3} are only used to determine $J^\mu$.
\section{Application of symmetries in relativistic MHD}\label{sec3}
\setcounter{equation}{0}
\subsection{General remarks}\label{sec3A}
Complicated equations of MHD may be simplified by symmetry considerations. A few decades ago, Bekenstein and Oron  showed that these equations can be significantly simplified using a temporal (stationary) and an axial symmetric flow \cite{beckenstein1978}. In this section, we generalize their treatment to the case of two arbitrary spatial symmetries.	
\par
Let us assume that there exists two vectors $\xi_{1}$ and $\xi_{2}$ that commute with the metric and every physical quantities that appear in the energy-momentum tensor,
\begin{eqnarray}\label{A1}
&&\hspace{-0.5cm}[\xi_{i},g_{\mu\nu}]=0,\nonumber\\
&&\hspace{-0.5cm}[\xi_{i},F_{\mu\nu}]=0,~[\xi_{i},u_{\mu}]=0,~\cdots,
~\mbox{for}~ i=1,2.
\end{eqnarray}
Based on the geometry and underlying physics of the system, symmetries are to be found. For simplicity, let us choose a coordinate system with $\xi_{i}=\partial_i, i=1,2$. Relations \eqref{A1} thus takes the form
\begin{equation}\label{A2}
\partial_{i} g_{\mu\nu}=0,~\partial_{i}F_{\mu\nu}=0,~\partial_{i}u_{\mu}=0,~\cdots,~\mbox{for}\quad i=1,2.
\end{equation}
As a consequence of \eqref{A2}, the homogeneous Maxwell equation \eqref{M2} reads
\begin{eqnarray}\label{A3}
\partial_0F_{12}&=&0,\nonumber\\
\partial_3F_{12}&=&0,\nonumber\\
\partial_3F_{02}+\partial_0F_{23}&=&0.
\end{eqnarray}
Assuming $F_{\alpha\beta}$ being zero at infinity, we arrive at
\begin{equation}\label{A4}
F_{12}=0.
\end{equation}
As aforementioned, in the ideal MHD limit, the electric field vanishes. Hence,  $F_{\alpha\beta}u^\beta=E_{\alpha}=0$ results in
\begin{eqnarray}\label{A5}
F_{01} u^1 + F_{02} u^2 + F_{03} u^3 &=& 0,\nonumber\\
F_{10} u^0 + F_{13} u^3 &=& 0,\nonumber\\
F_{20} u^0 + F_{23} u^3 &=& 0,\nonumber\\
F_{30} u^0 + F_{31} u^1 + F_{32} u^2 &=& 0.
\end{eqnarray}
These equations lead to a number of relations between electric and magnetic components of the field strength tensor,  $F_{0i}$ and $F_{ij}$,
\begin{eqnarray}\label{A6}
F_{01} &=& \frac{u^3}{u^0} F_{13},\nonumber\\
F_{02} &=& \frac{u^3}{u^0} F_{23},\nonumber\\
F_{03} &=& -\inv{u^0} (u^1 F_{13}  + u^2 F_{32}).
\end{eqnarray}
Plugging at this stage, \eqref{A6} into \eqref{A3}, we obtain
\begin{equation}\label{A7}
\tder{\log F_{13}} = \tder{\log F_{23}} = -u^0\partial_3\left(\frac{u^3}{u^0}\right).
\end{equation}
Two magnetic components $F_{13}$ and $F_{23}$ are thus related as
\begin{equation}\label{A8}
F_{13} = f(\vartheta)F_{23}.
\end{equation}
In \eqref{A8}, $\vartheta$ is a parameter that does not change through flow lines, i.e. $\tder{\vartheta}=0$.  It also respects the same symmetries as in \eqref{A2}, i.e. $\lieder{\xi_i}{\vartheta}=0, i=1,2$. Being a proper scalar, one can thus label flow lines with $\vartheta$. Let us notice that under certain circumstances, $\vartheta$ may also be regarded as a similarity variable \cite{rezolla-book}. Using  \eqref{A2}, the right hand side (rhs) of  \eqref{A7} reads
\begin{equation}\label{A9}
-u^0\partial_3\left(\frac{u^3}{u^0}\right)=-\partial_\alpha u^\alpha+\tder{\log(u^0)}.
\end{equation}
The first term on the rhs of \eqref{A9} may be written in a simpler form. To do this, let us consider a conserved current of type $Q^\mu = Q(x^0,x^3)u^\mu$ that satisfies
\begin{equation}\label{A10}
\cd{\alpha}{\left(Qu^\alpha\right)}=0.
\end{equation}
Physical examples of $Q$ include the conserved baryon number density $n$ and entropy density $s$. For following arguments, however, $Q$ is not required to be any physical quantity. It is merely a solution to \eqref{A10}. Using  \eqref{appA2}, \eqref{A10} gives rise to
\begin{equation}\label{A11}
\partial_\alpha u^\alpha = -\tder{\log(\sqrt{-g}Q)}.
\end{equation}
Plugging, at this stage, \eqref{A11} into \eqref{A9}, and then the resulting expression into \eqref{A7}, and also using  \eqref{A8}, the formal solution for the field strength tensor is found to be
\begin{eqnarray}\label{A12}
F_{13}&=&Q(x^0,x^3)u^0\sqrt{-g}f(\vartheta)h(\vartheta),\nonumber\\
F_{23}&=&Q(x^0,x^3)u^0\sqrt{-g}h(\vartheta),
\end{eqnarray}
with $Q$ satisfying (\ref{A10}). These relations are quite general, and are thus valid for different hydrodynamic flows. They will be used in the next sections to derive the evolution of magnetic fields from the $3+1$ dimensional self-similar and Gubser flows in the ideal MHD. Assuming the corresponding symmetries to these flows, we start with these four-velocity profiles, and solve \eqref{A10}. We show that the corresponding solutions are determined up to functions of the proper scalar, which respects the assumed symmetries. Using then the Euler equation \eqref{M17}, we determine these functions, and arrive at the final solutions of $B^{\mu}$ in each cases.
\subsection{Ideal transverse MHD}\label{sec3B}
To illustrate the approach described above, let us consider the Bjorken flow with a  transverse MHD setup \cite{rischke2015,shokri2017}. As it was recognized by Gubser in \cite{gubser2010-1}, the Bjorken velocity profile can be fixed by symmetry considerations alone. The symmetries that fix it are
	\begin{enumerate}
		\item\label{trans_inv} Translational invariance in the transverse $x$-$y$ plane.
		\item\label{rot_inv} Rotational invariance around the beamline, which is assumed to be in the $z$-direction.
		\item\label{boost_inv} Boost invariance along the beamline.
	\end{enumerate}
According to our arguments in \cite{shokri2017}, the translational invariance in the transverse plane leads automatically to the transverse MHD setup, where, in particular, $\boldsymbol{v}\cdot\boldsymbol{B}=0$.
Let us now parameterize the flat spacetime metric as
\begin{eqnarray}\label{A13}
{\rd{s}}^2 = -{\rd{\tau}}^2+{\rd{x}}^2+{\rd{y}}^2+\tau^2{\rd{\eta}}^2.
\end{eqnarray}
Here, $\tau\equiv\sqrt{t^2-z^2}$ is a combination of coordinates that respect all aforementioned symmetries.\footnote{The parameter $\tau$ is sometimes called spacetime proper time. In this work, we avoid this confusing terminology, and emphasize that, in general, $\tau$ is a coordinate that only in the Bjorken case identifies with the spacetime proper time.} The parameter $\eta\equiv \inv{2}\log\frac{t+z}{t-z}$ is defined so that $\partial_\eta$ is the Killing vector associated with a longitudinal boost. In terms of the coordinates of \eqref{A13}, the Killing vectors associated with translational and boost symmetries are thus ordinary partial derivatives with respect to $(x, y)$ and $\eta$, respectively. Remarkably, the Bjorken four-velocity turns out to be  given by \cite{dewolfe2014}
\begin{equation}\label{A14}
u_\mu = - \frac{\partial_\mu\tau}{\sqrt{-\partial_\mu\tau\partial^\mu\tau}}.
\end{equation}
In the metric \eqref{A13}, it reads $u^\mu=(1,0,0,0)$. Since there is no proper acceleration, the left hand side (lhs) of the Euler equation \eqref{M17} vanishes identically. Using the four-velocity profile from \eqref{A14}, let us now determine the magnetic field. In the transverse MHD setup, we choose the symmetries of electromagnetism to be
\begin{eqnarray}\label{A15}
\xi_{x}=\pderivv{x},\quad\xi_{y}=\pderivv{y}.
\end{eqnarray}
These are a subset of the Bjorken symmetries. Relaxing the boost invariance for quantities other than $u_{\mu}$, we can use $\eta$ to label flow lines. From \eqref{A6}, one notices that the electric components of the field strength tensors, $F_{0i}$, vanish. Moreover, according to \cite{shokri2017}, the solution to \eqref{A10} for Bjorken flow is given by
\begin{equation}\label{A16}
Q = Q_0 \frac{\tau_0}{\tau}.
\end{equation}
Plugging \eqref{A16} into  \eqref{A12}, we arrive at
\begin{eqnarray}\label{A17}
F_{x\eta}&=&Q_0\tau_0f(\eta)h(\eta),\nonumber\\
F_{y\eta}&=&Q_0\tau_0h(\eta).
\end{eqnarray}
The magnetic field $B^\mu$ is then immediately found by plugging \eqref{A17} into \eqref{M10}. It reads
\begin{equation}\label{A18}
B^\mu = Q_0 \frac{\tau_0}{\tau}h(\eta)\left(0,1,-f(\eta),0\right).
\end{equation}	
Here, functions $f$ and $h$ are found by plugging \eqref{A18} into \eqref{M17}, and solving the resulting equation. The latter can be simplified by symmetry arguments: The second term on the rhs of \eqref{M17}, i.e. $\cd{\rho}{\left(B_\nu B^\rho\right)}$, vanishes due to $\cd{\rho}{B^\rho}=0$ and the lack of connection between transverse and longitudinal directions. The transverse projector, i.e. $\Delta^{\mu\nu}$, vanishes for $\mu=\tau$. Moreover, for $\mu=x,y$, the rhs vanishes due to symmetries. The only nonvanishing component of  \eqref{M17} is thus in the $\eta$-direction. It reads
\begin{eqnarray}\label{A19}
\pderiv{B}{\eta}=0.
\end{eqnarray}
Using \eqref{A18}, this gives rise to
\begin{eqnarray}\label{A20}
\derivv{\eta}\left[h(\eta)^2\left(1+f(\eta)^2\right)\right]=0.
\end{eqnarray}
Let us write, without loss of generality, the solution to \eqref{A20} as
\begin{eqnarray}\label{A21}
h(\eta)=\inv{\sqrt{1+f^2(\eta)}}.
\end{eqnarray}
Plugging \eqref{A20} into \eqref{A18} leads to
\begin{eqnarray}\label{A22}	
B^\mu &=& B_0\frac{\tau_0}{\tau}\frac{1}{\sqrt{1+f^2(\eta)}}\left(0,1,-f(\eta),0\right),
\qquad
\mbox{and}\nonumber\\
B &=& B_0\frac{\tau_0}{\tau}.
\end{eqnarray}
While, according to \eqref{A19}, $B=\sqrt{B^\mu B_\mu}$ is forced to be boost invariant, it is not necessary for the individual components of $B^\mu$ to be so.\footnote{As a consequence of the $\eta$ dependence of $B^{\mu}$, the direction of the $\boldsymbol{B}$ field differs between two flow lines, but is frozen through each particular flow line.} Here, as in the case of the temperature and entropy density in self-similar flows \cite{csorgo2002,csorgo2002-2,shokri2017}, there is an arbitrariness in  \eqref{A22}. This arbitrariness disappears if one assumes $B^\mu$ to be boost invariant. Such an assumption has a crucial physical significance, as we show below.
\par
To do so, let us consider the electric current, $J^\mu$ from \eqref{M13}, where, according to the arguments in \cite{shokri2017},\footnote{See Appendix A 1 in \cite{shokri2017}.} the proper charge density $\rho_{e}$ vanishes. The electric current is thus given by $J^{\mu} = \sigma_{e} E^\mu$. It turns out to be ambiguous, because in the ideal MHD limit, as $E^\mu$ tends to zero, $\sigma_{e}$ goes to infinity. We are therefore left with a $0\times\infty$ product that cannot be naively set to zero. Plugging, nevertheless, \eqref{A22} into \eqref{M3}, the electric current is found to be
\begin{equation}\label{A23}
J^\mu= B_0\frac{\tau_0}{\tau^2}\frac{f^{\prime}(\eta)}{\left(1+f^2(\eta)\right)^{3/2}}\left(0,1,-f(\eta),0\right).
\end{equation}
Interestingly, the current vanishes if $B^{\mu}$ from
\eqref{A22} is assumed to be boost invariant. This assumption leads automatically to $f^{\prime}(\eta)=0$. Choosing, without loss of generality, $f=1$, a specific solution for $B^{\mu}$ from \eqref{A22} is given by
\begin{equation}\label{A24}
B^\mu=B_0\frac{\tau_0}{\tau}\frac{1}{\sqrt{2}}\left(0,1,-1,0\right).
\end{equation}
One may also notice that
\begin{eqnarray}\label{A25}
\frac{E}{B}\sim \frac{\sqrt{J^\mu J_\mu}}{\sigma_{e} B}=\inv{\sigma_{e}\tau}\frac{f'(\eta)}{\left(1+f(\eta)^2\right)}.
\end{eqnarray}
This heuristic result confirms our previous results presented from \cite{shokri2017}, where it was found that for $E\ll B$, $\sigma_{e}$ must be much larger than a typical value of $\tau$.
\subsection{General solutions of $B^{\mu}$ and $J^{\mu}$}\label{sec3C}
The rest of this work is devoted to flows that, in contrast to the transverse MHD flow, are not translational invariance in the transverse plane, and expand in transverse directions. Being motivated by the physics of the QGP in HICs, flows that we study share two symmetries, namely, boost invariance along and rotational invariance around the beamline. To reveal these symmetries, we parameterize the flat spacetime metric as
\begin{eqnarray}\label{A26}
{\rd{s}}^2 = -{\rd{\tau}}^2+r^2{\rd{\phi}}^2+\tau^2{\rd{\eta}}^2+{\rd{r}}^2.
\end{eqnarray}
For
\begin{eqnarray}\label{A27}
x^\mu=(\tau,\phi,\eta,r),
\end{eqnarray}
we thus have $r=\sqrt{x^2+y^2}$ and $\phi=\arctan\frac{y}{x}$. In terms of the  coordinates \eqref{A27}, the Killing vectors that are associated with these symmetries are
\begin{eqnarray}\label{A28}
\xi_{\phi}=\pderivv{\phi},\quad\xi_{\eta}=\pderivv{\eta}.
\end{eqnarray}
For the above metric \eqref{A26}, Christoffel symbols read
\begin{eqnarray}\label{A29}
\christoffel{\tau}{\eta}{\eta}&=&\tau,\quad\christoffel{\eta}{\tau}{\eta}=\christoffel{\eta}{\eta}{\tau}=\inv{\tau},\nonumber\\
	\christoffel{r}{\phi}{\phi}&=&-r,\quad\christoffel{\phi}{r}{\phi}=\christoffel{\phi}{\phi}{r}=\inv{r}.
\end{eqnarray}
Although the concrete form of the four-velocity profile is specific to each flow, longitudinal boost invariance together with $\boldsymbol{Z}_2$ symmetry under $\eta\to-\eta$ eliminates $u^\eta$ in any case \cite{gubser2010-1}.\footnote{This is because the $\boldsymbol{Z}_2$ symmetry demands $u^\eta(-\eta)=-u^\eta(\eta)$ and the longitudinal boost invariance demands the opposite.} If the system is nonrotating ($\phi$ independent and symmetric under $\phi\to -\phi$), we thus end up with
\begin{eqnarray}\label{A30}
u^\mu = \left(u^\tau,0,0,u^r\right).
\end{eqnarray}
Using  \eqref{A30} and  \eqref{A6}, we arrive, in particular, at
\begin{eqnarray}\label{A31}
F_{\tau r}=0.
\end{eqnarray}
Moreover, for the metric \eqref{A26}, \eqref{A12} turns out to be
\begin{eqnarray}\label{A32}
F_{\phi r}&=&r\tau Q(\tau,r)u^\tau f(\vartheta)h(\vartheta),\nonumber\\
F_{\eta r}&=&r\tau Q(\tau,r)u^\tau h(\vartheta).
\end{eqnarray}
Plugging first \eqref{A6} into \eqref{M10}, and using \eqref{A30}, the general solution of $B^{\mu}$ reads
\begin{equation}\label{A33}
B^\mu = \inv{u^\tau r\tau}\left(0,F_{\eta r},-F_{\phi r},0\right).
\end{equation}
Plugging then \eqref{A32} into \eqref{A33} leads to
\begin{equation}\label{A34}
B^\mu = Q(\tau,r)\left(0,h(\vartheta),-f(\vartheta)h(\vartheta),0\right).
\end{equation}
As concerns the electric current $J^\mu$, we use (\ref{M3}) to arrive at \begin{eqnarray}\label{A35}
	\hspace{-1cm}J^\mu &=& \inv{\sqrt{-g}}\partial_\nu\left(\sqrt{-g}F^{\mu\nu}\right)\nonumber\\
	\hspace{-1cm}&=& \left(\pderivv{\tau}+\inv{\tau}\right)F^{\mu \tau}+\left(\pderivv{r}+\inv{r}\right)F^{\mu r}.
	\end{eqnarray}
\section{The $\boldsymbol{3+1}$ dimensional self-similar flow in relativistic MHD}\label{sec4}
\setcounter{equation}{0}
A $3+1$ dimensional generalization of self-similar flows \cite{csorgo2002-2} was introduced in \cite{csorgo2002}. In this case, the four-velocity can be shown to respect rotational invariance around and boost invariance along the $x,y$ and $z$ directions. More restricting symmetries such as the spherical, cylindrical, and elliptical symmetries are introduced by the assumption of different similarity variables. In this section, we first present an alternative derivation of the self-similar flow from \cite{csorgo2002, csorgo2002-2}, that fits our purposes, and then implement it into ideal MHD, where, in particular, the self-similar solution of $B^{\mu}$ is presented.
\par		
To fix the four-velocity profile, we introduce the following similarity variable that commutes with the Killing vectors of \eqref{A28}
\begin{eqnarray}\label{S1}
\vartheta\equiv\frac{r}{\tau}.
\end{eqnarray} 	
Assuming the similarity variable $\vartheta$ to be proper (i.e. $\tder{\vartheta}=0$),  \eqref{A30} takes the form
\begin{eqnarray}\label{S2}
u^\mu = \left(\frac{\tau}{\sqrt{\tau^2-r^2}},0,0,\frac{r}{\sqrt{\tau^2-r^2}}\right).
\end{eqnarray}
A crucial point is that the combination $\varrho\equiv\sqrt{\tau^2-r^2}$ respects, apart from symmetries of \eqref{A28}, an extra symmetry represented by
\begin{eqnarray}\label{S3}
\xi_{03}=\tau\pderiv{}{r}+r\pderiv{}{\tau}.
\end{eqnarray}
Here, $\xi_{03}$ can be regarded as a boost in the radial $r$ direction (hereafter radial boost). Similar to \eqref{A14} for the Bjorken flow, the four-velocity profile can be written as,\footnote{One may call $\tau$ and $\varrho$ invariant scalars of the Bjorken and $3+1$ dimensional self-similar flows, respectively.}
\begin{equation}\label{S4}
u_\mu = - \frac{\partial_\mu\varrho}{\sqrt{-\partial_\mu\varrho\partial^\mu\varrho}}.
\end{equation}
Identifying $\varrho$ with the proper time, the proper acceleration $a^{\mu}=\tder u^{\mu}$ vanishes. Moreover, the covariant divergence of four-velocity is given by
\begin{eqnarray}\label{S5}
\cd{\mu}{u^\mu}=\frac{3}{\varrho}.
\end{eqnarray} 	
Physical quantities can be regarded as functions of $\varrho$ and $\vartheta$, instead of  $r$ and $\tau$. For a scalar function $f(r,\tau)$, the covariant proper time derivative is simply given by
\begin{eqnarray}\label{S6}
\tder{f}=\pderiv{f}{\varrho}.
\end{eqnarray}
Using \eqref{S5} and \eqref{S6}, the solution of \eqref{A10} for the $3+1$ dimensional self-similar flow is found to be
\begin{eqnarray}\label{S7}
Q(\tau,r) = Q_0 \left(\frac{\varrho_0}{\varrho}\right)^3 \mathcal{Q}(\vartheta).
\end{eqnarray}
Here, $Q_0\equiv Q(\tau_0,r_0)$, $\varrho_0\equiv \sqrt{\tau_0^2-r_0^2}$ and $\mathcal{Q}$ is an arbitrary differentiable function of $\vartheta$, referred to as the scaling function of $Q$ \cite{csorgo2002,csorgo2002-2}. Plugging, at this stage, \eqref{M18}, \eqref{S5} and \eqref{S6} into the energy equation \eqref{M15} gives rise to
\begin{eqnarray}\label{S8}
\pderiv{\epsilon}{\varrho}+ 3\frac{(1+\kappa)}{\kappa}\frac{\epsilon}{\varrho}=0.
\end{eqnarray}
The solution of this equation yields the $\varrho$ dependence of $\epsilon$. As concerns its $\vartheta$ dependence, we consider the Euler equation \eqref{M18}. Bearing in mind that $a^{\mu}$ on the lhs of \eqref{M18} vanishes, we obtain
\begin{eqnarray}\label{S9}
\Delta^{\mu\nu}\partial_\nu p = 0,
\end{eqnarray}
that requires $p$ and $\epsilon$ to be $\vartheta$ independent. The solution to  \eqref{S8} is thus given by
\begin{eqnarray}\label{S10}
\epsilon = \epsilon_0 \left(\frac{\varrho_0}{\varrho}\right)^{3(1+1/\kappa)},
\end{eqnarray} 	
as expected from \cite{csorgo2002,csorgo2002-2}.
\par
At this stage, we are in a position to implement the $3+1$ dimensional self-similar flow into ideal MHD. Plugging first \eqref{S7} into \eqref{A34} leads to
\begin{equation}\label{S11}
B^\mu = Q_0 \left(\frac{\varrho_0}{\varrho}\right)^3\left(0,h(\vartheta),-f(\vartheta)h(\vartheta),0\right),
	\end{equation}
where $\mathcal{Q}(\vartheta)$ is absorbed into $h(\vartheta)$. For further convenience, we introduce two constants $\mathcal{A}_1$ and $\mathcal{A}_2$ and two functions $\mathcal{H}(\vartheta)$ and $\mathcal{F}(\vartheta)$ as
\begin{eqnarray}\label{S12}
\mathcal{A}_1\sqrt{\mathcal{H}(\vartheta)}&\equiv& Q_0h(\vartheta),\nonumber\\ \mathcal{A}_2\sqrt{\mathcal{F}(\vartheta)}&\equiv& -f(\vartheta).
\end{eqnarray}
Using then these definitions, \eqref{S11} reads
\begin{equation}\label{S13}
B^\mu = \left(\frac{\varrho_0}{\varrho}\right)^3\mathcal{A}_1\sqrt{\mathcal{H}(\vartheta)}\left(0,1,\mathcal{A}_2\sqrt{\mathcal{F}(\vartheta)},0\right),
\end{equation}
and the magnitude of the $B^{\mu}$ field is given by,
\begin{equation}\label{S14}
B = \left(\frac{\varrho_0}{\varrho}\right)^3{\mathcal{A}_1}\sqrt{\mathcal{H}(\vartheta)\left(r^2+\tau^2\mathcal{A}_{2}^2\mathcal{F}(\vartheta)\right)}.
\end{equation}
Here, \eqref{M12} and the metric \eqref{A26} are used. To determine $\mathcal{H}(\vartheta)$ and $\mathcal{F}(\vartheta)$ in
\eqref{S13} and \eqref{S14}, we use, as in the case of transverse MHD, the Euler equation (\ref{M17}). For vanishing $a^{\mu}$ on the lhs of (\ref{M17}), we arrive first at
\begin{equation}\label{S15}
\inv{2}\left(u^\mu\pderivv{\varrho}+g^{\mu\nu}\partial_\nu\right)B^2+
\Delta^{\mu\nu}\christoffel{\alpha}{\beta}{\nu}B^\beta B_\alpha=0,
\end{equation}
where \eqref{S9} as well as $\cd{\rho}{B^\rho}=a^\rho B_\rho=0$ \cite{beckenstein1978,hernandez2017},  $B^\rho\partial_\rho=0$ and $u^\mu \partial_\mu = \partial_\varrho$ are used. The second term on the lhs of \eqref{S15} can be simplified using
\begin{equation}\label{S16}
\christoffel{\alpha}{\beta}{\nu}B^\beta B_\alpha = \christoffel{\phi}{\phi}{\nu}r^2\left(B^\phi\right)^2+\christoffel{\eta}{\eta}{\nu}\tau^2\left(B^\eta\right)^2.
\end{equation}	
Here, we can set $\alpha=\beta$, because there is no connection between longitudinal and transverse parts of the metric. Plugging, at this stage, \eqref{S16} into  \eqref{S15} gives rise to
\begin{eqnarray}\label{S17}
\lefteqn{\hspace{-1cm}
\inv{2}\left(u^\mu\pderivv{\varrho}+g^{\mu\nu}\partial_\nu\right)B^2}\nonumber\\
&&+\Delta^{\mu\nu}\left(\delta^r_\nu\,r\left(B^\phi\right)^2+
\delta^\tau_\nu\tau\left(B^\eta\right)^2\right)=0.
		\end{eqnarray}		
In the directions of symmetries, i.e. for $\mu=\{\eta,\phi\}$, \eqref{S17} turns out to be trivial. In the flow directions, i.e. for $\mu=\{r,\tau\}$,  however, we arrive at
\begin{equation}\label{S18}
\inv{2}\left(r\pderivv{\tau}+\tau\pderivv{r}\right)B^{2}+ r\tau\left[\left(B^\phi\right)^2+\left(B^\eta\right)^2\right]=0.
\end{equation}
Using the radial boost symmetry \eqref{S3}, we have
\begin{equation}\label{S19}
\left(r\pderivv{\tau}+\tau\pderivv{r}\right)f(\varrho)=0.
\end{equation}
Plugging \eqref{S19} into \eqref{S18}, and using \eqref{S13}, we obtain
\begin{eqnarray}\label{S20}
\lefteqn{\hspace{-2cm}\inv{2}\left(1-\vartheta^2\right)\left\{[\vartheta^2+
\mathcal{A}_2^2\mathcal{F}(\vartheta)]\deriv{\mathcal{H}}{\vartheta}+\mathcal{A}_2^2\mathcal{H}(\vartheta)\deriv{\mathcal{F}}{\vartheta}\right\}}\nonumber\\
&&\hspace{-1cm}+2\vartheta\mathcal{H}(\vartheta)[1+\mathcal{A}_2^2\mathcal{F}(\vartheta)]=0.
\end{eqnarray}
Similar to the case of transverse MHD, we are therefore left with one equation and two unknown functions ${\mathcal{H}}(\vartheta)$ and ${\mathcal{F}}(\vartheta)$. In contrast to the case of transverse MHD, however, they appear not only in $B^{\mu}$ from \eqref{S13}, but also in $B$ from \eqref{S14}.  Applying, as in the case of transverse MHD, the radial boost symmetry \eqref{S3} to remove the arbitrariness of these functions, we arrive at constant ${\mathcal{H}}$ and ${\mathcal{F}}$. Plugging these constant functions into \eqref{S20}, it reduces to
\begin{eqnarray}\label{S21}
2\vartheta\mathcal{H}\left(1+\mathcal{A}_2^2\mathcal{F}\right)=0.
\end{eqnarray}
For \eqref{S21} to hold, either $\mathcal{H}$ or $1+\mathcal{A}_2^2\mathcal{F}$ must vanish. The latter case is impossible, because, by \eqref{S12}, $\mathcal{F}$ is non-negative. For $\mathcal{H}=0$, we obtain $B^\mu=0$. We conclude that the radial boost symmetry  \eqref{S21} prohibits the existence of a magnetic field in any directions.
Similar to function $\mathcal{Q}$ in \eqref{S7}, $\mathcal{H}$ and $\mathcal{F}$ may also be considered as scaling functions. Although $\mathcal{Q}$ is arbitrary for conserved charges, the magnetic scaling functions are constrained by \eqref{S21}.
\par
In the rest of this section, we present two possible solutions to \eqref{S20}. The first one is found by assuming $J^\mu=0$, in the same spirit of \eqref{A24} in the transverse MHD case. In HICs, such a solution may be regarded as an approximation to  late time hydrodynamical expansion of the QGP, when induced currents are supposed to be exhausted. This solution is referred to as the stationary or ZCSSF solution. Another interesting solution is found by assuming $\mathcal{H}=1$. This assumption implies $B^\phi$, which translates into $B_y$ and $B_x$, to not change between flow lines at every fixed proper time. This solution, in contrast to the zero current one, turns out to be regular at $r=0$, and is thus referred to as the regular self-similar solution.
\subsection{Zero current self-similar solution}\label{sec4A}
Let us start by considering $J^{\mu}$ from \eqref{A35}. As it turns out, for self-similar flow $u^{\mu}$ from \eqref{S2} with vanishing $u^{\phi}$ and $u^{\eta}$, the $\tau$ and $r$ components of $J^{\mu}$ vanish. We are therefore left with its $\phi$ and $\eta$ components,
\begin{eqnarray}\label{S22}
\hspace{-1cm}J^\phi&=&\inv{r^2\tau}\left(r\pderivv{\tau}+\tau\pderivv{r}\right)F_{13}-\frac{F_{13}}{r^3},\nonumber\\
\hspace{-1cm}J^\eta&=&\inv{\tau^3}\left(r\pderivv{\tau}+\tau\pderivv{r}\right)F_{23}+\frac{\tau^2-2r^2}{r\tau^4}F_{23}.
\end{eqnarray}
The components of $F_{\mu\nu}$, arising in \eqref{S22}, are found by first plugging \eqref{S13} into \eqref{M8}, and then using \eqref{A8}. This results in
\begin{eqnarray}\label{S23}
\hspace{-1cm}J^\phi&=&-\frac{\mathcal{A}_1\mathcal{A}_2\varrho_0^3}{\varrho^4\sqrt{\mathcal{F}\mathcal{H}}}\left[2+\frac{1-\vartheta^2}{2\vartheta}\derivv{\vartheta}\left(\mathcal{F}\mathcal{H}\right)\right],\nonumber\\
\hspace{-1cm}J^\eta&=&\frac{\mathcal{A}_1\varrho_0^3}{\varrho^4\sqrt{\mathcal{H}}}\left[2\mathcal{H}+\inv{2}\vartheta(1-\vartheta^2)\deriv{\mathcal{H}}{\vartheta}\right].
\end{eqnarray}
The solution to \eqref{S20} that eliminates the current is thus given by
\begin{equation}\label{S24}
\mathcal{H}=\frac{(1-\vartheta^2)^2}{\vartheta^4},\quad\mathcal{F}=\vartheta^4.
\end{equation}
Plugging $\mathcal{H}$ and $\mathcal{F}$ from \eqref{S24} into \eqref{S13} and \eqref{S14}, we arrive at
\begin{eqnarray}\label{S25}
B^\mu =\mathcal{B} \left(\frac{\varrho_0}{\varrho}\right)^2\left(0,\frac{\varrho_0}{r^2},\mathcal{A}_2\frac{\varrho_0}{\tau^2},0\right),
\end{eqnarray}
and
\begin{eqnarray}\label{S26}
B = \mathcal{B} \left(\frac{\varrho_0}{\varrho}\right)^2\sqrt{\frac{\varrho_0^2}{r^2}+\mathcal{A}_{2}^2\frac{\varrho_0^2}{\tau^2}}.
\end{eqnarray}
Here, $\mathcal{B}\equiv\mathcal{A}_1\varrho_0$ is a constant with the dimension of a magnetic field. In the limit  $r\to 0$ and $r=\tau$, $B$ blows up. Let us notice that in a $3+1$ dimensional self-similar flow, other thermodynamic quantity such as the entropy and energy densities as well as the temperature are also proportional to $\rho^{-1}$ and  blow up at $r=\tau$.
Transforming $B^\mu$ from \eqref{S25} back into the Minkowski coordinate system $(t,x,y,z)$, it is given by
\begin{equation}\label{S27}
B^{\mu}=\mathcal{B} \left(\frac{\varrho_0}{\varrho}\right)^2\left(\mathcal{A}_2\frac{\varrho_0}{\tau}\sinh\eta,-\frac{y\varrho_0}{r^2},\frac{x\varrho_0}{r^2},\mathcal{A}_2\frac{\varrho_0}{\tau}\cosh\eta\right).
\end{equation}
Hence, $\mathcal{A}_2$ turns out to be proportional to $B_z/B$, with $B_z$ being the $z$ component of $\boldsymbol{B}$ from \eqref{M11} in the LRF of the fluid.
\subsection{Regular self-similar solution}\label{sec4B}
Let us now consider \eqref{S20} again. Plugging $\mathcal{H}=1$ into this equation gives rise to
\begin{equation}\label{S28}	
\inv{2}\mathcal{A}_2^2(1-\vartheta^2)\deriv{\mathcal{F}}{\vartheta}+2\vartheta\left(1+\mathcal{A}_2^2\mathcal{F}\right)=0,
\end{equation}
whose solution is given by
\begin{equation}\label{S29}
\mathcal{F}(\vartheta)=\mathcal{F}(0)(1-\vartheta^2)^2-\frac{\vartheta^2}{\mathcal{A}_2^2}(2-\vartheta^2).
\end{equation}	
Plugging \eqref{S29} into \eqref{S14} at the point $(\tau_0,r_0)=(\tau_0,0)$, we obtain
\begin{equation}\label{S30}
B_0=\mathcal{A}_1\mathcal{A}_2\varrho_0\mathcal{F}(0).
\end{equation}
Here, $\varrho_0=\sqrt{\tau_0^2-r_0^2}=\tau_0$ and $B_0=B(\tau_0,r_0=0)$. Let us assume, without loss of generality, $\mathcal{F}(0)=1$. We thus get $\mathcal{A}_1=\mathcal{A}_2\varrho_0/B_0$. Plugging $\mathcal{A}_{1}$ into
\eqref{S25}, we arrive at the regular self-similar solution for $B^{\mu}$
\begin{eqnarray}\label{S31}
\lefteqn{\hspace{-0.8cm}B^\mu = B_0 \left(\frac{\varrho_0}{\varrho}\right)^2}\nonumber\\
&&\hspace{-0.5cm}\times \left(0,\frac{\sqrt{a_0^2-1}}{\varrho},\inv{\tau^2}\sqrt{\varrho^2-\frac{r^2(\tau^2+\varrho^2)}{\varrho^2(a_0^2-1)}},0\right),
\end{eqnarray}
with
\begin{eqnarray}\label{S32}
B = B_0 \left(\frac{\varrho_0}{\varrho}\right)^2 \sqrt{1-a_0^2\left(\frac{r}{\tau}\right)^2}.
\end{eqnarray}
Here, $a_0^2 \equiv \frac{\mathcal{A}_2^2+1}{\mathcal{A}_2^2}$. The radial domain of above solution is $0\leq r < \frac{\tau}{a_0}$. On the other hand, $a_0$ is related to the relative strength of magnetic field in transverse directions compared to longitudinal ones,
\begin{equation}\label{S33}
a_0 = \sqrt{1+\left(\frac{B^\phi_0}{B^\eta_0}\right)^2}.
\end{equation}
Here, $B^\phi_0\equiv B^\phi(\tau_0,0)$ and $B^\eta_0\equiv B^\phi(\tau_0,0)$. As mentioned above, the radial domain of  \eqref{S31}, does not cover the whole radial domain of self-similar flow $r\leq\tau$.  At any value of $\tau$, the magnetic field exists only in a circle of radius $r^\star=\tau/a_0$. The value of $\int r\rd{r}B$ is constant within this circle. Moreover, $B$ exactly vanishes at $r=\tau/a_0$. It is also possible to show that $B_y=rB^\phi$ at point $(t,x,y,z)=(\tau,r,0,0)$. For the solution of  \eqref{S31}, $B_y=0$ at $r=0$. These kinds of properties are not relevant in the QGP context. We thus exclude this solution from the discussion in Sec. \ref{sec7}.
\par
Let us notice, at this stage, that other solutions can also be found for \eqref{S20}. For example, we may assume
\begin{eqnarray}\label{S34}
\mathcal{H}(\vartheta)=\inv{\vartheta^2}\exp\left(-\frac{\vartheta^2}{2b^2}\right).
\end{eqnarray}
Here, $b$ is a constant. Plugging  \eqref{S34} into  \eqref{S20}, one is able to find $\mathcal{F}$, which contains exponential integral functions, and becomes negative as $\vartheta$ tends to unity.
We may alternatively assume $\mathcal{F}$ to be unity, and find
\begin{eqnarray}\label{S35}
\mathcal{H}(\vartheta)=\frac{(1-\vartheta^2)^2}{(\mathcal{A}_2^2+\vartheta^2)^2}.
\end{eqnarray}
The radial domain of this solution is highly restricted too. Moreover, the corresponding electric current does not vanish, in contrast to $J^{\mu}$ arising from the ZCSSF solution \eqref{S25}. In the rest of this work we focus on this solution, which is nicely related to the Bjorken and Gubser solutions, whose corresponding currents also vanish.
\section{Gubser flow in relativistic MHD}\label{sec5}
\setcounter{equation}{0}
The Gubser flow was first introduced in \cite{gubser2010-1} and then, using a different approach, rederived in \cite{gubser2010-2}.
In this section, motivated by the approach presented in \cite{gubser2010-1}, we mainly focus on its realization in relativistic MHD.
In Sec. \ref{sec5A}, we first derive its symmetries in a rather different way than was presented in \cite{gubser2010-1}. These symmetries are then applied to MHD, and lead eventually to the evolution of the magnetic field in this setup (see Sec. \ref{sec5B}).
\subsection{Gubser flow and its symmetries}\label{sec5A}
As aforementioned, the Bjorken four-velocity \eqref{A14} can be fixed by considering three symmetries\footnote{For simplicity, the first vector is given in \eqref{A13} parameterization, while the second one in \eqref{A26}.}
\begin{eqnarray}\label{O1}
\xi_{x}=\pderivv{x},\quad\xi_{\phi}=\pderivv{\phi},\quad\xi_{\eta}=\pderivv{\eta}.
\end{eqnarray}
The translational invariance in the $y$-direction, i.e. $\pderivv{y}$, is found by commutating $\xi_{x}$ and $\xi_{\phi}$. Using the Jacobi identity, it can be shown to be  a symmetry as well \cite{generalrefs}. The assumption of translational invariance in the transverse $x$-$y$ plane, as in the Bjorken flow, implies the fluid transverse size to be infinitely large. Bjorken assumed that in the central rapidity region, where $\eta\approx 0$, hydrodynamic equations respect the symmetries of \eqref{O1}. In particular, he assumed that close to the center of collisions, there exists a region in the transverse plane where fluid elements are not affected by the finite size of the system \cite{bjorken1983}. This can be interpreted as if in this region the mean free path of fluid constituents are almost zero so that they are not \textit{aware} of the fluid finite size.  According to this picture, the size of this \textit{unawareness} region shrinks with the sound velocity as the system evolves. The mean free path thus increases with time, and the system becomes diluted. There are, however, experimental signals that suggest the early existence of a transverse expansion (for a discussion on these signals, see \cite{rajagopal2011} and the references therein). The Gubser flow takes this early transverse expansion into account.
\par
The Gubser's approach is to replace the Killing vectors associated with the translational invariance in \eqref{O1}, with \textit{weaker} symmetries that consider the finite transverse size. The Bjorken symmetries \eqref{O1} cover all Killing vectors of $M^{3,1}$ that may be appropriate in this context. To expand the number of available symmetries, one extends to the conformal group of $M^{3,1}$, and, instead of Killing vectors associated with the aforementioned translational invariance, considers appropriate conformal Killing vectors that satisfy conformal Killing equation \eqref{appA9} from Appendix \ref{appA}. In addition, such conformal Killing vectors must
	\begin{enumerate}
		\item\label{req_depend_on_q} depend on the typical transverse size of the system, $L$,
		\item\label{req_commute_with_eta} commute with $\partial_\eta$,
		\item\label{req_q_0_lim} reduce to $\partial_x$ and $\partial_y$ as $L\to\infty$,
		\item\label{req_x_0_lim} and, finally, reduce to $\partial_x$ and $\partial_y$ as $\tau\to 0$ and $r\to 0$.
	\end{enumerate}
Here, $r,\tau$ and $\eta$ are coordinates defined in \eqref{A27}.  For simplicity, we introduce  a quantity $q\sim L^{-1}$ having energy dimension. Let us consider $\xi_x\equiv\partial_x$ in \eqref{A27} coordinates
	\begin{equation}\label{O2}
	\xi_x = \cos\phi\pderivv{r}-\frac{\sin\phi}{r}\pderivv{\phi}.
	\end{equation}
	Let $\zeta$ be the conformal Killing vector that replaces $\xi_x$. By requirement  \ref{req_commute_with_eta}, components of $\zeta$ are found to be $\eta$ independent.
	Using  \eqref{appA9} with $\nu=\eta$, one immediately finds ${\zeta}^\eta=0$, and thus
\begin{equation}\label{O3}
\cd{\alpha}{\zeta^\alpha} = \frac{4}{\tau}\zeta^\tau.
\end{equation}
Equation \eqref{O3} and ${\zeta}^\eta=0$ ensures  \eqref{appA9} for metric components with $\mu=\eta$ and/or $\nu=\eta$. Using \eqref{appA7}, \eqref{appA9} is rewritten as
\begin{equation}\label{O4}
\cd{\mu}{\zeta_\nu}+\cd{\nu}{\zeta_\mu}=\frac{2}{\tau}\zeta^\tau g_{\mu\nu}.
\end{equation}
For $(\mu,\nu)=(\tau,\tau)$, \eqref{O4} gives rise to
\begin{equation}\label{O5}
\tau\pderiv{\zeta^\tau}{\tau}=\zeta^\tau,
\end{equation}
whose solution is given by
\begin{equation}\label{O6}
\zeta^\tau = \tau A(r,\phi).
\end{equation}
This leads immediately to $\zeta^\tau=0$ at $\tau=0$. Using the fact that $\zeta^\tau$ has also to vanish at $r=0$ and $q=0$, we arrive at $\zeta^{\tau}=q^2r\tau A(\phi)$, with $A(\phi)$ a function of $\phi$, which is to be determined. The other two components of $\zeta$, $\zeta^\phi$ and $\zeta^r$, must reduce to components of \eqref{O2} at $r=\tau=0$ and $q=0$. It is thus reasonable to assume
\begin{eqnarray}\label{O7}
\zeta&=&q^2 \tau r A(\phi)\pderivv{\tau}+[1+q^b B(\tau,r)]\cos\phi\pderivv{r}\nonumber\\
&&-[1+q^c C(\tau,r)]\frac{\sin\phi}{r}\pderivv{\phi},
	\end{eqnarray}
with $b$ and $c$ being positive constants, and functions $B(\tau,r)$ and $C(\tau,r)$ vanishing at $\tau=r=0$. These functions are determined by plugging \eqref{O7} into \eqref{O4}. For $(\mu,\nu)=(\tau,\phi)$ and $(\mu,\nu)=(\tau,r)$, we arrive  at
\begin{eqnarray}\label{O8}
q^2\tau A'(\phi) &=& -q^c\sin\phi\pderiv{C}{\tau},\nonumber\\ q^2\tau A(\phi) &=& q^b\cos\phi\pderiv{B}{\tau},
	\end{eqnarray}
respectively. An immediate result of \eqref{O8} is that $b=c=2$. Bearing in mind that $B(\tau,r)$ and $C(\tau,r)$ are functions of $\tau$ and $r$, and that $A(\phi)$ depends only on $\phi$, (\ref{O8}), we obtain
\begin{eqnarray}\label{O9}
A(\phi)&=&A\cos\phi,\nonumber\\
B(\tau,r)&=&\frac{A}{2}\tau^2+\bar{B}(r),\nonumber\\
C(\tau,r)&=&\frac{A}{2}\tau^2+\bar{C}(r),
\end{eqnarray}
with $A$ being a constant, and $\bar{B}$ as well as $\bar{C}$ two unknown functions depending only on $r$. They are determined by plugging \eqref{O9} into \eqref{O7}. This leads to
\begin{eqnarray}\label{O10}
\zeta&=&Aq^2 \tau r \cos\phi\pderivv{\tau}+(1+\frac{A}{2}q^2\tau^2+q^2B(r) )\cos\phi\pderivv{r}\nonumber\\
&&-(1+\frac{A}{2}q^2\tau^2+q^2C(r))\frac{\sin\phi}{r}\pderivv{\phi}.
\end{eqnarray}
Plugging, at this stage, \eqref{O10} into \eqref{O4} with $(\mu,\nu)=(r,r)$ and $(\mu,\nu)=(\phi,\phi)$, we then obtain
\begin{equation}\label{O11}
Ar^2-\bar{B}(r)+\bar{C}(r)=0,\quad \bar{B}'(r)=Ar.
\end{equation}
Solving \eqref{O11}, we finally end up with
\begin{eqnarray}\label{O12}
\zeta&=&Aq^2 \tau r \cos\phi\pderivv{\tau}+\left(1+\frac{A}{2}q^2(\tau^2+r^2) \right)\cos\phi\pderivv{r}\nonumber\\
&&-\left(1+\frac{A}{2}q^2(\tau^2-r^2)\right)\frac{\sin\phi}{r}\pderivv{\phi},
\end{eqnarray}
that satisfies \eqref{appA9}. Here, $A$ remains an arbitrary constant. It can be absorbed into $q$. Setting, however, $A=2$, the vector introduced in \cite{gubser2010-1} is found. It reads
\begin{eqnarray}\label{O13}
\zeta&=&2q^2\tau r\cos\phi\pderivv{\tau}+[1+q^2\left(\tau^2+r^2\right)]\cos\phi\pderivv{r}\nonumber\\
&&-\frac{[1+q^2\left(\tau^2-r^2\right)]}{r}\sin\phi\pderivv{\phi}.
\end{eqnarray}
Having $\zeta$ in hand, it is possible to determine the other symmetry of the Gubser flow, that replaces $\partial_y$ of the Bjorken flow. It is found from $\zeta'=[\zeta,\partial_\phi]$. It is then easy to check that $\partial_\phi$, $\zeta$ and $\zeta'$ satisfy the $SO(3)$ algebra $[\xi_i,\xi_j] \sim \xi_k$, and can be regarded as generators of this  group.
\par
In what follows, we use appropriate symmetry arguments, and show that the Gubser flow is given by
\begin{equation}\label{O14}
u^\mu = \left(\cosh\varTheta,0,0,\sinh\varTheta\right),
\end{equation}
with
\begin{equation}\label{O15}
\tanh\varTheta=\frac{2q^2r\tau}{1+q^2\left(\tau^2+r^2\right)}.
\end{equation}
To do this, let us consider
\begin{equation}\label{O16}
\lieder{\zeta}{\boldsymbol{X}}=-\frac{a_{\mbox{\tiny{$X$}}}}{4}\left(\cd{\lambda}{{\zeta^\lambda}}\right)\boldsymbol{X},
\end{equation}
with $\boldsymbol{X}$ being an arbitrary rank tensor with $\zeta$-weight equal to $a_{\mbox{\tiny{$X$}}}$ \cite{gubser2010-1}. Here, $a_{\mbox{\tiny{$X$}}}$ is a constant number. Moreover, for $\zeta$ from \eqref{O13}, we have
\begin{equation}\label{O17}
\cd{\lambda}{{\zeta^\lambda}} = 8q^2r\cos\phi.
\end{equation}
Before proving \eqref{O14}, let us first consider a number of relevant examples for the $\zeta$-weight $a_{\mbox{\tiny{$X$}}}$ of an arbitrary rank tensor $\boldsymbol{X}$. Plugging, for instance, the metric into \eqref{O16}, it turns out that it has a $\zeta$-weight equal to $a_g=-2$. Moreover, whereas the transverse coordinates, $r$ and $\phi$, do not have any well-defined $\zeta$-weights, the $\zeta$-weights for the longitudinal coordinates, $\tau$ and $\eta$, are given by $a_\tau=-1$ and $a_{\eta}=0$. They arise from
\begin{equation}\label{O18}
\lieder{\zeta}{\tau} = \frac{1}{4}\left(\cd{\lambda}{{\zeta^\lambda}}\right)\tau,\quad\mbox{and}\quad\lieder{\zeta}{\eta} =0,
\end{equation}
respectively. In addition to $\eta$, the only combination of coordinates with zero $\zeta$-weight turns out to be \cite{gubser2010-1}
\begin{equation}\label{O19}
G\equiv\frac{1-q^2\left(\tau^2-r^2\right)}{2q\tau}.
\end{equation}
Let us now turn back to the Gubser flow \eqref{O14}. To show it, one should bear in mind that $u_{\mu}$ is a hydrodynamical variable, and as such, it has a well-defined $\zeta$-weight. Moreover, it also respects boost and rotational symmetries. In other words, it  commutes with the Killing vectors from \eqref{A28}. It is thus constrained to be a function of $\tau$ and $G$.
To determine the $\zeta$-weight of $u^{\mu}$, let us remind that the transverse projector $\Delta_{\mu\nu}$ should have the same weight as the metric, i.e. $a_{\Delta}=-2$. This is satisfied if $u_\mu$ has a $\zeta$-weight equal to $a_{u_{\mu}}=-1$. From $u_\mu u^\mu = -1$, one then finds $u^\mu$'s $\zeta$-weight to be given by $a_{u^{\mu}}=+1$. Using, at this stage, \eqref{O16} for $u^{\mu}$ with $a_{u^{\mu}}=+1$ and the $\boldsymbol{Z}_2$ symmetry, we arrive at Gubser four-velocity (\ref{O14}). Alternatively, $u_{\mu}$ is given by \cite{dewolfe2014}
\begin{equation}\label{O20}
u_\mu=\frac{\partial_\mu G}{\sqrt{-\partial_\mu G\partial^\mu G}}.
\end{equation}	
Similar relations are also found for the Bjorken and self-similar flows in \eqref{A14} and  \eqref{S4}. In contrast to these flows, however, it turns out that one cannot introduce any proper similarity variable for the Gubser flow without destroying the corresponding symmetry constraints. This is because the only proper scalar in this setup,
\begin{equation}\label{O21}
\Lambda\equiv\frac{2qr}{1-q^2\left(r^2-\tau^2\right)},
\end{equation}
does not have any well-defined $\zeta$-weight, and cannot therefore be used as a proper similarity variable.
\par
Let us notice, at this stage, that using the general solution of (\ref{O16}),
\begin{equation}\label{O22}
\boldsymbol{X} = \frac{\tilde{\boldsymbol{X}}(G)}{\tau^a},
\end{equation}
with $G$ from \eqref{O19}), it is possible to determine the $(\tau,G)$ dependence of other hydro- and thermodynamical variables.\footnote{In the rest of this paper, quantities with ``tilde'' are defined to be functions of $G$.} Plugging, e.g., \eqref{O22} with $\boldsymbol{X}=\epsilon$ into \eqref{M15}, and using the EOS,\footnote{To respect conformal invariance the energy-momentum tensor must be traceless. This implies the EOS to be given by \eqref{O23}.}
\begin{equation}\label{O23}
\epsilon = 3p,
\end{equation}
leads to
\begin{equation}\label{O24}
\epsilon = \frac{\bar{\epsilon}_0}{\tau^4\left(1+G^2\right)^{4/3}},
\end{equation}
where $\bar{\epsilon}_0$  is an arbitrary integration constant.\footnote{Here, $\bar{\epsilon}_0\neq \epsilon(\tau_0,0)$.} Alternatively, we may use \eqref{appC18} for $\epsilon(\tau)$. By the magic of symmetries, the energy density and pressure automatically satisfy the Euler equation.
\subsection{Magnetic field from the Gubser flow in relativistic MHD}\label{sec5B}
Let us now turn back to the implementation of the Gubser flow into relativistic MHD. This eventually leads to a $(\tau,G)$ dependence of the magnetic field. Let us notice that such a formulation is possible, because the electromagnetic part of energy-momentum tensor \eqref{M7} is traceless. It thus respects the conformal invariance \cite{generalrefs}.
\par
Let us first consider
\begin{equation}\label{O25}
B\cdot\partial = B^\phi\pderivv{\phi}+B^\eta\pderivv{\eta},
\end{equation}
that arises from the application of rotational and boost symmetries from \eqref{A28}.
In the index free notation, the Lie derivative of the magnetic four-vector with respect to $\zeta$ from \eqref{O13} is thus given by
\begin{widetext}
\begin{eqnarray}\label{O26}\hspace{-1cm}
[\zeta,B\cdot\partial]&=& 2q^2r\tau B^\phi\sin\phi\pderivv{\tau}
+\frac{\cos\phi}{r}\left\{[1+q^2\left(\tau^2-r^2\right)]B^\phi+r\left([1+q^2\left(\tau^2+r^2\right)]\pderiv{B^\phi}{r}+2q^2r\tau\pderiv{B^\phi}{\tau}\right)\right\}\pderivv{\phi}\nonumber\\
&&+\cos\phi\left\{[1+q^2\left(\tau^2+r^2\right)]\pderiv{B^\eta}{r}+2q^2r\tau\pderiv{B^\eta}{\tau}\right\}\pderivv{\eta}
+[1+q^2\left(\tau^2+r^2\right)]B^\phi\sin\phi\pderivv{r}.
\end{eqnarray}
\end{widetext}
On the other hand, according to \eqref{O16}, we have $[\zeta, B\cdot \partial]\propto B\cdot\partial$ with $B\cdot\partial$ from \eqref{O25}. This implies the $\tau$ and $r$ components of \eqref{O26} to be vanishing. We therefore arrive at
\begin{equation}\label{O27}
B^\phi = 0.
\end{equation}
Using, at this stage, the general solution of \eqref{O16} from \eqref{O22}, the formal solution of $B^\mu$ is given by
\begin{equation}\label{O28}
B^\mu = \frac{1}{\tau^{a_{\mbox{\tiny{$B$}}}}}(0,0,\tilde{B}^\eta(G),0),
\end{equation}
where $a_{\mbox{\tiny{$B$}}}$ is the $\zeta$-weight of the magnetic four-vector $B^{\mu}$ and $\tilde{B}^\eta$ is a scalar function of $G$ from (\ref{O19}). To determine $a_{\mbox{\tiny{$B$}}}$, let us consider the total energy-momentum tensor $T^{\mu\nu}$ from \eqref{M5}. Using \eqref{O24}, the $\zeta$-weight of the energy density turns out to be $a_{\epsilon}=+4$. Bearing in mind that $a_{u^{\mu}}=+1$, the $\zeta$-weight of $T^{\mu\nu}$ turns out to be $a_{\mbox{\tiny{$T$}}}=+6$ [see \eqref{M6}].
Plugging, on the other hand, $F^{\mu\nu}\propto B_{\rho}$ from \eqref{M8} with $E^{\mu}=0$ into \eqref{M7}, it turns out that $B^\mu$ shows up in a $B^\mu B^\nu$ combination in the energy-momentum tensor. Its $\zeta$-weight is thus given by $a_{\mbox{\tiny{$B$}}}=+3$. This immediately leads to
\begin{equation}\label{O29}
B^\mu = \frac{1}{\tau^3}(0,0,\tilde{B}^\eta(G),0).
\end{equation}
Comparing, at this stage, \eqref{O29} with the general solution \eqref{A34} for the $B^{\mu}$ field, and bearing in mind that the Gubser's setup does not  comprise any similarity variable, the functions $f$ and $h$ in \eqref{A34} turn out to be constant. Hence, the only nonvanishing component of $B^{\mu}$ is $B^{\eta}$, and (\ref{A34}) thus reduces to
	\begin{equation}\label{O30}
	B^\eta = - \mathcal{A} Q,
	\end{equation}
with $\mathcal{A}$ being a constant and $Q$ the solution to \eqref{A10} for the Gubser flow \eqref{O14}. To find $Q$, we first notice that, according to \eqref{O30},  $a_{\mbox{\tiny{$Q$}}}=a_{\mbox{\tiny{$B$}}}=+3$. We thus have
\begin{equation}\label{O31}
Q=\frac{1}{\tau^3}\tilde{Q}(G).
\end{equation}
Plugging  \eqref{O31} into \eqref{A10}, and using  \eqref{appA2} as well as \eqref{O14}, we arrive first at
	\begin{equation}\label{O32}
	\pderivv{\tau}\left(r\cosh\varTheta\frac{\tilde{Q}(G)}{\tau^2}\right)+\pderivv{r}\left(r\sinh\varTheta\frac{\tilde{Q}(G)}{\tau^2}\right)=0,
	\end{equation}
that leads to
\begin{equation}\label{O33}
2G\tilde{Q}(G)+(1+G^2)\deriv{\tilde{Q}}{G}=0.
\end{equation}
Here, the definition of $G$ from (\ref{O19}) is used. Solving \eqref{O33} results in
\begin{equation}\label{O34}
Q = \frac{\bar{Q}_0}{\tau^3(1+G^2)},
\end{equation}
where $\bar{Q}_{0}$ is an arbitrary integration constant. Plugging finally \eqref{O34} into  \eqref{O30} leads to
\begin{equation}\label{O35}
B^\mu = (0,0,B^{\eta},0),\quad\mbox{with}\quad B^{\eta}=-\frac{\bar{B}_0}{\tau^3(1+G^2)}.
\end{equation}
Here, $\bar{B}_0\equiv \mathcal{A}\bar{Q}_0$ and $G$ is given in (\ref{O19}). Using  \eqref{O35}, \eqref{O19}, and the metric \eqref{A26}, the magnitude of magnetic field thus reads
\begin{equation}\label{O36}
B = \frac{4q^2\bar{B}_0}{1+2q^2(\tau^2+r^2)+q^4(\tau^2-r^2)^2}.
\end{equation}
As concerns the Euler equation, plugging \eqref{O35} and \eqref{O36} into  \eqref{M17}, and using  \eqref{O23} as well as \eqref{O24}, it turns out to be automatically satisfied.
\par
At this stage, a number of remarks are in order. As we have shown, in the above method of the MHD realization of the Gubser flow, only the longitudinal $z$ component of the magnetic field survives.\footnote{According to \eqref{appC3}, in the LRF of the fluid  $B_x=0, B_y=rB^{\phi}=0$ and $B_{z}=\tau B^{\eta}\neq 0$, with $B^{\eta}$ given in \eqref{O35}.} In HICs, however, the created magnetic field is believed to be aligned in the transverse $x$-$y$ directions, while its longitudinal components are reported to be small \cite{warringa2007,Huang-review,skokov2009, zakharov2014}. Although the elimination of transverse components in $B^{\mu}$ from \eqref{O35} is not a feature of HICs, one may get some insights about the longitudinal component of the magnetic field in this approach. The first point is that the existence of a longitudinal component is controlled by the finiteness of the transverse size, i.e. by taking the limit $q\to 0$ or $L\to \infty$, $B_z$ and $B$ automatically vanish. The second point is that if we consider the ratio $\varsigma\equiv B(\tau,0)/B(\tau,1/q)$, we obtain
\begin{equation*}
\varsigma=\frac{4+q^4\tau^4}{(1+q^2\tau^2)^2}.
\end{equation*}
Whereas at $\tau=0$ we have $\varsigma=4$, $\varsigma$ reduces to a minimum of $4/5$ at $\tau=2/q$, and then asymptotically tends to unity as $\tau\to\infty$. This indicates that $B_z$ becomes spatially homogeneous in late times. The question whether these features are of any relevance for the magnetic fields produced in HICs remains, however, open.
\par
In Secs. \ref{sec3} and \ref{sec4}, we introduced a proper similarity variable, and relaxed at least one of the symmetries of the flow. According to our arguments in the present section, however, such a  similarity variable cannot be defined for the Gubser flow without destroying the corresponding symmetry constraints. We notice that without an appropriate similarity variable, we have to apply all symmetries from RHD to the relativistic MHD. It is exactly this full set of symmetries that prohibits the magnetic field \eqref{O35} to possess transverse components.
In the next section, we slightly modify the alternative approach to the Gubser flow from \cite{gubser2010-2}, and implement it into relativistic MHD. This modification enables us to define an appropriate similarity variable, and relax at least one of the symmetries of the flow. We show that apart from longitudinal components, nonvanishing transverse components of the magnetic field also arise, and, at the same time, the corresponding flow remains preserved. The results presented in the next section are supposed to be more relevant for the magnetic fields created in HICs.
\section{Conformal MHD}\label{sec6}
\setcounter{equation}{0}
In this section, we first start with a brief review of the method presented in  \cite{gubser2010-2}. Then, relaxing a number of symmetries in this setup, we introduce an appropriate similarity variable. We finally generalize our arguments to relativistic MHD, and determine the spacetime dependence of the magnetic field.
\par
Let us consider two spacetimes that are related through a Weyl rescaling
\begin{equation}\label{U1}
{\rd{s}}^2=\Omega(x)^2{\rd{\hat{s}}}^2.
\end{equation}
A physical quantity $\boldsymbol{X}$, being an arbitrary rank tensor, is said to have a conformal weight of $w_{\mbox{\tiny{$X$}}}$ if
\begin{equation}\label{U2}
\boldsymbol{X}=\Omega(x)^{-w_{\mbox{\tiny{$X$}}}}\hat{\boldsymbol{X}}.
\end{equation}
Here, $\hat{\boldsymbol{X}}$ is the tensor in the spacetime associated with ${\rd{\hat{s}}}^2$ from \eqref{U1}.\footnote{In the rest of this paper, quantities with ``hat'' are in the spacetime associated with ${\rd{\hat{s}}}^2$ from \eqref{U1}.} Following the standard practice, we denote the conformal weight of $\boldsymbol{X}$ with $[\boldsymbol{X}]$. All hydrodynamical degrees of freedom, in particular, the four-velocity, have definite conformal weights \cite{gubser2010-2}.\footnote{This is a similar concept like the $\zeta$-weight in Sec. \ref{sec5}.} In contrast, the acceleration $a_{\mu}$ does not have any definite conformal weight. This is why, a nonaccelerating flow in a conformally flat spacetime can transform to an accelerated one in the flat spacetime. Using $\Omega=\tau$ in \eqref{U1}, the four-dimensional Minkowski spacetime is transformed into $dS^{3}\times E^1$ \cite{gubser2010-2}. The corresponding metric is then parameterized as
\begin{equation}\label{U3}
{\rd{\hat{s}}}^2=
-{\rd{\rho}}^2+\cosh^{2}\rho~\sin^{2}\theta~{\rd{\phi}}^2+{\rd{\eta}}^2+
\cosh^{2}\rho~{\rd{\theta}}^2.
\end{equation}
Here,
\begin{eqnarray}\label{U4}
\sinh\rho&=&-\frac{1-q^2\left(\tau^2-r^2\right)}{2q\tau},\nonumber\\
\tan\theta&=&\frac{2qr}{1+q^2\left(\tau^2-r^2\right)}.
\end{eqnarray}
Comparing with the definitions of $G$ and $\Lambda$ in \eqref{O19} and \eqref{O20}, one notices that $G=-\sinh\rho$ and $\Lambda=\tan\theta$.
The corresponding Christoffel symbols to \eqref{U3} read
\begin{eqnarray}\label{U5}
\begin{array}{rclcrcl}
\christoffel{\rho}{\theta}{\theta}&=&\cosh\rho\sinh\rho,&\quad&\christoffel{\rho}{\phi}{\phi}&=&\cosh\rho\sinh\rho\sin^{2}\theta,\\
\christoffel{\theta}{\rho}{\theta}&=&\tanh\rho,&\quad&\christoffel{\theta}{\phi}{\phi}&=&-\cos\theta\sin\theta,\nonumber\\
\christoffel{\phi}{\rho}{\phi}&=&\tanh\rho,&\quad&\christoffel{\phi}{\phi}{\theta}&=&\cot\theta.
\end{array}\\
\end{eqnarray}	
According to the arguments in \cite{gubser2010-2}, a stationary fluid with $SO(3)\times SO(1,1)\times\boldsymbol{Z}_2$ symmetry in $dS^{3}\times E^1$ transforms into the Gubser flow in the Minkowski spacetime $M^{3,1}$. As in previous sections, the Killing vector associated with the $SO(1,1)$ subgroup is $\partial_\eta$. Moreover, it is known  that the $SO(3)$ subgroup of $SO(3)\times SO(1,1)\times\boldsymbol{Z}_2$, which acts on the $S^2$ part of $dS^{3}$ with a  constant $\rho$, contains $\partial_\phi$ and \cite{generalrefs}
\begin{eqnarray}\label{U6}
\xi_{1}&=&\sin\phi\pderivv{\theta}+\cot\theta\cos\phi\pderivv{\phi},\nonumber\\
\xi_{2}&=&\cos\phi\pderivv{\theta}-\cot\theta\sin\phi\pderivv{\phi}.
\end{eqnarray}
Comparing \eqref{U6} with \eqref{O2}, reveals $\xi_{1}$ and $\xi_{2}$ being translations in $S^2$. Using \eqref{U6} and \eqref{A28}, physical quantities are thus functions of $\rho$, that plays the role of $\tau$ in the Bjorken flow. A comparison with the Bjorken case \eqref{A14} gives rise to
\begin{equation}\label{U7}
\hat{u}_\mu = -\frac{\partial_\mu\rho}{\sqrt{-\partial_\mu\rho\partial^\mu\rho}} =(-1,0,0,0).
\end{equation}
Hence, in the coordinates presented by \eqref{U3}, the fluid is stationary. The Gubser flow in the flat spacetime, \eqref{O14}, can be rederived from \eqref{U7} with an appropriate Weyl rescaling together with a coordinate transformation \cite{gubser2010-2}.
\par
At this stage, instead of applying \eqref{U6} and \eqref{A28} to the field strength tensor, and arriving at a constant magnetic field, we relax \eqref{U6}, and introduce a proper similarity variable $\vartheta=\vartheta(\theta)$. This leads to the same velocity profile \eqref{U7}, but physical quantities, in particular, the electromagnetic field strength tensor may acquire $\theta$ dependence. Following the arguments presented in \cite{csorgo2002-2, shokri2017}, it turns out that the $\theta$ dependence of the energy density and pressure are eliminated. However, certain scaling function for the temperature remains (see Appendix \ref{appB} for some more details). In this way, the hydrodynamics of the Gubser flow remain essentially the same as presented in \cite{gubser2010-1,gubser2010-2}.
\par
To generalize the above arguments to MHD, let us bear in mind that the equations of MHD \eqref{M1} and \eqref{M3} are conformal invariant, and that a solution can be transformed between two conformally related spacetimes. Inspecting electromagnetic terms in $T^{\mu\nu}$, one finds
\begin{equation}\label{U8}
[F^{\alpha\beta}] = \frac{d+4}{2},\quad [F_{\alpha\beta}] = \frac{d-4}{2}.
\end{equation}
For $d=4$, we have, in particular, $[F_{\alpha\beta}]=0$. On the other hand, as we have argued before, $B^\mu$ appears in a $B^\mu B^\nu$ combination in the energy-momentum tensor.  Hence,
\begin{equation}\label{U9}
[B^\mu]=\frac{d+2}{2},\quad[B_\mu]=\frac{d-2}{2},\quad[B]=\frac{d}{2}.
\end{equation}
We therefore have
\begin{equation}\label{U10}
B^\mu(x) = \Omega^{-\frac{d+2}{2}}\pderiv{x^\mu}{\hat{x}^\nu}\hat{B}^\nu(\hat{x}).
\end{equation}
In what follows, we first determine the components of $B^{\mu}$ in $dS^{3}\times E^{1}$, where, according to \eqref{U7}, the fluid turns out be stationary. Using \eqref{U10}, we then transform them back into the Minkowski spacetime. To solve MHD equations in the \eqref{U3} spacetime, let us consider the Killing vectors \eqref{A28}, this time in the $dS^3\times E^1$ spacetime. Similar to the transverse MHD setup, since the fluid is stationary, the electric component of the field strength tensor vanish, i.e. $\hat{F}_{0i}=\hat{F}_{i0}=0$, and \eqref{A3} thus leads to
\begin{equation}\label{U11}
\pderiv{\hat{F}_{13}}{\rho}=0,\quad\pderiv{\hat{F}_{23}}{\rho}=0.
\end{equation}
We use the following ansatz for two nonvanishing components of $F_{\mu\nu}$, which turn out to be functions of $\vartheta$,
\begin{equation}\label{U12}
\hat{F}_{23}=\mathcal{A}_1 \sqrt{\mathcal{F}(\vartheta)},\quad \hat{F}_{13} = -\sin\theta \mathcal{A}_2\sqrt{\mathcal{H}(\vartheta)}.
\end{equation}
Here, similar to \eqref{S12}, $\mathcal{A}_1$ and $\mathcal{A}_2$ are constants, and ${\cal{F}}(\vartheta)$ and ${\cal{H}}(\vartheta)$ two unknown scaling functions. Plugging, at this stage, \eqref{U12} into \eqref{M10}, we arrive first at
\begin{equation}\label{U13}
\hat{B}^\mu=\inv{\cosh^{2}\rho\sin\theta}\left(0,\mathcal{A}_1 \sqrt{\mathcal{F}(\vartheta)},\sin\theta \mathcal{A}_2\sqrt{\mathcal{H}(\vartheta)},0\right).
\end{equation}	
To determine the scaling functions ${\cal{F}}(\vartheta)$ and ${\cal{H}}(\vartheta)$, let us consider the Euler equation \eqref{M17}. Here, similar to the case of the Bjorken flow, the fluid is not accelerated. We thus have $\hat{a}^\mu=0$. Being merely a function of $\rho$, the pressure $\hat{p}$ satisfies
\begin{equation}\label{U14}
\hat{\Delta}^{\mu\nu}\partial_\nu\hat{p}=0.
\end{equation}
The Euler equation \eqref{M17} thus reduces to
\begin{eqnarray}\label{U15}
\lefteqn{\frac{1}{2}\hat{\Delta}^{\mu\nu}\partial_{\nu}\hat{B}^{2}=\hat{\Delta}^{\mu\nu}\nabla_{\rho}
\left(\hat{B}_{\nu}\hat{B}^{\rho}\right)}\nonumber\\
&=&-\hat{\Delta}^{\mu\nu}\christoffel{\beta}{\alpha}{\nu}\hat{B}^\alpha\hat{B}_\beta\nonumber\\
&=&-\cosh^{2}\rho~\sin^{2}\theta~\hat{\Delta}^{\mu\nu}\christoffel{\phi}{\phi}{\nu}\left(\hat{B}^\phi\right)^{2}\nonumber\\
&=&-\cosh^{2}\rho~\sin^{2}\theta~\hat{\Delta}^{\mu\nu}\left(\tanh\rho~\delta^\rho_\nu+
\cot\theta~\delta^\theta_\nu\right)\left(\hat{B}^\phi\right)^{2}.\nonumber\\
\end{eqnarray}
Plugging \eqref{U13} into \eqref{U15}, we thus arrive at
\begin{widetext}
\begin{equation}\label{U16}
\inv{2}\hat{\Delta}^{\mu\nu}\partial_\nu\left[\left(\inv{\cosh^{2}\rho\sin\theta}
\right)^2\left(r^2\mathcal{A}_1^2\mathcal{F}(\vartheta)+\tau^2\mathcal{A}_2^2\sin^{2}\theta~\mathcal{H}(\vartheta)\right)\right]=-\frac{r^2\mathcal{A}_1^2\mathcal{F}(\vartheta)}{\cosh^{2}\rho}~\hat{\Delta}^{\mu\nu}\left(\tanh\rho~\delta^\rho_\nu+
\cot\theta~\delta^\theta_\nu\right).
\end{equation}
\end{widetext}
Bearing in mind that the fluid in $dS^3\times E^1$ is stationary, it turns out that $\hat{\Delta}^{i\nu}=g^{i\nu}$ for the spatial directions $i=\theta,\phi,\eta$, while it vanishes in the temporal direction $\rho$. Hence, \eqref{U16} becomes trivial for $\mu=\rho$. For $\mu=\{\phi,\eta\}$, the lhs of \eqref{U16} vanishes because of \eqref{A28}, and the rhs because of Kronecker $\delta$s. Setting $\mu=\theta$, \eqref{U16} thus reads	
\begin{equation}\label{U17}
\mathcal{A}_1^{2}\cosh^{2}\rho~\left(\inv{2}\deriv{\vartheta}{\theta}\deriv{\mathcal{F}}{\vartheta}+\cot\theta\mathcal{F}(\vartheta)\right)+\inv{2}\mathcal{A}_2^2\deriv{\vartheta}{\theta}\deriv{\mathcal{H}}{\vartheta}=0.
\end{equation}
Let us notice that \eqref{U17} must be satisfied for any value of $\rho$. In addition, the solutions are to be independent of a particular choice for $\vartheta$. Hence, without loss of generality, we let $\vartheta=\theta$. At $\rho=0$, we thus obtain
\begin{equation}\label{U18}
\mathcal{A}_1^{2}\left(\inv{2}\deriv{\mathcal{F}}{\theta}+\cot\theta\mathcal{F}(\theta)\right)=-\inv{2}\mathcal{A}_2^2\deriv{\mathcal{H}}{\theta}.
\end{equation}
Plugging \eqref{U18} back into \eqref{U17} leads to
\begin{equation}\label{U19}
\left(\inv{2}\deriv{\mathcal{F}}{\theta}+\cot\theta\mathcal{F}(\theta)\right)=0,\quad\mbox{and}\quad\deriv{\mathcal{H}}{\theta}=0.
\end{equation}
The solutions to \eqref{U19} read\footnote{For simplicity, integration constants are chosen to be unity.}
\begin{equation}\label{U20}
\mathcal{F}(\theta)=\inv{\sin^{2}\theta},\quad\mbox{and}\quad\mathcal{H}(\theta)=1.
\end{equation}
Plugging, at this stage, \eqref{U20} into \eqref{U13} leads to
\begin{equation}\label{U21}
\hat{B}^{\phi}=\frac{\mathcal{A}_1}{\cosh^{2}\rho~\sin^{2}\theta},\quad\mbox{and}\quad\hat{B}^{\eta}=\frac{\mathcal{A}_2}{\cosh^{2}\rho}.
\end{equation}
To transform $\hat{B}^{\mu}$ back into the Minkowski spacetime, we use \eqref{U10}. Using the fact that the coordinates $\eta$ and $\phi$ are the same in $dS^3\times E^1$ and $M^{3,1}$, \eqref{U10} with $\Omega=\tau$ and $d=4$ reduces to
\begin{equation}\label{U22}
B^\mu = \frac{\hat{B}^\mu}{\tau^3}.
\end{equation}	
Plugging \eqref{U21} into (\ref{U22}), and using (\ref{U4}) as well as the relation $r=\tau\cosh\rho\sin\theta$, the components of the magnetic field in conformal MHD read
\begin{eqnarray}\label{U23}
\hspace{-0.3cm}
B^\phi &=& \frac{\mathcal{A}_1}{r^2\tau},\nonumber\\
\hspace{-0.3cm}B^\eta &=& \inv{\tau}\frac{4q^2\beta_0\mathcal{A}_1}{[1+q^4(\tau^2-r^2)^2+2q^2(\tau^2+r^2)]},
\end{eqnarray}
with $\beta_0\equiv\mathcal{A}_2/\mathcal{A}_1$. In the limit of $\tau\to 0$, $\beta_{0}$ reduces, for $qr=1$, to
\begin{equation}\label{U24}
\beta_0=\lim_{\tau\to 0}\frac{B^\eta}{B^\phi}\bigg|_{qr=1}.
\end{equation}	
Hence, $B^{\mu}$ in the Minkowski spacetime is given by
\begin{eqnarray}\label{U25}
B^{\mu}=\left(0,B^{\phi},B^{\eta},0\right),
\end{eqnarray}
with $B^{\phi}$ and $B^{\eta}$ from \eqref{U23}. Taking the covariant square root of \eqref{U25}, the magnitude of the magnetic field given by
\begin{eqnarray}\label{U26}
\lefteqn{\hspace{-1cm}B=\mathcal{A}_1\bigg[\frac{1}{r^2\tau^2}
}\nonumber\\
&&\hspace{-1cm}+\beta_0^2\left(\frac{4q^2}{[1+q^4(\tau^2-r^2)^2+2q^2(\tau^2+r^2)]}\right)^2\bigg]^{1/2}.
\end{eqnarray}
Let us now consider the solutions \eqref{U23} and \eqref{U26} for $B^{\mu}$ and $B$ in the Minkowski spacetime. As it turns out, in contrast to the longitudinal coordinate of the magnetic field $B^{\eta}$, its transverse one $B^{\phi}$ is independent of the system transverse size $L\sim q^{-1}$. In addition, $B$ exhibits a full symmetry under exchange of $\tau$ and $r$. The first term is, however, singular in $r$ and $\tau$. Neglecting the longitudinal term including $q$,  and defining $B_{0}$ to be the value of $B$ at some arbitrary point $(r_0,\tau_0)$, $B$ is then given by
	\begin{equation}\label{U27}
	B = B_0\frac{r_0\tau_0}{r\tau}.
	\end{equation}
For any fixed radius $r^{\star}$, \eqref{U27} is $B=\mathbb{B}_0\frac{\tau_{0}}{\tau}$, which is the same as the transverse MHD result \eqref{A22}, with $B_0$ replaced with $\mathbb{B}_{0}\equiv B_0r_0/r^{\star}$ and fixed $r^{\star}$. Same scaling behavior occurs for the radial evolution of the magnetic field, because of the aforementioned symmetry under the exchange of $r$ and $\tau$.
\par
We close this section with the computation of the electric current in this conformal MHD setup. Using \eqref{M4} and the $dS^3\times E^1$ metric \eqref{U3}, we first arrive at
\begin{eqnarray}\label{U28}
\hat{J}^\mu &=& \inv{\sqrt{-g}}\partial_\nu\left(\hat{F}^{\mu\nu}\sqrt{-g}\right)\nonumber\\
&=&\inv{\sqrt{-g}}\left[\partial_\rho\left(\hat{F}^{\mu\rho}\sqrt{-g}\right)+\partial_\theta\left(\hat{F}^{\mu\theta}\sqrt{-g}\right)\right]\nonumber\\
&=&\inv{\cos^{2}\rho~\sin\theta}\partial_\theta\left(\hat{F}^{\mu\theta}\cos^{2}\rho~\sin\theta\right)\nonumber\\\hspace{-1cm}
&=&\left(\pderivv{\theta}+\cot\theta\right)\hat{F}^{\mu\theta}.
\end{eqnarray}
Plugging then \eqref{U20} into \eqref{U12},  we obtain
\begin{eqnarray}\label{U29}
\hat{F}^{\phi\theta}&=&-\frac{\beta_0\mathcal{A}_0}{\cosh^{4}\rho~\sin\theta},\nonumber\\
\hat{F}^{\eta\theta}&=&\frac{\mathcal{A}_0}{\cosh^{2}\rho~\sin\theta}.
\end{eqnarray}
We finally arrive at
\begin{eqnarray}\label{U30}
\hat{J}^\mu = 0,
\end{eqnarray}
by plugging \eqref{U29} into \eqref{U28}. Transforming \eqref{U30} back into the Minkowski space, we obtain\footnote{We notice that such a transformation is only allowed when a quantity has a definite conformal weight. According to \cite{gubser2010-2} the conformal weight of the current is $[J^{\mu}]=4$.}
\begin{eqnarray}\label{U31}
J^\mu = 0.
\end{eqnarray}
This result is in contrast to the cases of transverse MHD and self-similar flow in Secs. \ref{sec3} and \ref{sec4}. In transverse MHD setup in Sec. \ref{sec3}, the electric current vanishes if $B^{\mu}$ from \eqref{A22} is assumed to be boost invariant. In the self-similar flow in Sec. \ref{sec4A}), we assumed $J^\mu=0$, and found a specific solution to \eqref{S20}. In both cases, we could, in principle, use the heuristic relation $E/B\sim J/(B\sigma_{e})$ to study the consistency of the ideal MHD limit. However, since $J^\mu$ identically vanishes, such argument does not hold in the present case.
\section{Comparison of solutions}\label{sec7}
\setcounter{equation}{0}
In this section, we compare different features of the solutions presented in previous sections. In particular, we focus on the ZCSSF solution from \eqref{S25} and \eqref{S26} as well as the CMHD solution from \eqref{U23} and \eqref{U26}. To do this, we define (see Appendix \ref{appC} for more details),
\begin{eqnarray}\label{D1}
B_{\text{init.}}\equiv|\boldsymbol{B}(\tau_0,r_0)|,
\end{eqnarray}
with $\boldsymbol{B}=\left(0,rB^{\phi},\tau B^{\eta}\right)$ from \eqref{appC3}, and examine the evolution of the dimensionless quantity $B/B_{\text{init.}}$ for $B$ being ZSCCF and CMHD solutions.
Let us notice that studying  $B/B_{\text{init.}}$, instead of $B$, enables us to compare these solutions independent of $B_{\text{init.}}$, that cannot be determined in the MHD framework. In this way, we measure, without loss of generality, any local quantity at a point $(t,x,y,z)=(\tau,r,0,0)$. In the context of HICs, where the $z$-axis is identified with the beam direction, $(\tau,r,0,0)$ turns out to be on the axis of the impact parameter characterized by $\phi=0$ in the mid-rapidity $\eta=0$. Apart from $B_{\text{init.}}$, we also define
\begin{equation}\label{D2}
\alpha=\frac{B_z}{B_y}\at[\Big]{\phi=0,\eta=0},
\end{equation}
as the ratio of the longitudinal and transverse components of the magnetic field. Here, $\alpha$ is the cotangent of the angle between $\boldsymbol{B}$ and the beamline.
Using $\alpha$, and, in particular,  $\alpha_{0}$ gives us the possibility to express different solution-dependent parameters with a single free parameter (see Appendix \ref{appC} for more details).\footnote{In general, $X_0$ denotes the quantity $X$ as measured at the initial point, i.e. $X_{0}\equiv X(\tau_{0},r_{0})$.} In what follows, the point $(\tau_0,r_0)$ is referred to as the initial point. We assume $\tau_0=0.5\fmc$ and $r_0=c_s\tau_0$, where $c_s=1/\sqrt{3}$ is the speed of sound. Whereas, the value of $r_0$ is arbitrary, and can thus be chosen as small as desired, $\tau_0$ is roughly equal to the thermalization time \cite{kolb2003}. Another useful quantity is $\lambda$, defined by
\begin{figure*}[hbt]	
\includegraphics[width=7cm,height=7.5cm]{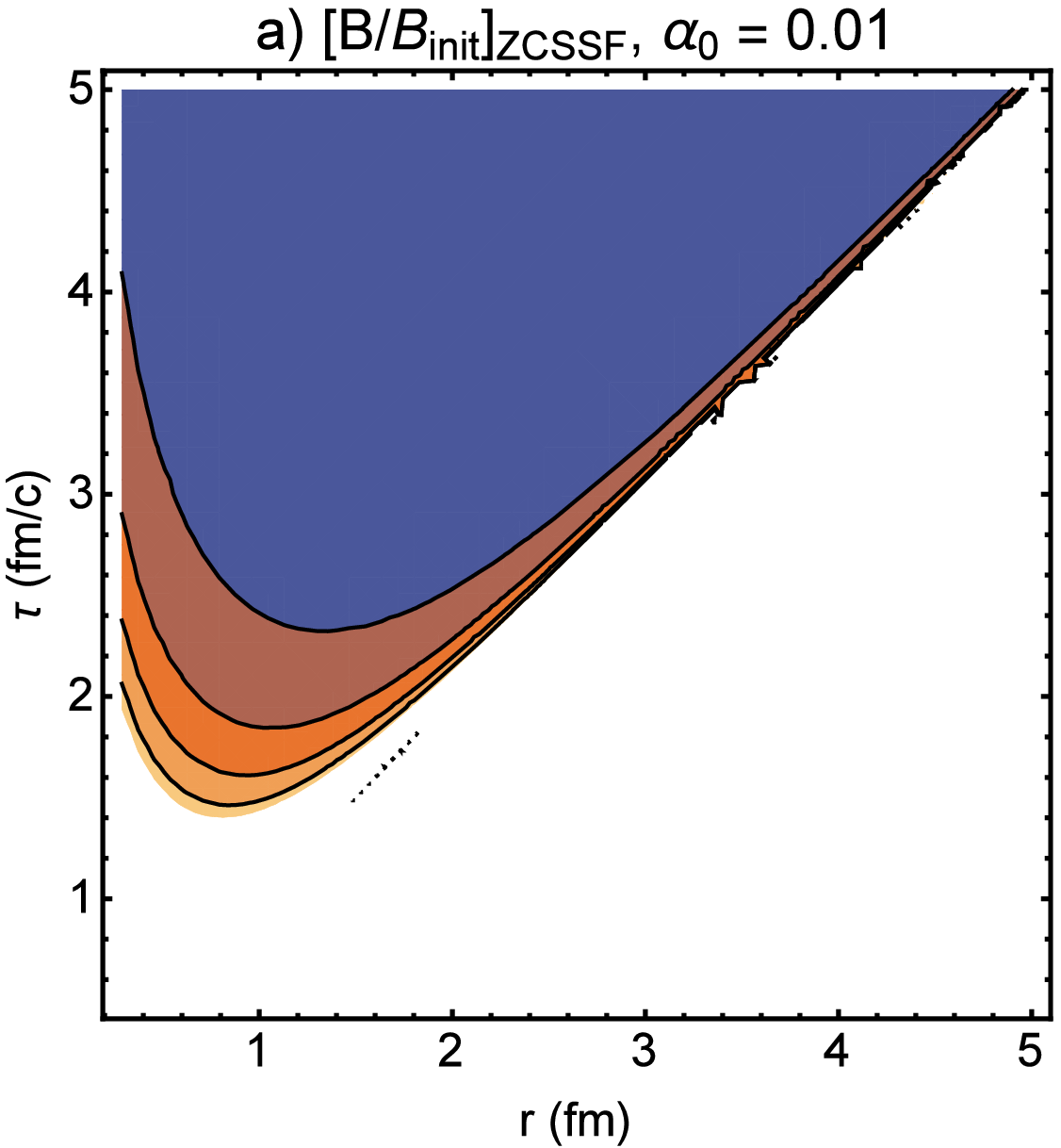}
\includegraphics[width=1cm,height=6cm]{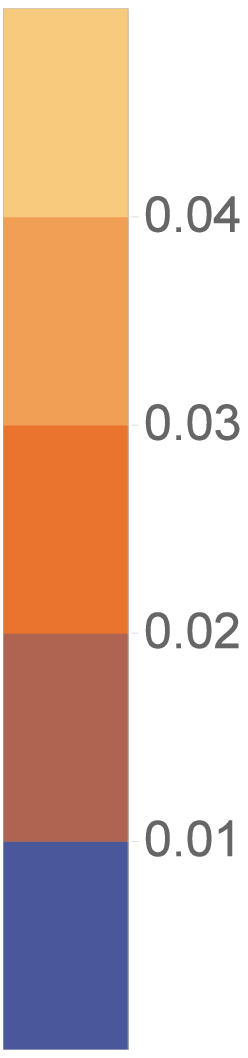}
\hspace{0.3cm}
\includegraphics[width=7cm,height=7.5cm]{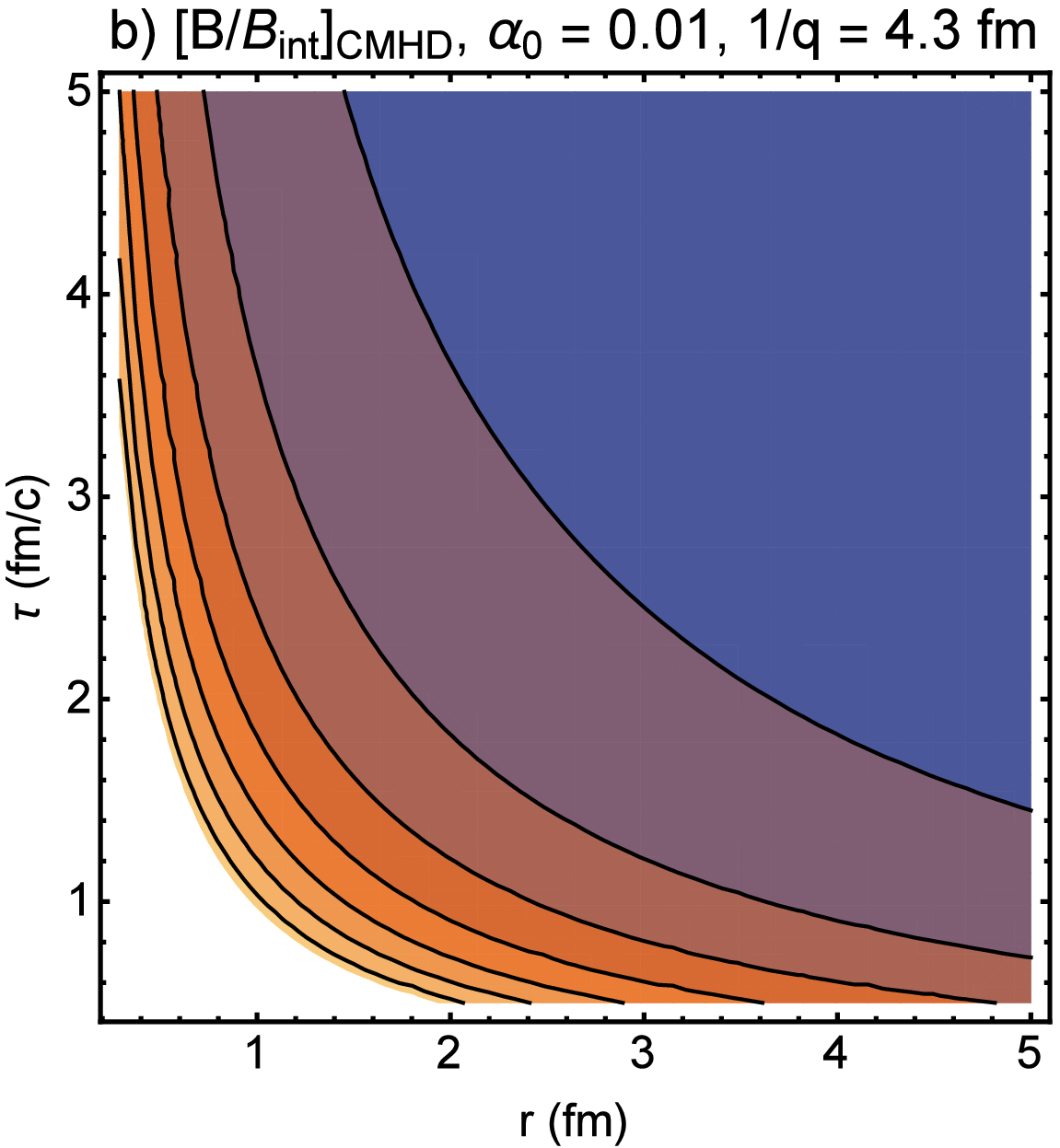}
\includegraphics[width=1.3cm,height=6cm]{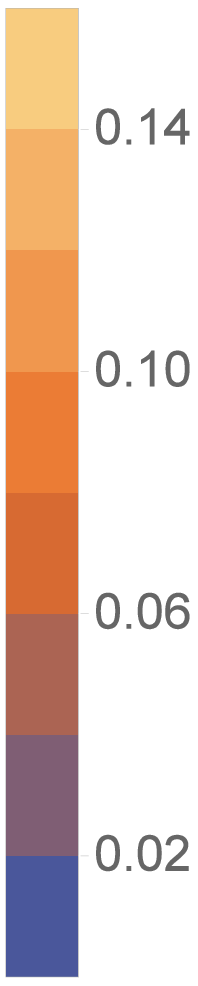}
\caption{(color online). (a) The $(\tau,r)$ dependence of $B/B_\text{init.}$ for the ZCSSF solution (\ref{appC9}) is plotted for $\alpha_0=0.01$. The magnetic field turns out to be restricted to the domain $r\leq\tau$.  (b) The $(\tau,r)$ dependence of $B/B_\text{init.}$  for the CMHD solution \eqref{appC15} is plotted for $\alpha_0=0.01$ and $1/q=4.3\fm$ \cite{gubser2010-1}. In contrast to the ZCSSF case, there is no restriction on the radial domain of the CMHD solution.}\label{fig1}
\end{figure*}	
\begin{equation}\label{D3}
\lambda \equiv \frac{B_z}{B}\at[\Big]{\phi=0,\eta=0}.
\end{equation}	
Here, $B_{z}=\tau B^{\eta}$ is the longitudinal magnetic field component and $B=|\boldsymbol{B}|$, with $\boldsymbol{B}$ from \eqref{appC3}.
We demonstrate the evolution of the dimensionless quantity $\lambda/\lambda_0$ for the ZCSSF and CMHD solutions with respect to $\tau$ and $r$. Similar to $\alpha$,  $\lambda$ turns out to be the cosine of the angle between $\boldsymbol{B}$ and the beamline.
To have a measure for the strength of the magnetic field, we also study the ratio of magnetic field energy $B^{2}$ over the fluid energy density $\epsilon$,
\begin{equation}\label{D4}
\sigma \equiv \frac{B^2}{2\epsilon}.
\end{equation}
This quantity is, in particular, related to the Alfv$\acute{\mbox{e}}$n wave velocity \cite{gedalin1993}
\begin{equation}\label{D5}
v_A \equiv \sqrt{\frac{2\sigma}{2\sigma+1+c_s^2}},
\end{equation}
that goes to the speed of light $c=1$, as $\sigma$ tends to infinity.
\par
\begin{figure}[b]
\begin{center}
\includegraphics[width=7cm,height=7.5cm]{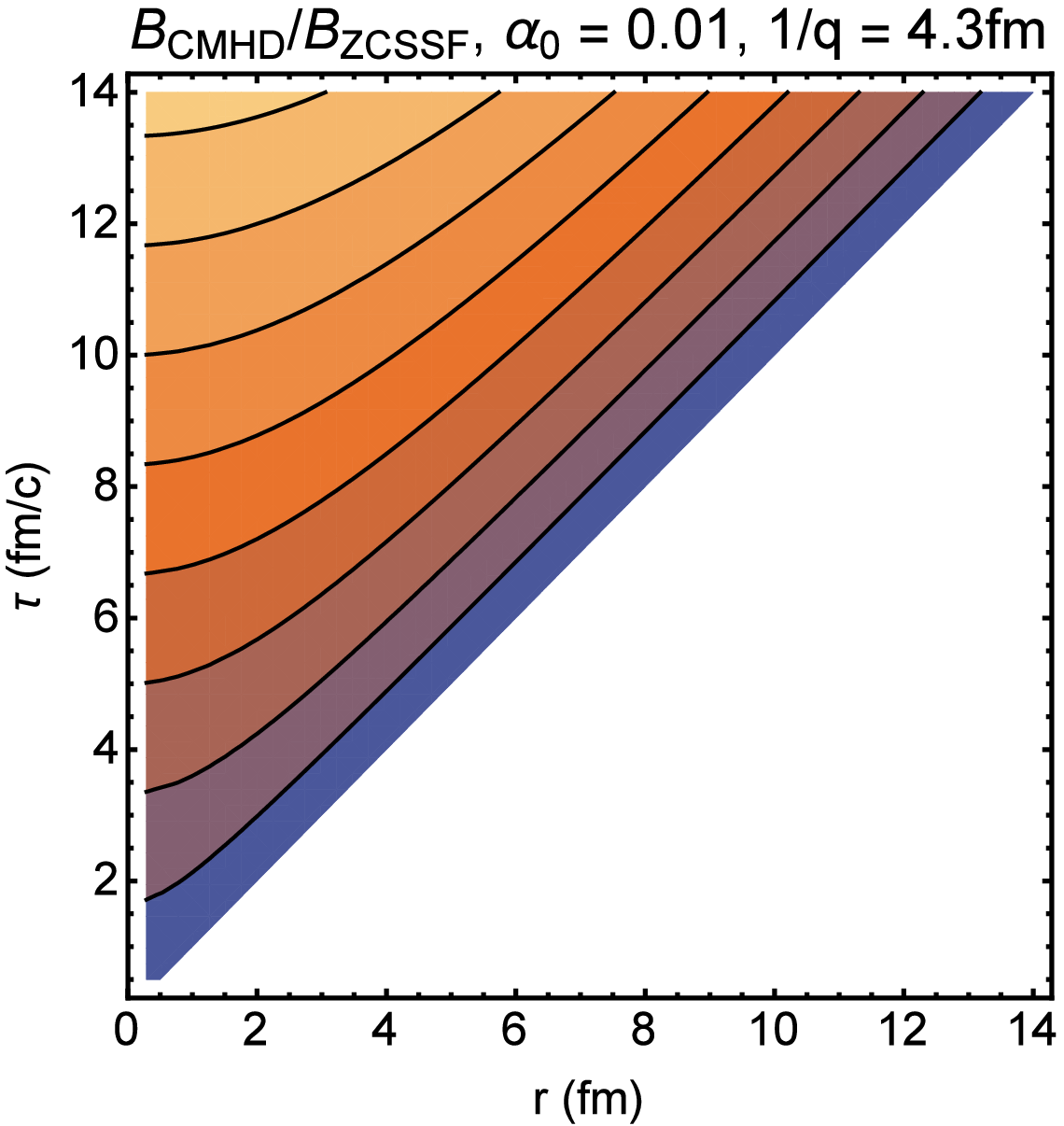}
\includegraphics[width=1cm,height=6cm]{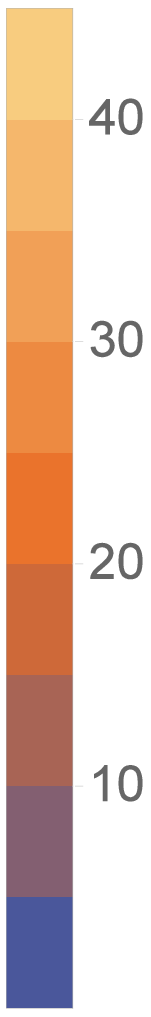}
\end{center}
\caption{(color online). The $(\tau,r)$ dependence of the ratio $B_{\text{\tiny{ZCSSF}}}/B_{\text{\tiny{ZCSSF}}}$ is plotted for $\alpha_{0}=0.01$ and $1/q=4.3$ fm. Both magnetic fields are comparable around $\tau=r$ line. For $\tau>r$ and any fixed value of $r$, the CMHD solution becomes significantly larger than the ZCSSF one.}\label{fig2}
\end{figure}
\begin{figure*}[hbt]
\includegraphics[width=8cm,height=6cm]{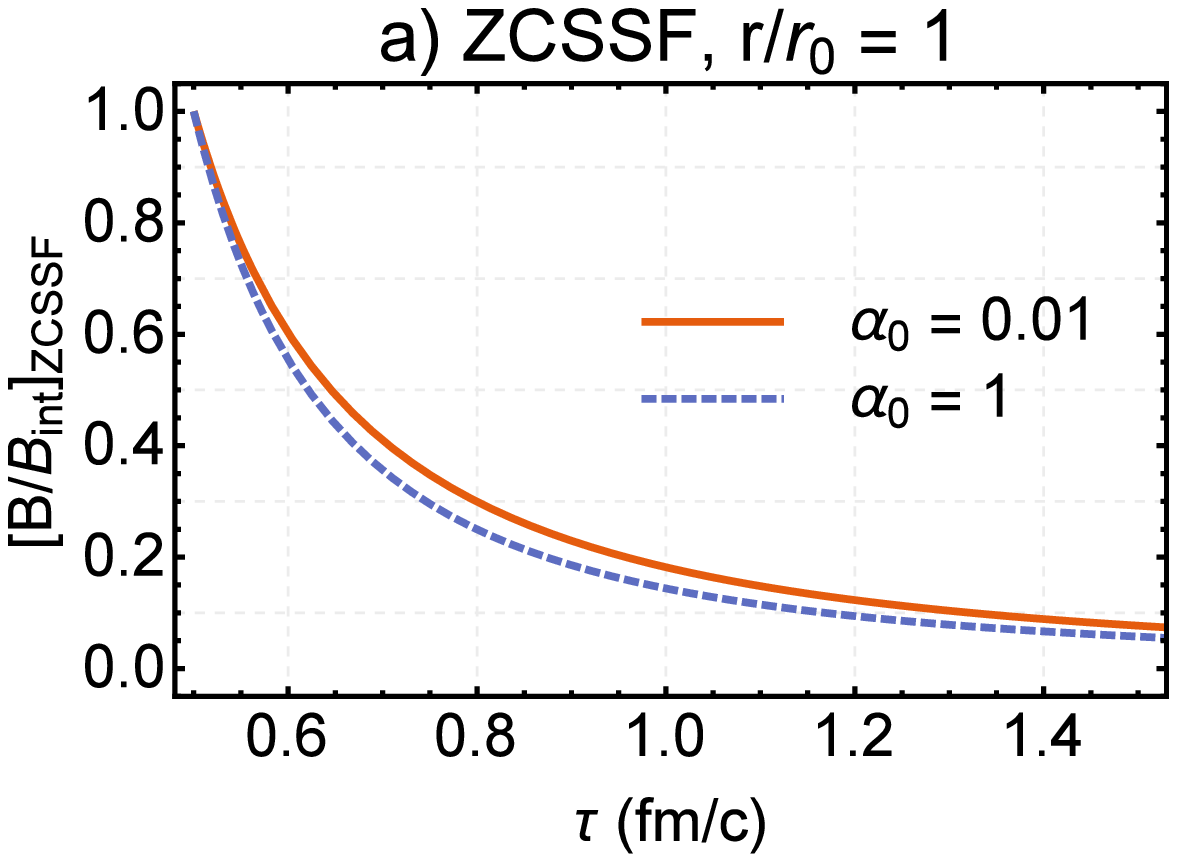}\hspace{0.3cm}
\includegraphics[width=8cm,height=6cm]{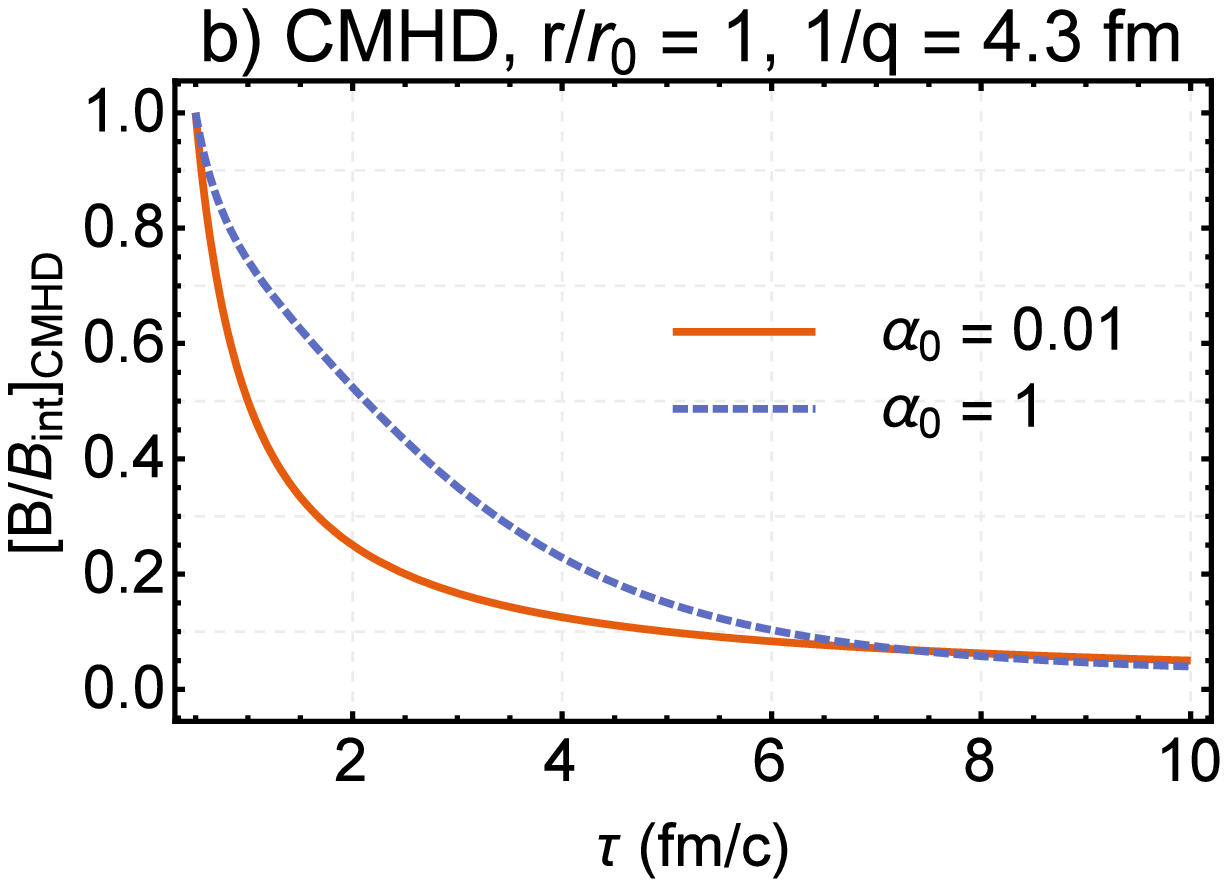}
\caption{(color online). The $\tau$ dependence of $B/B_\text{init.}$ for the ZCSSF solution (panel a) and CMHD solution (panel b) is plotted at $r=r_0$ and for $\alpha_0=0.01$ (solid orange curve) and $\alpha_{0}=1$ (dashed blue curve). For the CMHD solution $q$ is chosen to be $q=1/4.3$ fm$^{-1}$.  Whereas $B/B_\text{init.}$ for the ZCSSF solution drops below $0.1$ around $\tau\sim 1\fmc$, $B$ for the CMHD solution becomes $0.1 B_\text{init.}$ at $\tau\sim 6\fmc$. In contrast to $B_{\text{CMHD}}$, the decay of the ZCSSF field turns out to become faster, if the initial magnetic field has a large component along the beamline. The latter is characterized with larger $\alpha_0$.}\label{fig3}
\end{figure*}

The corresponding $B$ fields to the ZCSSF and CMHD solutions turn out to be only functions of $r$ and $\tau$, and to have no radial components.\footnote{In \eqref{S25} and \eqref{U26}, $r=\sqrt{x^2+y^{2}}$ is the length of $\boldsymbol{r}=(x,y)$, which is defined to be in the transverse $x$-$y$ plane.} In general, at a fixed value of $r$, the $\tau$-dependency of the $B$ field describes its evolution with time, whereas the $r$ dependence at a fixed value of $\tau$ gives the spatial distribution of the magnetic field in the transverse plane. As it turns out, the evolution and spatial distribution in both cases are sensitive to $\alpha_0=\alpha(\tau_{0},r_{0})$ with $\alpha$ defined in \eqref{D2}.
Despite this similarity, two solutions are different in many aspects. The first difference is in their radial domain of validity. Whereas the CMHD solution covers the whole domain $[r_0,\infty)$ for any fixed value of $\tau$, the ZCSSF solution merely covers $r\leq \tau$. This is demonstrated in Fig. \ref{fig1}, where the $(\tau,r)$ dependence of $B/B_{\text{init.}}$ for the ZCSSF and CMHD solutions from \eqref{appC9} and \eqref{appC15} are plotted in Fig. \ref{fig1}(a) and Fig. \ref{fig1}(b), respectively. Moreover, the magnetic field turns out to be generally stronger in the CMHD solution than the ZCSSF one [see Fig. \ref{fig2}, where the $(\tau,r)$ dependence of $B_{\text{\tiny{CMHD}}}/B_{\text{\tiny{ZCSSF}}}$ is plotted].\footnote{Here, $B_{\text{\tiny{sol}}}$ with sol $=\{$ZCSSF,CMHD$\}$ are defined by $B=|\boldsymbol{B}|$ from \eqref{appC3}, with $(B^{\phi},B^{\eta})$ from \eqref{S25} for the ZCSSF solution and from \eqref{U23} for the CMHD solution (see Appendix  \ref{appC} for more details).} As the system evolves, the ZCSSF solution significantly lags behind the CMHD one. For the CMHD solution, $B$ is not far from $B_\text{init.}$ for a significant timescale $\tau\simeq 5$-$6$ fm/c, that covers most of the hydrodynamical expansion near the center of the collision. On the other hand, for the ZCSSF solution, the magnetic field becomes one order of magnitude smaller at a very short timescale. Let us notice that if $B_\text{init.}$ is sufficiently large to have measurable quantum effects, then significant physical differences will arise. These properties are demonstrated in Fig. \ref{fig3}, where the $\tau$ dependence of $[B/B_{\text{init.}}]_{\text{{sol}}}$, with sol $=\{$ZCSSF,CMHD$\}$, is plotted for $\alpha_{0}=0.01$ (solid orange curves) and $\alpha_{0}=1$ (dashed blue curves).
\par
\par
\par
\begin{figure*}[hbt]
\includegraphics[width=8cm,height=6cm]{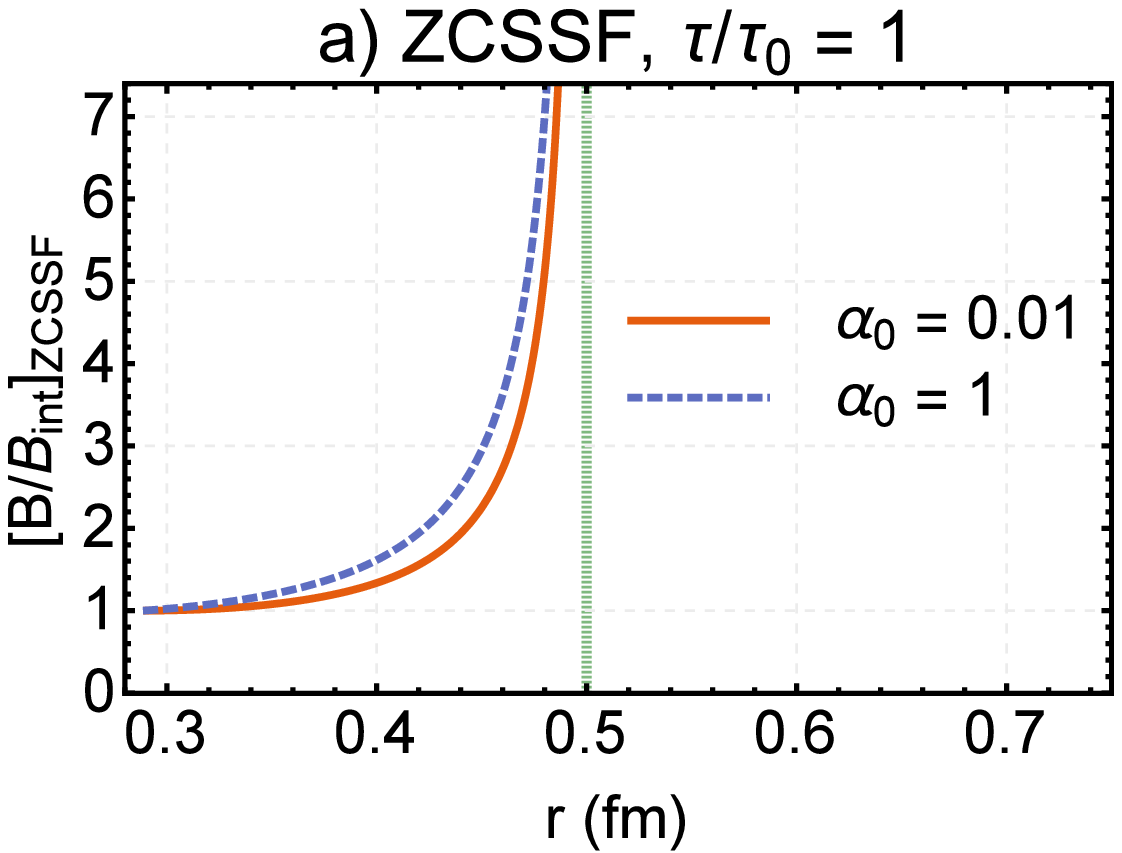}\hspace{0.3cm}
\includegraphics[width=8cm,height=6cm]{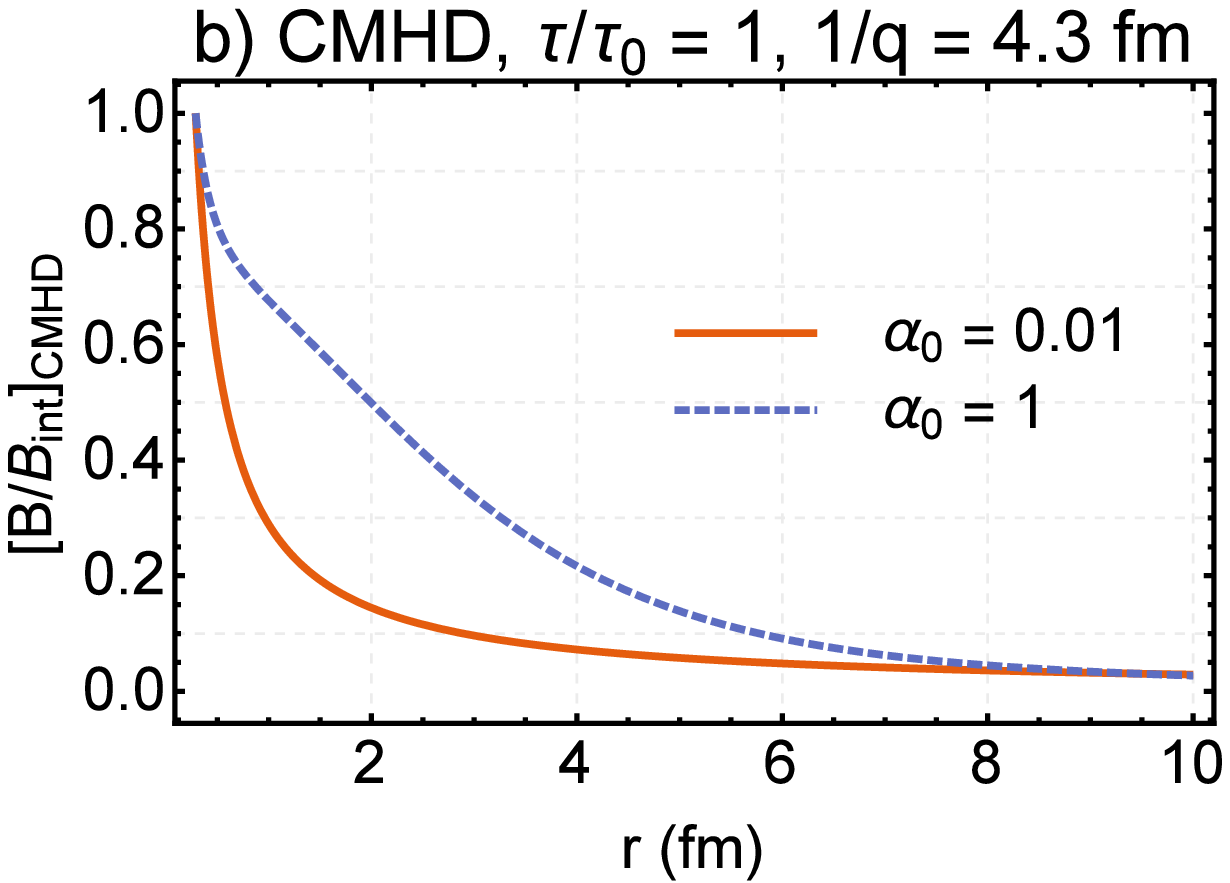}		
\caption{(color online). The $r$ dependence of $B/B_\text{init.}$ for the ZCSSF solution (panel a) and the CMHD solution (panel b) is plotted at $\tau=\tau_0$ and for $\alpha_0=0.01$ (solid orange curves) and $\alpha_{0}=1$ (dashed blue curves). For the CMHD solution $q$ is chosen to be $q=1/4.3$ fm$^{-1}$. Whereas $B_{\text{\tiny{ZCSSF}}}$ blows up as $r\to \tau$ (this is indicated by the vertical green line in panel a), $B_{\text{\tiny{CMHD}}}$ is finite in the whole range of $r$. For the latter case, the decay of $B$ is slower for the initial magnetic field having a larger component along the beamline (larger $\alpha_0$). }\label{fig4}
\end{figure*}
\par
In Fig. \ref{fig4}, the $r$ dependence of $B/B_{\text{init.}}$ for the ZCSSF and CMHD solutions is demonstrated for $\alpha_{0}=0.01$ (solid orange curve) and $\alpha_{0}=1$ (dashed blue curve). A comparison of the radial and temporal dependence of $B_{\text{\tiny{ZCSSF}}}$ in Figs. \ref{fig4}(a) and \ref{fig3}(a) shows that the radial dependence of $B_{\text{\tiny{ZCSSF}}}$ is quite different from its temporal evolution. As concerns its radial dependence, the magnetic field tends to infinity as $r/\tau\to 1$. In particular, it does not exist for $r>\tau$. The vertical green line in Fig. \ref{fig4}(a) indicates the $r=\tau$ validity borderline for the ZCSSF solution (here, $\tau=\tau_{0}=0.5$ fm/c). We notice that, mathematically, this inherent feature of the ZCSSF solution arises from the factor $1/\varrho$ in \eqref{appC9}. In contrast to the ZCSSF solution, the radial dependence of the magnetic field for the CMHD solution, demonstrated in Fig. \ref{fig4}(b) is very similar to its temporal evolution from Fig. \ref{fig3}(b). Similar to its temporal dependence, $B\lesssim B_{\text{init.}}$ for a relatively large distance $r\sim 4$-$5$ fm.
\par
\begin{figure*}[hbt]
\includegraphics[width=8cm,height=6cm]{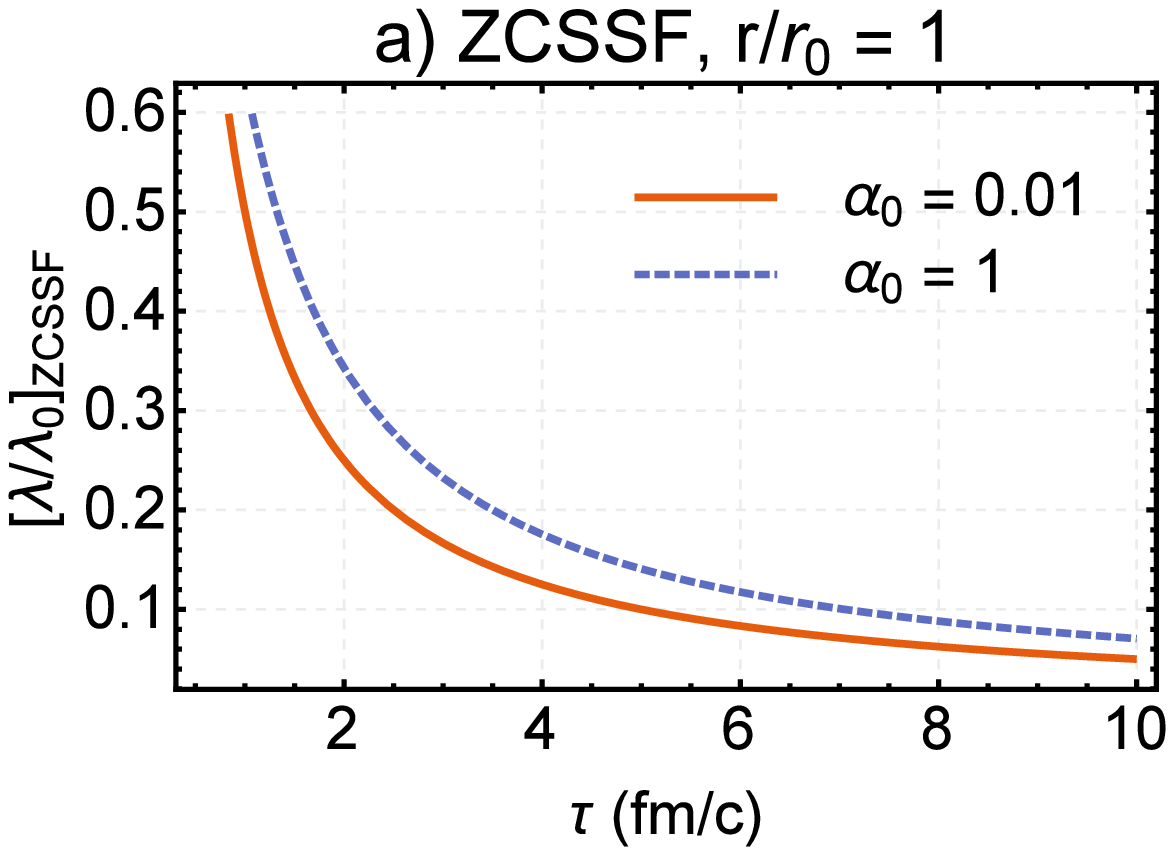}\hspace{0.3cm}
\includegraphics[width=8cm,height=6cm]{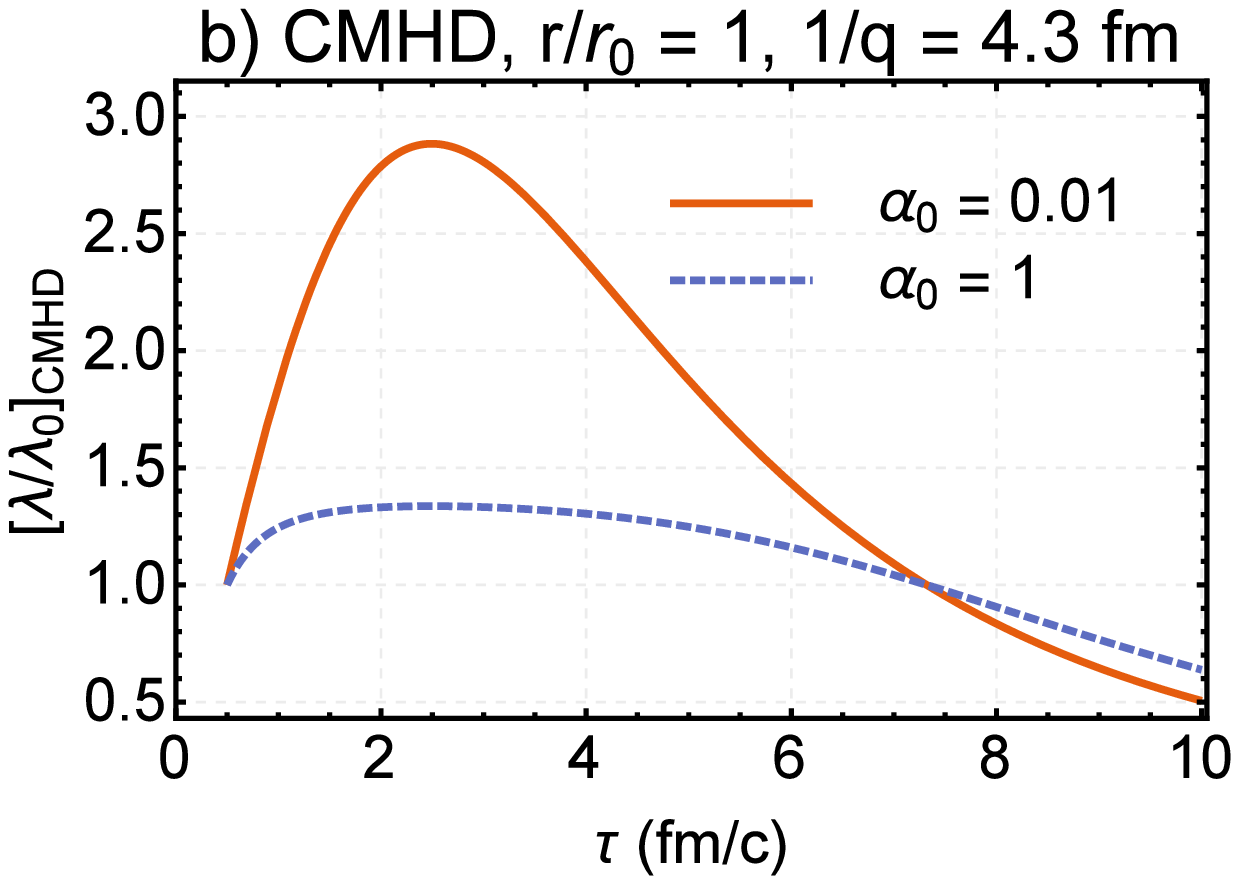}
\caption{(color online).
The $\tau$ dependence of $\lambda/\lambda_{0}$ for the ZCSSF solution (panel a) and the CMHD solution (panel b) is plotted at $r=r_0$ and for $\alpha_0=0.01$ (solid orange curves) and $\alpha_{0}=1$ (dashed blue curves). For the CMHD solution $q$ is chosen to be $q=1/4.3$ fm$^{-1}$. Whereas $[\lambda/\lambda_{0}]_{\text{\tiny{ZCSSF}}}$ decreases with increasing $\tau$,
$[\lambda/\lambda_{0}]_{\text{\tiny{CMHD}}}$ exhibits a maximum at $\tau\sim 4\tau_{0}$, and then slowly decreases with increasing $\tau$. This maximum becomes larger, the smaller $\alpha_{0}$ is chosen.}\label{fig5}		
\end{figure*}
\par
\begin{figure*}[hbt]
\includegraphics[width=8cm,height=6cm]{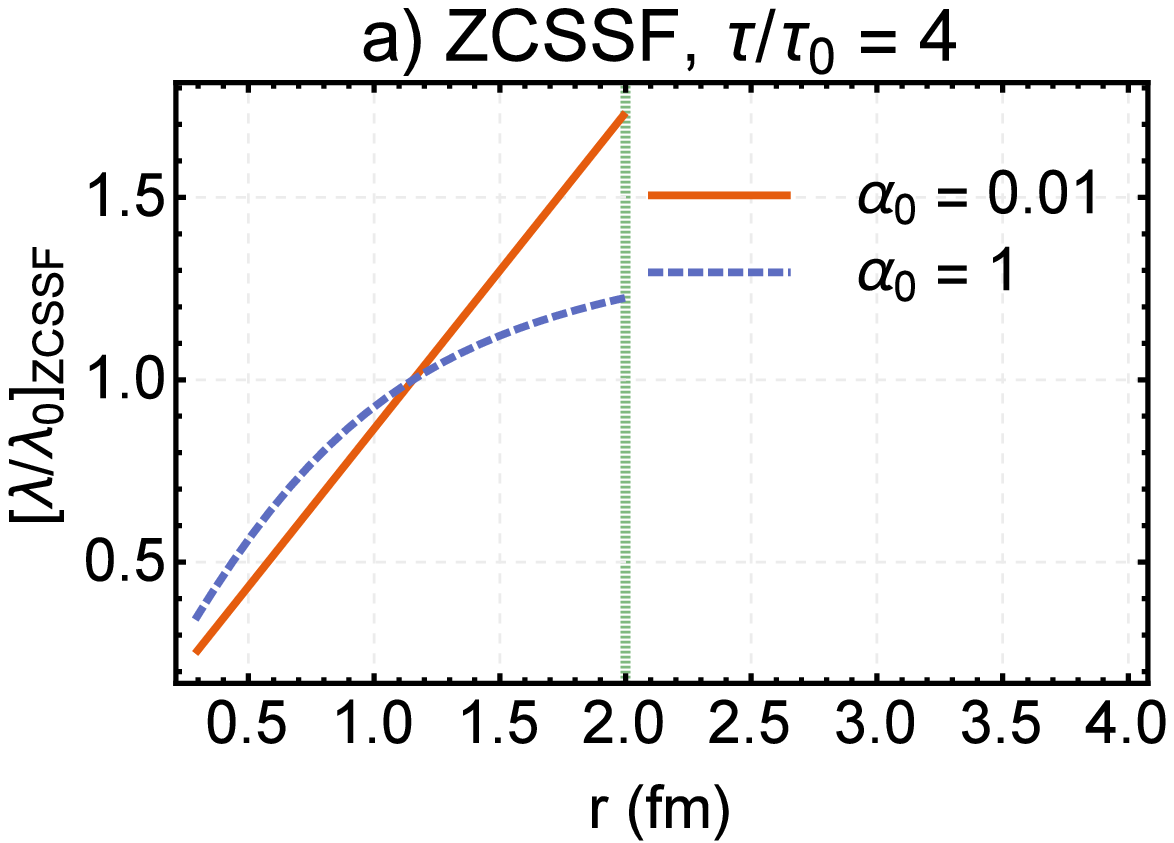}\hspace{0.3cm}
\includegraphics[width=8cm,height=6cm]{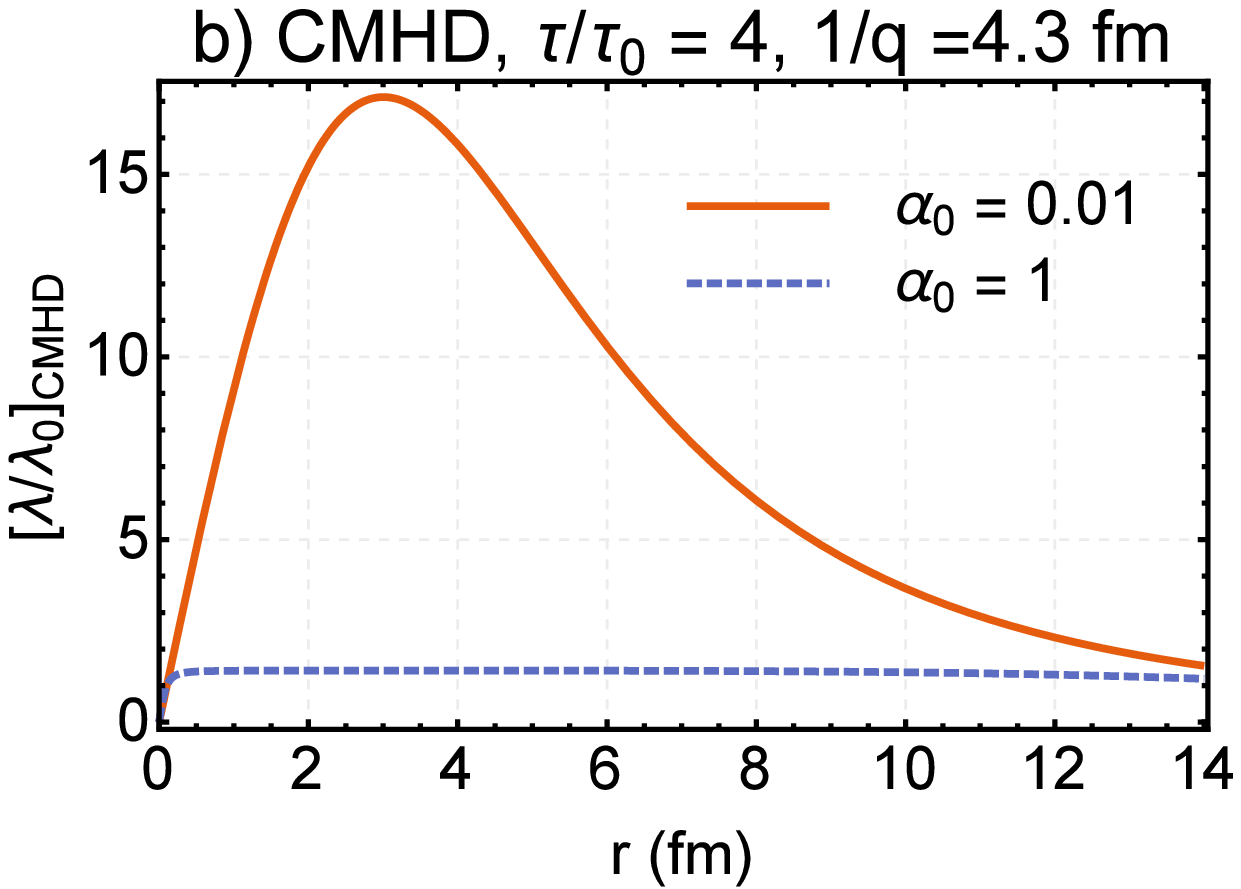}		
\caption{(color online).
The $r$ dependence of  $\lambda/\lambda_0$ for the ZCSSF solution (panel a) and the CMHD solution (panel b) is plotted at $\tau=4\tau_0$ and for $\alpha_0=0.01$ (solid orange curves) and $\alpha_{0}=1$ (dashed blue curves). For the CMHD solution $q$ is chosen to be $q=1/4.3$ fm$^{-1}$. Whereas $[\lambda/\lambda_0]_{\text{\tiny{ZCSSF}}}$ increases with increasing $r$, and it is slightly suppressed once larger values of $\alpha_{0}$ are chosen, the $r$ dependence  of $[\lambda/\lambda_0]_{\text{\tiny{CMHD}}}$ share the same properties with its $\tau$ dependence, i.e. for $\alpha_{0}$ being small enough, a relatively large maximum appears that then slowly decays. For larger values of $\alpha_{0}$, $[\lambda/\lambda_0]_{\text{\tiny{CMHD}}}$ remains small.}\label{fig6}
\end{figure*}
In HIC experiments, the longitudinal component of the magnetic field, $B_z$, is generally reported to be small \cite{Huang-review}. However, in our solutions, $B_{z}$ does not vanish neither in the ZCSSF nor in the CMHD cases.\footnote{Let us notice that $B_{z}$ is not forced to be zero. But, if it is zero at the initial point, it remains zero during the evolution.} Moreover, as it is demonstrated in Figs. \ref{fig3} and \ref{fig4} for the $\tau$ and $r$ dependence of the magnetic field, $B/B_{\text{init.}}$ is sensitive to $\alpha_{0}$. As it turns out, the ZCSSF and CMHD solutions behave differently for various choices of $\alpha_0$. Whereas the ZCSSF solution decays faster for larger values of $\alpha_0$, the CMHD solution lives longer for the initial magnetic field having larger component along the beamline (see the plots in Fig. \ref{fig3}, and compare the $\tau$ dependence of the ZCSSF and CMHD solutions for $\alpha_{0}=0.01$ and $\alpha_{0}=1$). Moreover, the CMHD solution turns out to be significantly more sensitive to $\alpha_0$ comparing to the ZCSSF solution. As concerns the radial dependence of $B/B_{\text{init.}}$ for the ZCSSF and CMHD solutions, whereas $B_{\text{\tiny{ZCSSF}}}$ increases with a larger slope, the decay of $B_{\text{CMHD}}$ becomes slower for larger values of $\alpha_0$ [see Fig. \ref{fig4}(b)], so that for larger values of $\alpha_{0}$, $B_{\text{\tiny{CMHD}}}$ is relatively strong in larger radial distances with respect to the centrum of the collision at $r=0$.
\par
The evolution of $[\lambda/\lambda_0]_{\text{\tiny{ZCSSF}}}$ and $[\lambda/\lambda_0]_{\text{\tiny{CMHD}}}$ in the temporal $\tau$ and radial $r$ directions is presented in Figs. \ref{fig5} and \ref{fig6}, respectively. For the ZCSSF solution, $\lambda/\lambda_{0}$ always decreases as the system evolves, although for an initially large $\alpha_{0}$, it decreases at a slower pace [see Fig. \ref{fig5}(a) and compare the evolution of $[\lambda/\lambda_{0}]_{\text{\tiny{ZCSSF}}}$ for $\alpha_{0}=0.01$ (solid orange curve) and $\alpha_{0}=1$ (dashed blue curve)].
The evolution of $\lambda/\lambda_0$ for the CMHD solution is rather different. It experiences an initial rise to a peak, and then mildly tends to zero at infinity. Interestingly, a smaller initial $\alpha_{0}$ enhances  $[\lambda/\lambda_{0}]_{\text{\tiny{CMHD}}}$ significantly stronger than a larger one [see Fig. \ref{fig5}(b), and compare the evolution of $[\lambda/\lambda_{0}]_{\text{\tiny{CMHD}}}$ for $\alpha_{0}=0.01$ (solid orange curve) and $\alpha_{0}=1$ (dashed blue curve)].
As it is demonstrated in Fig. \ref{fig6}(a), the radial distribution of $[\lambda/\lambda_0]_{\text{\tiny{ZCSSF}}}$ is approximately linear, and becomes larger with increasing $r$ up to the validity borderline of this solution at $r=\tau$,  demonstrated with a vertical green line (here $\tau=4\tau_0=2$ fm/c). The radial dependence of $[\lambda/\lambda_{0}]_{\text{\tiny{CMHD}}}$ in Fig. \ref{fig6}(b) shares the same properties with its temporal evolution from Fig. \ref{fig5}(b). In view of  the above qualitative results, it would be interesting to further explore the role playing by the longitudinal component of the magnetic fields created in HIC experiments.
\par
\begin{figure}[hbt]
\includegraphics[width=8cm,height=6cm]{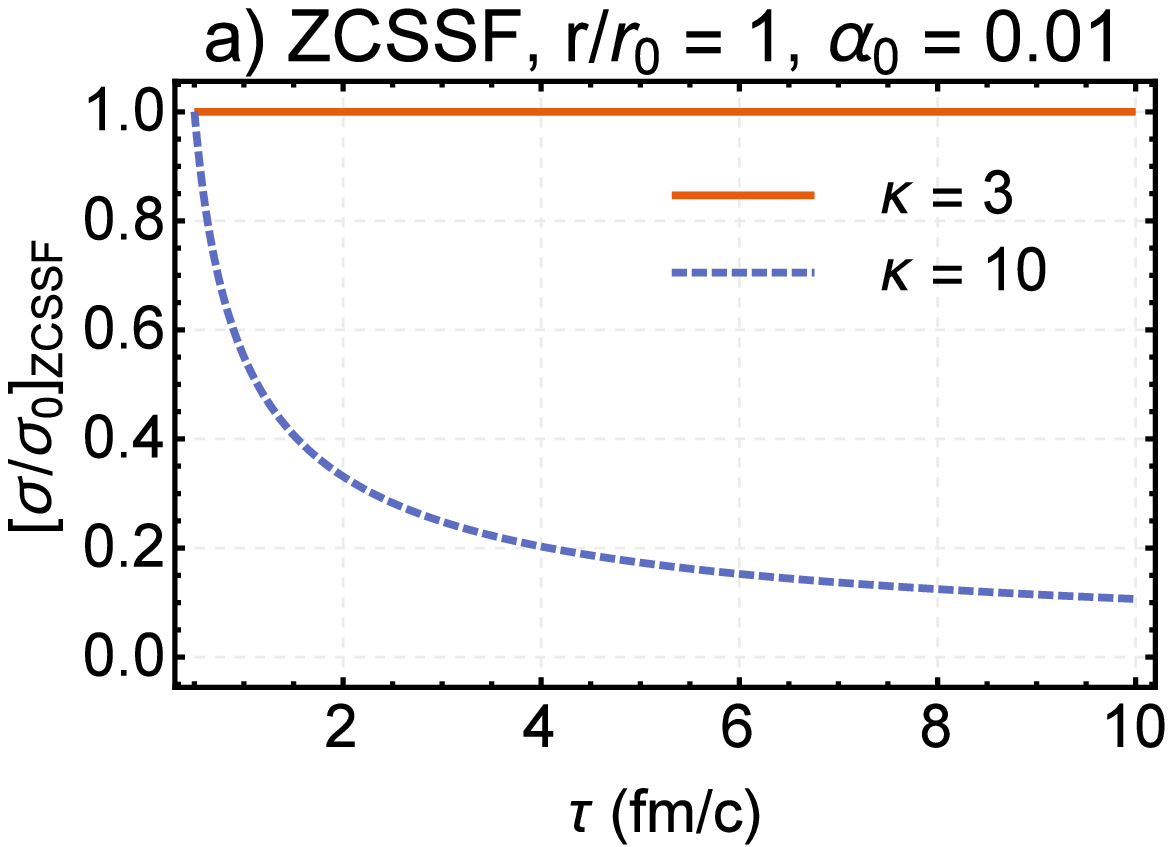}		
\includegraphics[width=8cm,height=6cm]{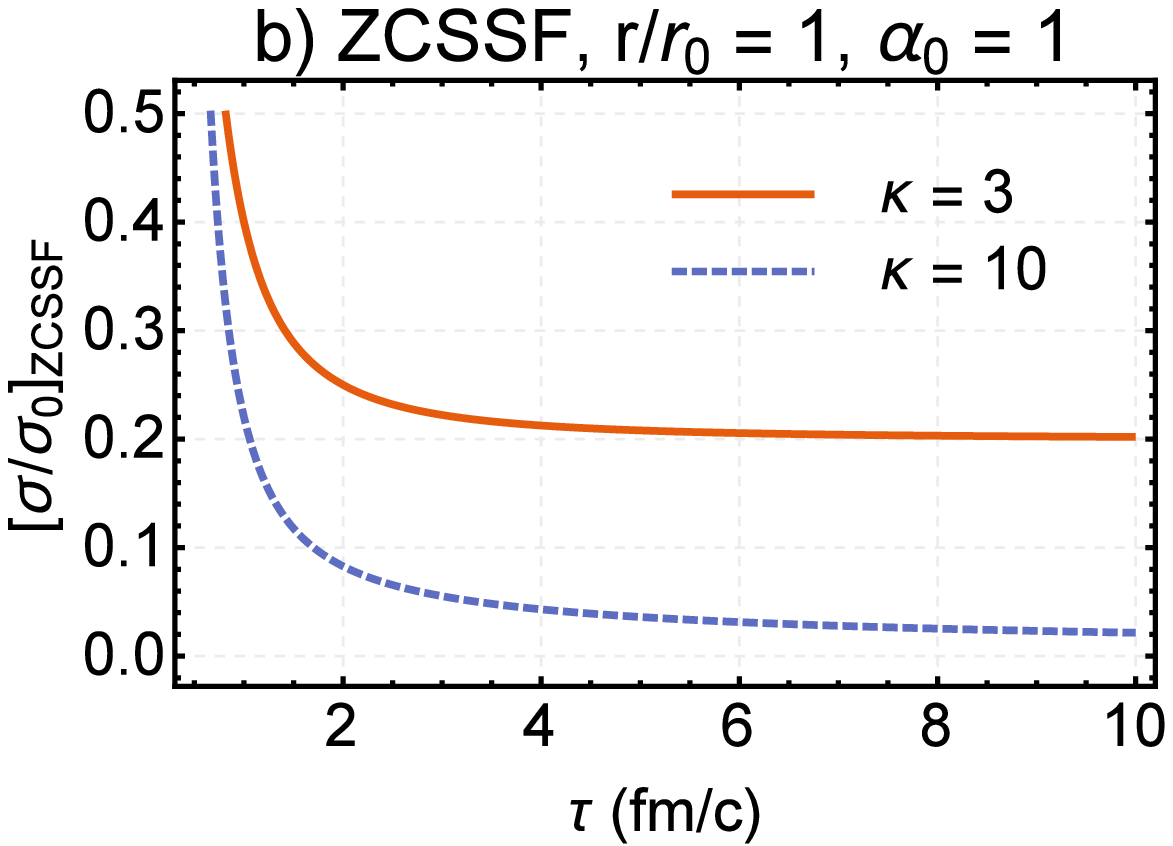}
\caption{(color online). The $\tau$ dependence of $[\sigma/\sigma_0]_{\text{\tiny{ZCSSF}}}$ is plotted for $r=r_0$, $\alpha_0=0.01$ (panel a) and $\alpha_{0}=1$ (panel b) as well as different values for $\kappa=3$ (solid orange curves) and $\kappa=10$ (dashed blue curves). For small values of $\alpha_{0}$, $[\sigma/\sigma_0]_{\text{\tiny{ZCSSF}}}$ remains almost constant
for $\kappa=3$, that characterizes the conformal EOS. For a smaller speed of sound, e.g. $\kappa=c_{s}^{-2}=10$, $[\sigma/\sigma_0]_{\text{\tiny{ZCSSF}}}$ decays as the system evolves.}\label{fig7}
\end{figure}

Let us now consider $\sigma$ defined in (\ref{D4}). The ratio $\sigma/\sigma_{0}$ for the ZCSSF and CMHD solutions are presented in \eqref{appC12} and \eqref{appC19}, respectively.\footnote{See Appendix \ref{appC} for a derivation of these two expressions in \eqref{appC12} for the ZCSSF and \eqref{appC19} for the CMHD solution.} Whereas for  $[\sigma/\sigma_{0}]_{\text{\tiny{ZCSSF}}}$ the parameter $\kappa$ turns out to be a free parameter, $\sigma/\sigma_{0}$ arising from the conformal solution for the magnetic field \eqref{U23} as well as the corresponding energy density \eqref{O24}, or equivalently \eqref{appC18}, are restricted to possess a conformal EOS $\epsilon=\kappa p$ with $\kappa=3$. Let us first consider $[\sigma/\sigma_{0}]_{\text{\tiny{ZCSSF}}}$ from \eqref{appC12}. In Fig. \ref{fig7}, we have plotted the $\tau$ dependence of this quantity for fixed $r=r_0$ and $\alpha_{0}=0.01$ [Fig. \ref{fig7}(a)] and $\alpha_{0}=1$ [Fig. \ref{fig7}(b)] for two different $\kappa=3$ (solid orange curves) and $\kappa=10$ (blue dashed curves).\footnote{Nonconformal values for $\kappa$ in $\epsilon=\kappa p$ are also used in \cite{kasza2018}.} Whereas for $\kappa=3$ and $\alpha=0.01$ $[\sigma/\sigma_{0}]_{\text{\tiny{ZCSSF}}}$ remains almost constant, for $\alpha_0=1$, it decreases very fast in the early stages after the collision, and then becomes saturated to a constant value $0.2$ in late times. In contrast, for $\kappa=10$, $[\sigma/\sigma_{0}]_{\text{\tiny{ZCSSF}}}$ decreases for both $\alpha_{0}=0.01$ and $\alpha_{0}=1$.
\par
\begin{figure*}[hbt]
\includegraphics[width=8cm,height=6cm]{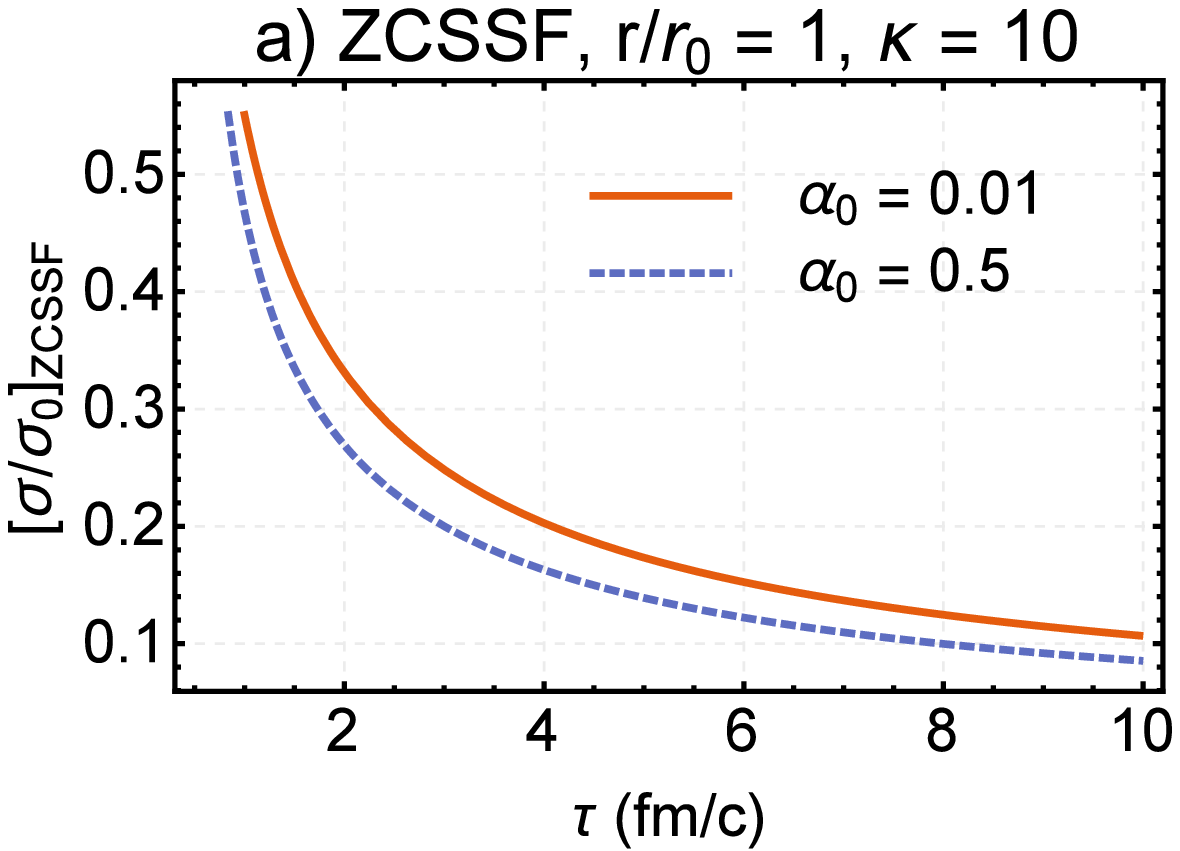}\hspace{0.3cm}
\includegraphics[width=8cm,height=6cm]{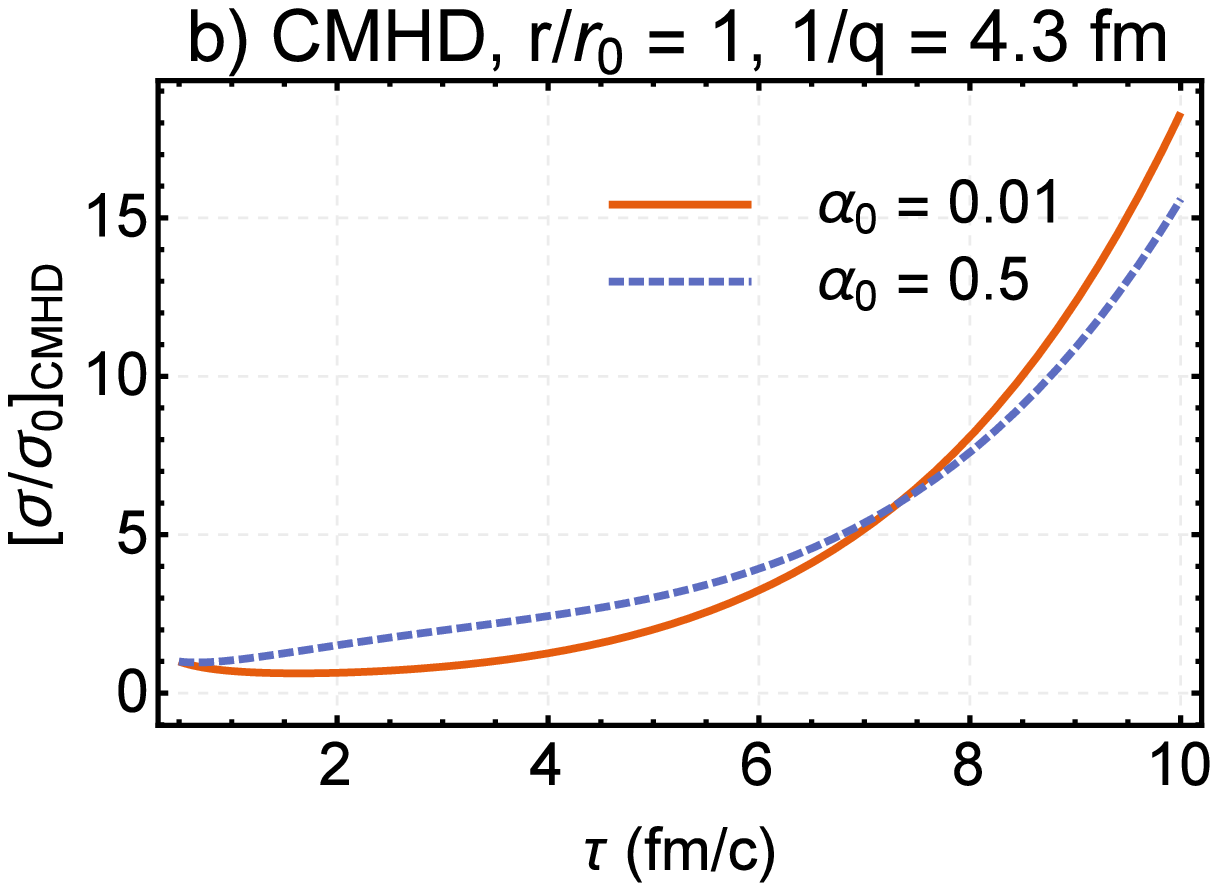}	
\caption{(color online).
The $\tau$ dependence of $[\sigma/\sigma_{0}]$ for the ZCSSF solution (panel a) and the CMHD solution (panel b) is plotted at $r=r_{0}$ and for $\alpha_{0}=0.01$ (solid orange curves) and $\alpha_{0}=0.5$ (dashed blue curves). For the CMHD solution $q$ is chosen to be $q=1/4.3$ fm$^{-1}$. The parameter $\kappa$, appearing in the EOS $\epsilon=\kappa p$ is chosen to be $\kappa=10$ for the ZCSSF and $\kappa=3$ for the CMHD solution. In contrast to $[\sigma/\sigma_{0}]_{\text{\tiny{ZCSSF}}}$, $[\sigma/\sigma_{0}]_{\text{\tiny{CMHD}}}$ increases with increasing $\tau$. Different choices for $\alpha_{0}$ affect the evolution of $\sigma/\sigma_0$.}\label{fig8}
\end{figure*}
\par		
\begin{figure*}[hbt]
\includegraphics[width=8cm,height=6cm]{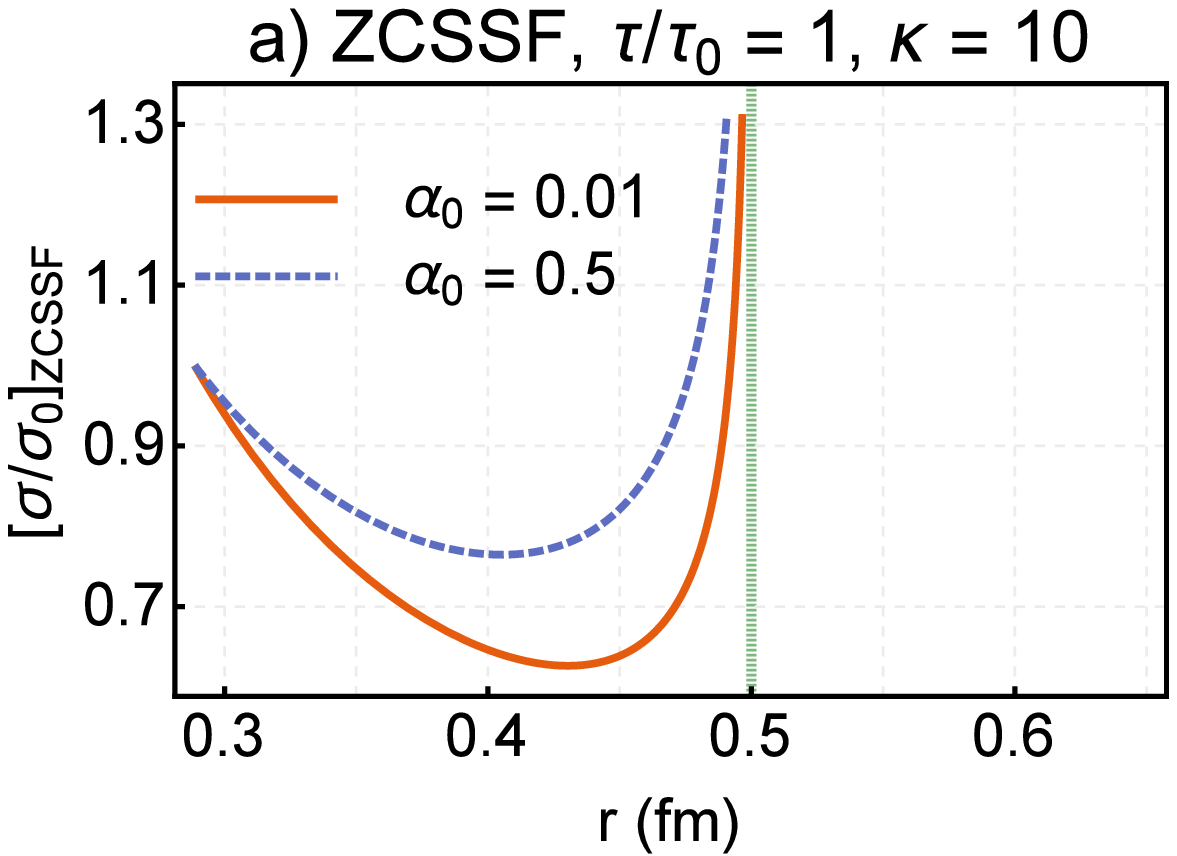}\hspace{0.3cm}
\includegraphics[width=8cm,height=6cm]{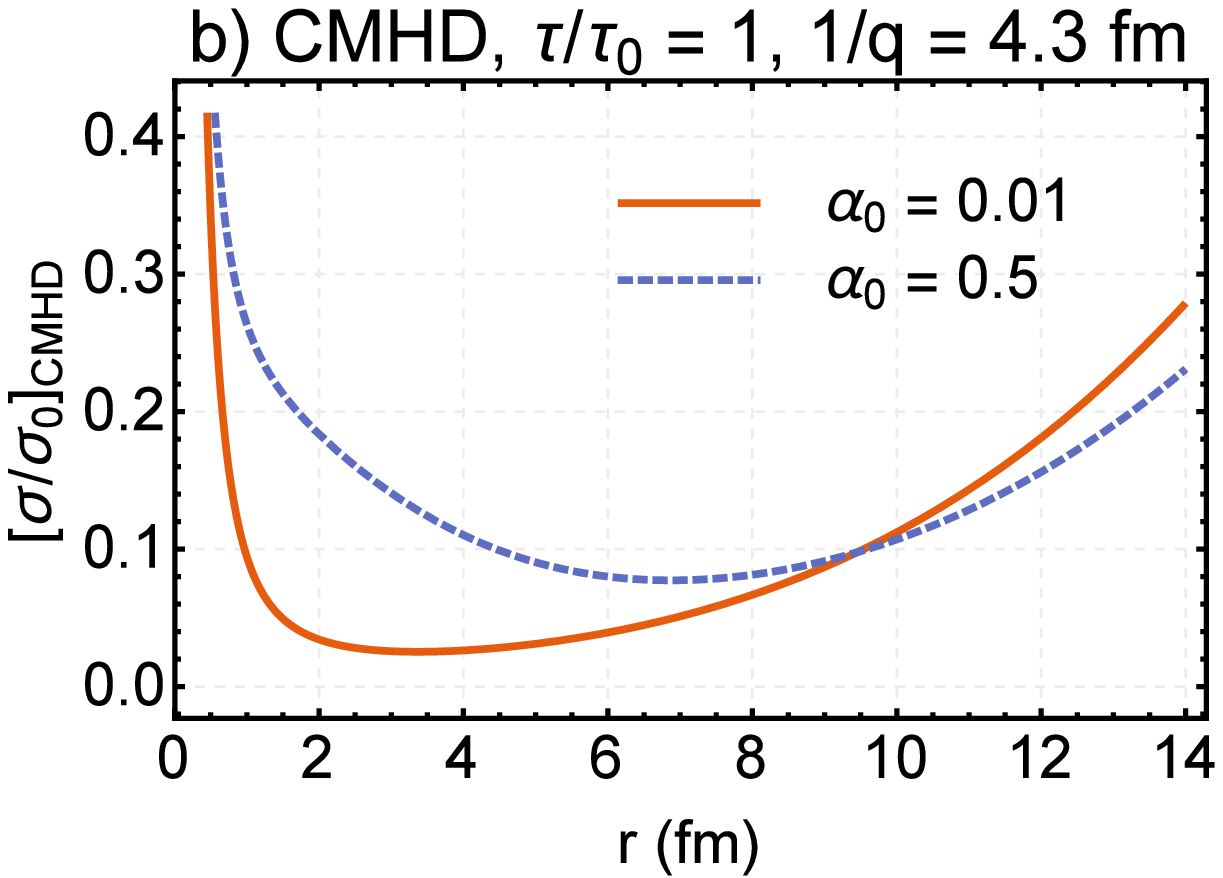}	
\caption{(color online).
The $r$ dependence of $[\sigma/\sigma_{0}]$ for the ZCSSF solution (panel a) and the CMHD solution (panel b) is plotted at $\tau=\tau_{0}$ and for $\alpha_{0}=0.01$ (solid orange curves)  and $\alpha_{0}=0.5$ (dashed blue curves). For the CMHD solution $q$ is chosen to be $q=1/4.3$ fm$^{-1}$. The parameter $\kappa$, appearing in the EOS $\epsilon=\kappa p$ is chosen to be $\kappa=10$ for the ZCSSF and $\kappa=3$ for the CMHD solution. The validity borderline for the ZCSSF solution at $\tau=r$ is demonstrated by a vertical green line. Both $[\sigma/\sigma_{0}]_{\text{\tiny{ZCSSF}}}$ and $[\sigma/\sigma_{0}]_{\text{\tiny{CMHD}}}$ decrease with increasing $r$ at early stages after the collision, and then, after passing a minimum, they increase with increasing $r$. Different choices for $\alpha_{0}$ significantly affect the evolution of $\sigma/\sigma_0$.}\label{fig9}
\end{figure*}	
\par		
\begin{figure*}[hbt]
\includegraphics[width=8cm,height=6cm]{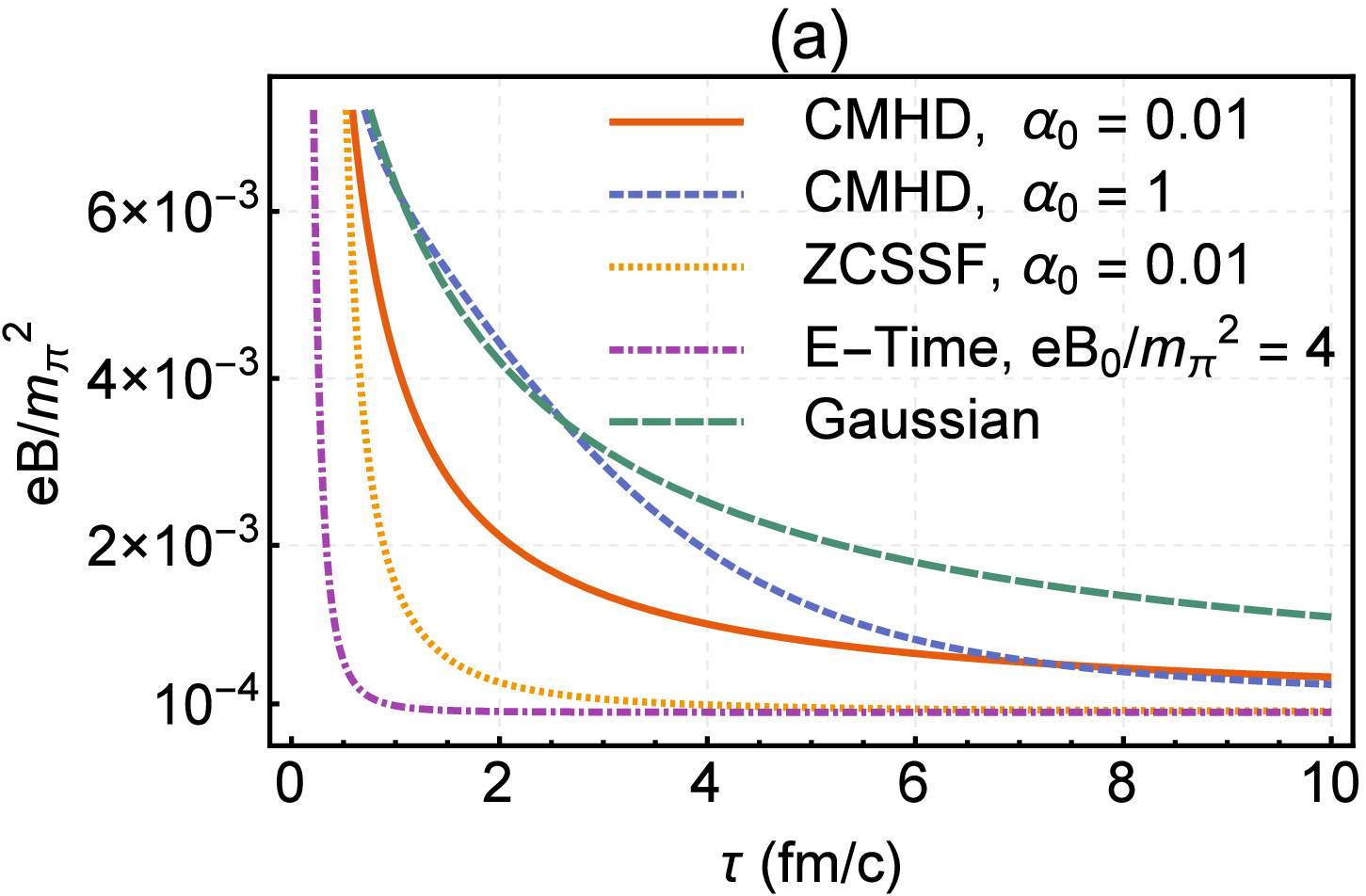}\hspace{0.3cm}		
\includegraphics[width=8cm,height=6cm]{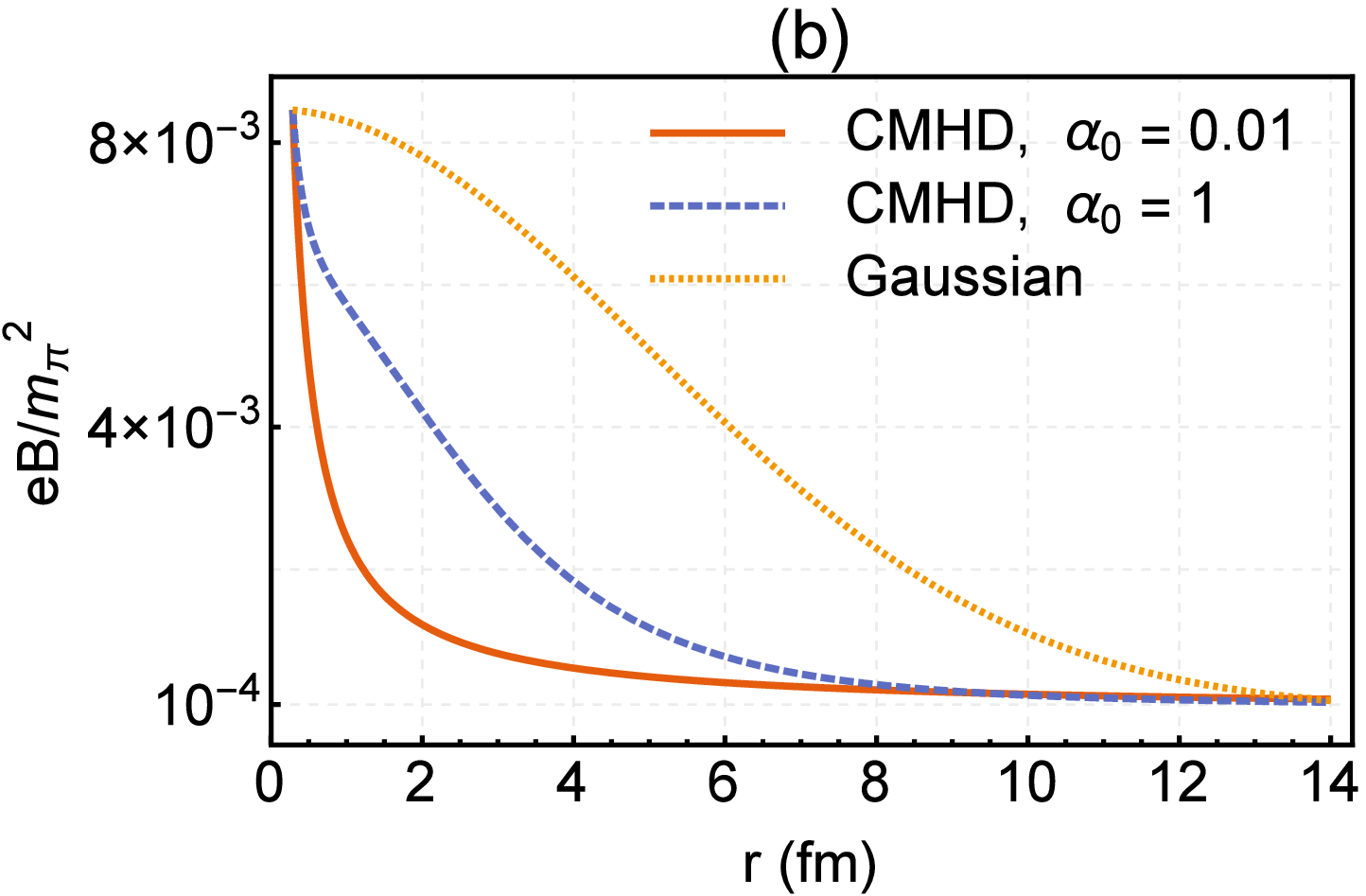}
\caption{(color online). (a) The $\tau$ dependence of $eB/m_{\pi}^2$ is plotted for the CMHD solution with $\alpha_{0}=0.01, 1$ as well as $1/q=4.3$ fm, the ZCSSF solution for $\alpha_{0}=0.01$, the early time dynamics \eqref{D6} (denoted by ``E-Time'') for $eB_{0}=4 m\pi^{2}$, and the phenomenological Gaussian ansatz for $r=r_0$ (denoted by ``Gaussian'') \eqref{D7}. The $\tau$ dependence of \eqref{D7} is roughly $\tau^{-1}$, while \eqref{D6} ones is approximately $\tau^{-3}$. The CMHD solution transmits between $\tau^{-1}$ at early times to $\tau^{-3}$ at late times. The CMHD solution with larger $\alpha_0$ turns out to be more similar to the Gaussian ansatz. The ZCSSF solution is more similar to the early time dynamics, and changing $\alpha_0$ does not significantly modify this behavior (not shown). (b) The $r$ dependence of $eB/m_{\pi}^2$ is plotted for the CMHD solution with $\alpha_{0}=0.01, 1$ and $1/q=4.3$ fm as well as for the Gaussian ansatz for $\tau=\tau_0$. Although for larger $\alpha_{0}$, the magnetic field survives up to larger distances to the origin, but the $r$ dependence of the CMHD solution remains quite different from the $r$ dependence of the Gaussian ansatz (\ref{D7}). }\label{fig10}	
\end{figure*}
\par
The effect of different choices of $\alpha_{0}$ on the evolution of  $[\sigma/\sigma_{0}]_{\text{\tiny{ZCSSF}}}$ is also demonstrated in Fig. \ref{fig8}(a), where the $\tau$ dependence of $[\sigma/\sigma_{0}]_{\text{\tiny{ZCSSF}}}$ is plotted for $r=r_{0}$, $\kappa=10$, and two different $\alpha_{0}=0.01$ (solid orange curve) and $\alpha_{0}=0.5$ (dashed blue curve). As it turns out, larger values of $\alpha_{0}$ suppress the evolution of $[\sigma/\sigma_{0}]_{\text{\tiny{ZCSSF}}}$. As concerns the evolution of  $[\sigma/\sigma_{0}]_{\text{\tiny{CMHD}}}$, it is plotted in Fig. \ref{fig8}(b) for $r=r_{0}, 1/q=4.3$ fm, and two different $\alpha_{0}=0.01$ (solid orange curve) and $\alpha_{0}=0.5$ (dashed blue curve).
In contrast to $[\sigma/\sigma_{0}]_{\text{\tiny{ZCSSF}}}$,$[\sigma/\sigma_{0}]_{\text{\tiny{CMHD}}}$ increases with increasing $\tau$.
\par
The $r$ dependence of  $[\sigma/\sigma_{0}]_{\text{\tiny{sol}}}$ for sol $=\{$ZCSSF,CMHD$\}$ is plotted in \ref{fig9} for two different $\alpha_{0}=0.01$ (solid orange curves) and $\alpha_{0}=0.5$ (dashed blue curves).
Neglecting the blow up at the $r=\tau$ validity borderline, demonstrated by the vertical green line in Fig. \ref{fig9}(a), $[\sigma/\sigma_{0}]_{\text{\tiny{ZCSSF}}}$ decreases with $r$, and, in contrast to its temporal evolution, it is enhanced for larger values of $\alpha_0$. The same is also true for the $r$ dependence of $[\sigma/\sigma_{0}]_{\text{\tiny{CMHD}}}$ from Fig. \ref{fig9}(b), which is maximized in regions where $\tau/r$ is far from unity. Comparing the $\tau$ and $r$ dependence of $[\sigma/\sigma_{0}]_{\text{\tiny{CMHD}}}$ from Figs. \ref{fig8}(b) and \ref{fig9}(b), it turns out that at any fixed value of $r$, $\sigma_{\text{\tiny{CMHD}}}$ increases significantly with $\tau$, while for fixed values of $\tau$, $\sigma$ starts with a sharp decline to a minimum at some $r>r_0$. It then increases with increasing $r$. The difference between the temporal evolution and radial distribution for the CMHD solution, is because of the breakdown of $r\leftrightarrow\tau$ symmetry in $\epsilon$ from \eqref{O24}, or equivalently from \eqref{appC18}.
\par	
Let us notice, at this stage, that in \cite{roy2015} the value of $\sigma_0$ is reported to be of order $10^{-2}$ in central HICs. This is a small value that makes the effects arising from magnetic fields much inferior than that from the hydrodynamical expansion. However, if in a particular event the initial value of $\sigma$ is not very small,
in late times or far from the center of the collision, the magnetic energy density $B^{2}$ may compete with fluid energy density $\epsilon$. This may be a motivation for studying the effects arising from magnetic fields on the acceleration of the fluid.
\par
We close this section with a qualitative comparison of our solutions with few numerical results from the literature. The magnitude of the magnetic field generated in HIC experiments is usually reported in the form $eB/m_\pi^2$. According to \cite{warringa2007,Huang-review,skokov2009, zakharov2014,roy2015}, at RHIC center of mass energies $\sqrt{s_{NN}}=200\text{GeV}$, $eB/m_\pi^2$ is estimated to be of order $eB\sim 5m_{\pi}^{2}$. The aforementioned value is the event-by-event average value of $B_y$, and, one should bear in mind that these values correspond to very early stages of the collision. External sources quickly vanish, and the magnetic field declines in a nonconductive gluon dominated medium. A formula for the early time dynamics of the magnetic field is given by (see \cite{Huang-review} and the references therein),
\begin{equation}\label{D6}
eB_y(\tau) = \frac{eB_y\left(0\right)}{\left(1+\tau^2/t_B^2\right)^{3/2}},
\end{equation}
where $t_B=0.065$ fm/c at RHIC top energies. Using \eqref{D6}, one arrives for $eB_{y}(0)=5 m_\pi^{2}$ and $\tau_{B}=0.065$ fm/c at $eB_y \sim 10^{-2}m_\pi^2$ for $\tau=0.5\fmc$. In \eqref{D6}, there is no information about spatial distribution of $B$, and it thus cannot be used for fixing $B_\text{init.}$. In \cite{roy2017,zakharov2014}, another useful ansatz is suggested for magnetic fields arising from near-central collisions
\begin{equation}\label{D7}
\frac{eB(\tau,r)}{m_{\pi}^2} = \inv{a_1+b_1\tau}\exp\left(-\frac{r^2}{\sigma_r^2}\right).
\end{equation}
Here, numerical parameters are given by $a_1=78.2658$, $b_1=79.5457\fm^{-1}$ and $\sigma_r=3.5\fm$ for a zero impact parameter $b=0$.
In Fig. \ref{fig10}(a), the $\tau$ dependence of  $eB/m_{\pi}^{2}$ is plotted for $eB$ arising from the CMHD solution with $\alpha_{0}=0.01$ (solid orange curve) and $\alpha_{0}=1$ (dashed blue curve), and from the ZCSSF solution with $\alpha_0=0.01$ (dotted yellow curve). Moreover, $eB/m_{\pi}^{2}$ is plotted for the early time magnetic field from \eqref{D6} with $eB_{y}(0)=4 m_{\pi}^{2}$ (dotted-dashed magenta curve) and the Gaussian ansatz \eqref{D7} (green dashed curve). To arrive at $eB_{y}(0)=4 m_{\pi}^{2}$ for \eqref{D6}, we compared $eB$ from \eqref{D7} at $\tau=\tau_0=0.5$ fm/c and $r=r_{0}=0.5 c_s$ with $eB$ from \eqref{D6} at $\tau=\tau_0=0.5$ fm/c, and arrived at $eB_{0}\approx 4 m_{\pi}^{2}$.
As it turns out, the early time dynamics \eqref{D6} is a decay of $\tau^{-3}$ type leading to a fast decay of the magnetic field. On the other hand,  \eqref{D7} is a combination of a $\tau^{-1}$ temporal decay with a Gaussian radial distribution, and turns out to be slower than that arising from the Bjorken MHD, i.e. $B\propto \tau^{-1}$ from \eqref{A22}. The ZCSSF solution for small values of $\alpha_{0}$ is very close to the early time dynamics, the CMHD solution transmits between a $\tau^{-1}$ decay of Gaussian type magnetic field from \eqref{D7} at early times to a $\tau^{-3}$ from \eqref{D6} decay at late times [see Fig. \ref{fig10}(a)]. This is consistent with the phenomenological picture of the QGP evolution, that it transmits from an early Bjorken flow to a later Hubble expansion \cite{kolb2003}. The $3+1$ dimensional self-similar flow, being a Hubble expansion, may be considered as an effective picture in the later stages of the QGP spacetime history. By virtue of these results, one may conclude that the CMHD solution probably gives the best qualitative picture of the magnetic field evolution in all stages of the QGP evolution from $\tau_{0}\sim 0.5$ fm to $\tau_{f}\sim 10$ fm/c.
In Fig. \ref{fig10}(b), the $r$ dependence of $eB/m_{\pi}^{2}$ is plotted for $B_{\text{\tiny{CMHD}}}$ with $\alpha_{0}=0.01$ (solid orange curve), $\alpha_{0}=1$ (dashed blue curve) and the Gaussian ansatz \eqref{D7} (dotted yellow curve). Although for $\alpha_{0}=1$,
 the magnetic field survives in larger distances from the origin, but the $r$ dependence of the CMHD solution remains quite different from the $r$ dependence of the Gaussian ansatz (\ref{D7}).
	\section{Concluding remarks}\label{sec8}
	\setcounter{equation}{0}
In this paper, we studied the evolution of magnetic fields within an infinitely conductive fluid, using, in particular,  the $3+1$ dimensional self-similar and Gubser flows. We followed a systematic procedure, and derived the corresponding flows and magnetic fields to these setups. This procedure is mainly based on the application of appropriate spacetime symmetries, and can be summarized as follows: In general, a solution to RHD may be obtained by considering a set of isometries $\mathcal{I}$. This set must at least contain three independent isometries to fix the four-velocity $u^{\mu}$. There may also exist a scalar, $\Gamma$, that is invariant under all isometries in $\mathcal{I}$.
If this is the case, the partial differential equations of RHD reduce to ordinary differential equations with $\Gamma$ being an independent variable. In addition, the four-velocity may be proportional to partial derivatives of $\Gamma$ with respect to given coordinates.
For the Bjorken, $3+1$ dimensional self-similar and Gubser flows, $\Gamma$ was found to be $\Gamma=\{\tau,\varrho,G\}$, respectively [see \eqref{A13}, \eqref{S1}, and  \eqref{O19}]. It is also possible to derive the same velocity four-vectors by relaxing some of the aforementioned isometries in $\mathcal{I}$. This can be done by introducing a proper scalar $\vartheta$, that must respect the remaining set of isometries $\mathcal{S}$. In this way, apart from the four-velocity, the energy density remains also the same as is obtained by application of $\mathcal{I}$.
In the case of an ideal fluid, this is because the Euler equation prevents the pressure and energy density to obtain $\vartheta$ dependence. As a simple example, we considered the spacetime rapidity $\eta$ as the proper scalar in the Bjorken flow. Here, although the reduced set of isometries did not include the boost invariance, the Euler equation forced the pressure and energy density to be boost invariant.\footnote{This is the essence of the $1+1$ self-similar flow \cite{csorgo2002-2}.}  We showed that this trick is indeed crucial for the generalization of RHD solutions to ideal MHD. If $\mathcal{S}$ contains at least two independent isometries, the homogeneous Maxwell equations can significantly be simplified. Assuming the ideal MHD limit, one is then able to determine field strength tensor, and eventually the magnetic field, up to two unknown functions of $\vartheta$. These functions, that are referred to as scaling functions, can be found by solving the MHD Euler equation.
\par
To set a benchmark for this procedure, we reproduced the previous results on the transverse MHD from  \cite{rischke2015,shokri2017}. In addition to these results, we found that the induced current vanishes if the boost invariance is not relaxed. We then applied this procedure to the case of $3+1$ dimensional self-similar and Gubser flows in order to study the consequences of the QGP transverse expansion on the lifetime of the magnetic field. Here, in contrast to the Bjorken $1+1$ dimensional case, the dependence of the magnitude of magnetic field on the proper scalar $\vartheta$ did not vanish. This was because of additional terms in the MHD Euler equation, that did not appear in the Bjorken case. In the $3+1$ dimensional self-similar flow, the aforementioned $\vartheta$ dependence turned out to be mandatory. In addition, the Euler equation transformed into one equation for two unknown functions. We found a physically acceptable solution by assuming that the induced current vanishes (the ZCSSF solution).
\par
As concerns the implementation of the Gubser flow into relativistic MHD, it turned out that in the flat space, it is impossible to introduce a proper scalar that respects the desired symmetries. It was this lack of a proper scalar that led to the elimination of the magnetic field in transverse directions with respect to the beamline. This makes the corresponding magnetic field unappropriate for the purpose of HICs. We could resolve this problem by exploiting the technique of Weyl transformations from \cite{gubser2010-2}. The resulting solution was referred to as the CMHD solution. According to our numerical results from Sec. \ref{sec7}, the CMHD solution transforms from an early time Bjorken ideal MHD solution to a late time $3+1$ self-similar MHD solution as the system evolves. Moreover, the corresponding induced current automatically vanishes. This is consistent with the numerical results from \cite{zakharov2014}. Let us notice that if the induced current vanishes, the electric conductivity of the QGP becomes almost irrelevant for the magnetic field evolution. This is also in agreement with our previous results from \cite{shokri2017}, where we showed that in nonideal transverse MHD the magnetic field is not significantly modified by the finite electric conductivity, and that a finite electric conductivity does not necessarily lead to a deviation from the ideal MHD. It is noteworthy to mention that in order for the magnetic field to deviate from the ideal regime, the boost invariance must also be broken. In the nonideal transverse MHD, although the magnitude of the magnetic field remains boost invariant, but its direction explicitly depends on the spacetime rapidity $\eta$ \cite{shokri2017}.
\par
According to our results from Sec. \ref{sec6}, in any fixed distances from the origin, the evolution of the transverse component of the CMHD solution is similar to that of the Bjorken solution from the ideal transverse MHD [see  \eqref{U23}]. This is also consistent with the numerical results from \cite{zakharov2014,inghirami2018}. We also showed that, once  a small longitudinal component for the magnetic field is assumed, the radial expansion of the fluid does not significantly modify the magnetic field evolution. Let us, however, notice that the longitudinal component of the magnetic field vanishes only if the fluid transverse size is assumed to be infinitely large. In a more realistic setup, however, even an initially small longitudinal component can be enhanced in certain regions of the spacetime (see Sec. \ref{sec7}). We thus conclude that there may be an unexplored role of the longitudinal component of the magnetic field in HICs, which deserves to be taken into account.
\par
At this stage, let us notice that the ideal MHD limit, which is used in the present work, is based on the assumption of an infinitely large electric conductivity of the QGP. This leads, however, to a large magnetic Reynolds number $R_m\equiv \sigma_{e} Lu$. Here, $L$ and $u$ are a typical size and velocity of the fluid. For $L\sim 10$ fm and $u\sim 0.5$, we arrive for $R_{m}\gg 1$ at $\sigma_{e}\gg 40$ MeV. Such a large electric conductivity is much larger than typical lattice QCD results for $\sigma_{e}$ from, e.g., \cite{skellerud2015}. It would be thus useful to extend the present work to nonideal MHD, and look for a generalization of self-similar and Gubser solutions of ideal MHD to a resistive fluid (nonideal MHD), as is already performed for the $1+1$ dimensional Bjorken flow in \cite{shokri2017}. Another important extension is related to the assumed rotational invariance around the beamline, which turns out to be a poor approximation if the collision is not near-central. Interestingly, a novel analytical solution for off-central HICs is recently introduced in \cite{romatschke2018}, that calls for a generalization to ideal and noideal MHD, using the symmetry arguments presented in this paper.
\section{Acknowledgments}
The authors thank S. M. A. Tabatabaee for useful discussions. M.~S. thanks A.~Yousefi Sostani, S.~Pu and M.~Csan$\grave{\mbox{a}}$d for private communications. N.S. thanks N.~Brambilla for the hospitality during her stay in the theoretical physics department of the Technical University of Munich (TUM), where the final stage of this work is performed. Her visit is supported by the DFG cluster of excellence `Origin and Structure of the Universe'.\footnote{www.universe-cluster.de}
\begin{appendix}
		\section{Useful definitions}\label{appA}
In this appendix, we present a quick review of mathematical concepts used in this work (see \cite{generalrefs} for more details).
\par
The covariant derivative of a vector is given by
\begin{eqnarray}\label{appA1}
\cd{\mu}{W^\nu}&=&\partial_\mu W^\nu +\christoffel{\nu}{\mu}{\rho}W^\rho,\nonumber\\
\cd{\mu}{W_\nu}&=&\partial_\mu W_\nu -\christoffel{\rho}{\mu}{\nu}W_\rho.
\end{eqnarray}
For an arbitrary rank tensor $Q^{\mu_1\mu_2\cdots}_{\nu_1\nu_2\cdots}$, the covariant derivative is found by assuming a multiplication of vectors $W^{\mu_1}V^{\mu_2}U_{\nu_1}X_{\nu_2}\cdots$. Some useful identities are
\begin{eqnarray}\label{appA2}
\cd{\mu}{W^\mu}&=&\inv{\sqrt{-g}}\partial_\mu W^\mu,\\
\cd{\mu}{Y^{\mu\nu}}&=&\inv{\sqrt{-g}}\partial_\mu\left(\sqrt{-g}Y^{\mu\nu}\right)+\christoffel{\nu}{\mu}{\rho}Y^{\mu\rho}.\label{appA3}
\end{eqnarray}
For the antisymmetric tensor $F^{\mu\nu}$, we have, in particular,
\begin{equation}\label{appA4}
\cd{\mu}{F^{\mu\nu}}=\inv{\sqrt{-g}}\partial_\mu\left(\sqrt{-g}F^{\mu\nu}\right).
\end{equation}
The Lie derivative of a vector with respect to $\xi^\mu$ is given by
\begin{eqnarray}\label{appA5}
\lieder{\xi}{W^\mu}&=&\xi^\nu\partial_\nu W^\mu - W^\nu\partial_\nu \xi^\mu,\nonumber\\
\lieder{\xi}{V_\mu}&=&\xi^\nu\partial_\nu V_\mu + V_\nu\partial_\mu \xi^\nu.
\end{eqnarray}
In the index free notation, i.e. $W=W^{\mu}\partial_\mu$, the Lie derivative is replaced by the Lie bracket as
\begin{equation}\label{appA6}
\lieder{\xi}{W^\mu}=[\xi,W]^\mu.
\end{equation}
The Lie derivative of the metric is given by
\begin{equation}\label{appA7}
\lieder{\xi}{g_{\mu\nu}}=\cd{\mu}{\xi_\nu}+\cd{\nu}{\xi_\mu}.
\end{equation}
A Killing vector $\xi$ is a vector that satisfies the Killing equation
\begin{equation}\label{appA8}
\lieder{\xi}{g_{\mu\nu}}=0.
\end{equation}
In the same spirit, in four-dimensional spacetime, a conformal Killing vector $\xi$ satisfies
\begin{equation}\label{appA9}
\lieder{\xi}{g_{\mu\nu}}=\inv{2}(\nabla.\xi) g_{\mu\nu}.
\end{equation}
The $dS^3$ is the set of points in $M^{3,1}$ that satisfies
\begin{equation}\label{appA10}
-(X^0)^2+(X^1)^2+(X^2)^2+(X^3)^2=L^2.
\end{equation}
Assuming $L=1$, for the sake of simplicity, the $X^\mu$ coordinates can be re-parameterized using
\begin{eqnarray}\label{appA11}
\hspace{-1cm}
X^0 &=& \sinh\rho,\quad X^3=\cosh\rho\cos\theta,\nonumber\\
\hspace{-1cm}
X^1 &=& \cosh\rho\sin\theta\cos\phi,\quad X^2= \cosh\rho\sin\theta\sin\phi.
\end{eqnarray}
In these coordinates, we have ${\rd{s}}^2=-{\rd{X^0}}^2+{\rd{X^1}}^2+{\rd{X^2}}^2+{\rd{X^3}}^2$ takes the form \eqref{U3}.
\section{Scaling functions in the conformal RHD}\label{appB}
In this appendix, we briefly comment on scaling functions in the conformal RHD. As it was mentioned in Sec.  \ref{sec6}, one may assume $\theta$ to be a proper similarity variable. This also allows the temperature to possess a $\theta$ dependency as
\begin{eqnarray}\label{appB1}
T=\frac{\hat{T}_0}{\tau}\left(\cosh\rho\right)^{-2/3}\mathcal{T}(\theta).
\end{eqnarray}
Here, $\mathcal{T}(\theta)$ is an arbitrary scaling function. Using  \eqref{O23}, we obtain for a baryon-free quark matter,
\begin{eqnarray}\label{appB2}
Ts=\epsilon+p=4p.
\end{eqnarray}
For an ideal nondissipative fluid the entropy density satisfies \eqref{A10}. This leads to
\begin{eqnarray}\label{appB3}
s=\frac{\hat{s}_0}{\tau^3}\left(\cosh\rho\right)^{-2}\mathcal{S}(\theta).
\end{eqnarray}
According to \eqref{M18}, $p$ from \eqref{appB2} is not function of $\theta$. We thus have $\mathcal{T}(\theta)\mathcal{S}(\theta)=1$.
\vspace{0.5cm}
\section{Matching free parameters in the ZCSSF and  CMHD solutions}\label{appC}
In Sec. \ref{sec7}, we compared different features of the ZCSSF and CMHD solutions from \eqref{S25} and \eqref{U26}. To do this, we had to bring the free parameters appearing in these solutions into connection. In this appendix, we explain how free parameters in these solutions are matched. Before starting, let us remind that two parameters appear in each of these solutions. We thus need two equations to fix them. Here, we use the magnitude of the magnetic field at some fixed point and the ratio $B_y/B_z$ at the same point, and reexpress free parameters in terms of these quantities. To find them, we first find a relation between the corresponding magnetic four-vectors and the local magnetic three-vectors to these solutions. Let us consider $B^{\mu}$ in ordinary Minkowski coordinates $(t,x,y,z)$,
\begin{equation}\label{appC1}
B^\mu = \left(zB^\eta,-yB^\phi,xB^\phi,tB^\eta\right).
\end{equation}
Performing an appropriate boost to the LRF of the fluid at $(t,x,y,z)=(\tau,r,0,0)$, we arrive for the ZCSSF and Gubser flows
\begin{widetext}
\begin{equation}\label{appC2}
u^\mu_{\tiny{\mbox{ZCSSF}}}=x^\mu/\varrho,\quad
u^\mu_{\tiny{\mbox{Gubser}}}=\left(\cosh\varTheta\cosh\eta,\sinh\varTheta\cos\phi,\sinh\varTheta\sin\phi,\cosh\varTheta\sinh\eta\right),
\end{equation}
\end{widetext}
at
\begin{equation}\label{appC3}
\boldsymbol{B} = (0, rB^\phi,\tau B^\eta).
\end{equation}	
The Lorentz transformation tensor associated with this boost reads
\begin{equation}\label{appC4}
\Lambda^0_\nu = -u_\nu,\quad \Lambda^i_j = \delta^i_j+\frac{u^iu_j}{1+u^0}.
\end{equation}
At the initial point $(\tau_{0},r_{0})$, \eqref{appC3} gives rise to
\begin{equation}\label{appC5}
\boldsymbol{B}_{\text{init.}} = (0, r_0B^\phi_0,\tau_0B^\eta_0),
\end{equation}		
where $B_{0}^{\phi/\eta}=B^{\phi/\eta}(\tau_{0},r_{0})$. Plugging at this stage, \eqref{appC3} into $\alpha$ from \eqref{D2}, we arrive first at
\begin{equation}\label{appC6}
\alpha=\frac{\tau B^\eta}{r B^\phi}.
\end{equation}
At the initial point, \eqref{appC6} then reads
\begin{equation}\label{appC7}
\alpha_0 = \frac{\tau_0B^\eta_0}{r_0B^\phi_0}.
\end{equation}
We are now in a position to use $\alpha_0$ from \eqref{appC7} to fix free parameters $\mathcal{A}_2$, $a_0$ and $\beta_0$ in  \eqref{S25}, \eqref{S31} and  \eqref{U26}.
\par
For the ZCSSF solution from \eqref{S25}, we get
\begin{equation}\label{appC8}
\mathcal{A}_2 = \alpha_0\frac{\tau_0}{r_0}.
\end{equation}
Plugging \eqref{appC8} into \eqref{S25}, and using  \eqref{appC5}, we arrive at
\begin{equation}\label{appC9}
\bigg[\frac{B}{B_{\text{init.}}}\bigg]_{\text{\tiny{ZCSSF}}}=\frac{1}{\sqrt{1+\alpha_0^2}}\left(\frac{\varrho_0}{\varrho}\right)^2\sqrt{\left(\frac{r_0}{r}\right)^2+\alpha_0^2\left(\frac{\tau_0}{\tau}\right)^2},
\end{equation}
with $B=|\boldsymbol{B}|$ and $B_{\text{init.}}=|\boldsymbol{B}_{\text{init.}}|$, where $\boldsymbol{B}$ and $\boldsymbol{B}_{\text{init.}}$ are from (\ref{appC3}) and (\ref{appC5}), respectively. Plugging first \eqref{S25} into $\lambda$ from \eqref{D3}, and using \ref{appC8}, leads to
\begin{equation}\label{appC10}
\lambda_{\text{\tiny{ZSSF}}}=\alpha_0\left(\frac{\tau_0}{\tau}\right)\left(\frac{r}{r_0}\right)\left[1+\alpha_0^2\left(\frac{\tau_0}{\tau}\right)^2\left(\frac{r}{r_0}\right)^2\right]^{-1/2}.
\end{equation}
Using \eqref{appC10}, we then arrive at
\begin{equation}\label{appC11}
\bigg[\frac{\lambda}{\lambda_0}\bigg]_{\text{\tiny{ZCSSF}}}=\frac{\tau_0}{\tau}\frac{r}{r_0}\sqrt{\frac{1+\alpha_0^2}{1+\alpha_0^2\left(\frac{\tau_0}{\tau}\right)^2\left(\frac{r}{r_0}\right)^2}}.
\end{equation}
Finally, using \eqref{S10} and \eqref{appC5} for the ZCSSF solution (\ref{S25}), $\sigma$ from \eqref{D4} is given by
\begin{equation}\label{appC12}
\sigma_{\text{\tiny{ZCSSF}}} = \frac{\sigma_0}{1+\alpha_0^2}\left(\frac{\varrho_0}{\varrho}\right)^{1-3/\kappa}\left[\left(\frac{r_0}{r}\right)^2+\alpha_0^2\left(\frac{\tau_0}{\tau}\right)^2\right],
\end{equation}
with $\sigma_0\equiv\sigma(\tau_0,r_0)\equiv \frac{B_{\text{init.}}}{2\epsilon_{0}}$. 	
\par
Let us now consider the CMHD solution \eqref{U23} [or equivalently \eqref{U22} with $\hat{B}^{\mu}$ from \eqref{U21}]. Plugging \eqref{U22} into $\alpha_0$ from \eqref{appC7}, we first arrive  at
\begin{equation}\label{appC13}
\alpha_0=\frac{\tau_0}{r_0}\frac{\hat{B}^\eta_0}{\hat{B}^\phi_0}=\frac{\tau_0}{r_0}\beta_0\sin^{2}\theta_0,
\end{equation}
where $\theta=\theta(\tau_0,r_0)$. Using then \eqref{U4}, \eqref{appC13} yields
\begin{equation}\label{appC14}
\beta_0=\alpha_0\frac{[1+q^4\left(\tau_0^2-r_0^2\right)^2+2q^2\left(\tau_0^2+r_0^2\right)]}{4q^2r_0\tau_0}.
\end{equation}
Using \eqref{appC14}, $B/B_\text{init.}$ is found for the CMHD solution. In terms of $(\rho,\theta)$ coordinates appearing in \eqref{U3} and \eqref{U4}, it is given by
\begin{eqnarray}\label{appC15}
\lefteqn{\hspace{-1cm}\bigg[\frac{B}{B_{\text{init.}}}\bigg]_{\text{\tiny{CMHD}}}=\frac{1}{\sqrt{1+\alpha_{0}^{2}}}\left(\frac{\tau_{0}}{\tau}\right)^{2}\left(\frac{\cosh\rho_{0}}{\cosh\rho}\right)^{2}
}\nonumber\\
&&\times \left[\alpha_0^2+\left(\frac{\tau_0}{\tau}\right)^2\left(\frac{r}{r_0}\right)^2\left(\frac{\sin^{4}\theta_0}{\sin^{4}\theta}\right)\right]^{1/2}.
\end{eqnarray}
Plugging \eqref{U23} [or equivalently \eqref{U22} with $\hat{B}^{\mu}$ from \eqref{U21}] into $\lambda$ from \eqref{D3}, we get
\begin{eqnarray}\label{appC16}
\lambda_{\text{\tiny{CMHD}}}=\frac{\alpha_0 r_0\tau\sin^{2}\theta}{\sqrt{\alpha_0^2r_0^2\tau^2\sin^{4}\theta+r^2\tau_0^2\sin^{4}\theta_0}}.
\end{eqnarray}
Here, $\beta_{0}=\frac{\alpha_{0}r_{0}}{\tau_{0}\sin^{2}\theta_{0}}$ from (\ref{appC13}) is used. From  \eqref{appC16}, we obtain
\begin{equation}\label{appC17}
\bigg[\frac{\lambda}{\lambda_0}\bigg]_{\text{\tiny{CMHD}}}=r_0\tau\sin^{2}\theta\sqrt{\frac{1+\alpha_0^2}{\alpha_0^2r_0^2\tau^2\sin^{4}\theta+r^2\tau_0^2\sin^{4}\theta_0}}.
\end{equation}		
To determine $\sigma$ from \eqref{D4} for the CMHD solution, let us first consider $\epsilon$ from \eqref{O24}. Bearing in mind that in the coordinates appearing in (\ref{U3}) and \eqref{U4}, $1+G^{2}=\cosh^{2}\rho$, and defining $\epsilon_{0}\equiv\frac{\bar{\epsilon}_{0}}{\tau_{0}^{4}\left(\cosh^{2}\rho_{0}\right)^{4/3}}$, we arrive at
\begin{eqnarray}\label{appC18}
\epsilon=\epsilon_{0}\left(\frac{\tau_{0}}{\tau}\right)^{4}\left(\frac{\cosh\rho_{0}}{\cosh\rho}\right)^{8/3}.
\end{eqnarray}
Using then \eqref{appC15}, we finally obtain
\begin{eqnarray}\label{appC19}
\hspace{-1.5cm}\sigma_{\text{\tiny{CMHD}}}&=&\frac{\sigma_0}{1+\alpha_0^2}\left(\frac{\cosh\rho_0}{\cosh\rho}\right)^{4/3}
\nonumber\\
&&\hspace{-0.5cm}\times\left[\alpha_0^2+\left(\frac{\tau_0}{\tau}\right)^2\left(\frac{r}{r_0}\right)^2\left(\frac{\sin^{4}\theta_0}{\sin^{4}\theta}\right)\right],
\end{eqnarray}
with $\sigma_{0}\equiv \frac{B^{2}_{\text{\tiny{init.}}}}{2{\epsilon}_{0}}$.
\end{appendix}


\begin{thebibliography}{9}
\bibitem{rajagopal2018}
U.~G\"ursoy, D.~Kharzeev, E.~Marcus, K.~Rajagopal and C.~Shen,
\textit{Charge-dependent flow induced by magnetic and electric fields in heavy ion collisions},
arXiv:1806.05288 [hep-ph].
\bibitem{Huang-review}
X.~G.~Huang,
\textit{Electromagnetic fields and anomalous transports in heavy ion collisions --- A pedagogical review,}
Rept.\ Prog.\ Phys.\  {\bf 79}, no. 7, 076302 (2016),
arXiv:1509.04073 [nucl-th].
\bibitem{warringa2007}
D.~E.~Kharzeev, L.~D.~McLerran and H.~J.~Warringa,
\textit{The effects of topological charge change in heavy ion collisions: 'Event by event P and CP violation'},  Nucl.\ Phys.\ A {\bf 803}, 227 (2008),
arXiv:0711.0950 [hep-ph].
\bibitem{skokov2009}
V. Skokov, A. Y. Illarionov and V. Toneev, \textit{Estimate of
the magnetic field strength in heavy ion collisions}, Int.\ J.\ Mod.\ Phys.\ A {\bf 24}, 5925 (2009),
arXiv:0907.1396 [nucl-th].
\bibitem{zakharov2014}
		B.~G.~Zakharov,
		\textit{Electromagnetic response of quark-gluon plasma in heavy ion collisions},
		Phys.\ Lett.\ B {\bf 737}, 262 (2014),
arXiv:1404.5047 [hep-ph].
		\bibitem{hatsuda-book}
		K.~Yagi, T.~Hatsuda and Y.~Miake,
		\textit{Quark-gluon plasma: From big bang to little bang},
Cambridge University Press, Cambridge, 2005 and Camb.\ Monogr.\ Part.\ Phys.\ Nucl.\ Phys.\ Cosmol.\  {\bf 23}, 1 (2005).
		\bibitem{romatschke2017}
		P.~Romatschke and U.~Romatschke,
		\textit{Relativistic fluid dynamics in and out of equilibrium -- Ten years of progress in theory and numerical simulations of nuclear collisions},
		arXiv:1712.05815 [nucl-th].
		\bibitem{rischke2015}
		V.~Roy, S.~Pu, L.~Rezzolla and D.~Rischke,
		\textit{Analytic Bjorken flow in one-dimensional relativistic magnetohydrodynamics},
		Phys.\ Lett.\ B {\bf 750}, 45 (2015),
		arXiv:1506.06620 [nucl-th].
		\par
		S.~Pu, V.~Roy, L.~Rezzolla and D.~H.~Rischke,
		\textit{Bjorken flow in one-dimensional relativistic magnetohydrodynamics with magnetization}, Phys.\ Rev.\ D {\bf 93}, 074022 (2016),
		arXiv:1602.04953 [nucl-th].		
		\bibitem{bjorken1983}
		J.~D.~Bjorken,
		\textit{Highly relativistic nucleus-nucleus collisions: The central rapidity region},
		Phys.\ Rev.\ D {\bf 27}, 140 (1983).
		\bibitem{shokri2017}		
		M.~Shokri and N.~Sadooghi,
		\textit{Novel self-similar rotating solutions of nonideal transverse magnetohydrodynamics},
		Phys.\ Rev.\ D {\bf 96}, no. 11, 116008 (2017),
		arXiv:1705.00536 [nucl-th].
		\bibitem{magnetic-rest}
  E.~Stewart and K.~Tuchin,
  \textit{Magnetic field in expanding quark-gluon plasma},
  Phys.\ Rev.\ C {\bf 97}, no. 4, 044906 (2018),
  arXiv:1710.08793 [nucl-th].
\par
  V.~Roy, S.~Pu, L.~Rezzolla and D.~H.~Rischke,
  \textit{Effect of intense magnetic fields on reduced-MHD evolution in $\sqrt{s_{\rm NN}}$ = 200 GeV Au+Au collisions},
  Phys.\ Rev.\ C {\bf 96}, no. 5, 054909 (2017),
  arXiv:1706.05326 [nucl-th].
\par
  A.~Das, S.~S.~Dave, P.~S.~Saumia and A.~M.~Srivastava,
  \textit{Effects of magnetic field on plasma evolution in relativistic heavy-ion collisions},
  Phys.\ Rev.\ C {\bf 96}, no. 3, 034902 (2017),
  arXiv:1703.08162 [hep-ph].
\par
  L.~G.~Pang, G.~Endr\"odi and H.~Petersen,
  \textit{Magnetic-field-induced squeezing effect at energies available at the BNL Relativistic Heavy Ion Collider and at the CERN Large Hadron Collider},
  Phys.\ Rev.\ C {\bf 93}, no. 4, 044919 (2016),
  arXiv:1602.06176 [nucl-th].
\par
S.~Pu and D.~L.~Yang,
\textit{Transverse flow induced by inhomogeneous magnetic fields in the Bjorken expansion},
Phys.\ Rev.\ D {\bf 93}, no. 5, 054042 (2016),
arXiv:1602.04954 [nucl-th].	
		\bibitem{kolb2003}
		P.~F.~Kolb and U.~W.~Heinz,
		\textit{Hydrodynamic description of ultrarelativistic heavy ion collisions},
		In Hwa, R.C. (ed.) et al.: Quark gluon plasma 634-714,
arXiv:nucl-th/0305084.
		\bibitem{rajagopal2011}
  J.~Casalderrey-Solana, H.~Liu, D.~Mateos, K.~Rajagopal and U.~A.~Wiedemann,
 \textit{Gauge/String duality, hot QCD and heavy ion collisions},
 Cambridge University Press, Cambridge, 2014,
 arXiv:1101.0618 [hep-th].
\par
  W.~Busza, K.~Rajagopal and W.~van der Schee,
  \textit{Heavy ion collisions: The big picture, and the big questions},
  arXiv:1802.04801 [hep-ph].
		\bibitem{csorgo2002}
		T.~Cs\"org\"o, F.~Grassi, Y.~Hama and T.~Kodama,
		\textit{Simple solutions of relativistic hydrodynamics for cylindrically symmetric systems},
		Acta Phys.\ Hung.\ A {\bf 21}, 63 (2004),
		arXiv:hep-ph/0204300.
		\par
		T.~Cs\"org\"o, L.~P.~Csernai, Y.~Hama and T.~Kodama,
		\textit{Simple solutions of relativistic hydrodynamics for systems with ellipsoidal symmetry},
		Acta Phys.\ Hung.\ A {\bf 21}, 73 (2004),
arXiv:nucl-th/0306004.
		\bibitem{csorgo2002-2}
		T.~Cs\"org\"o, F.~Grassi, Y.~Hama and T.~Kodama,
		\textit{Simple solutions of relativistic hydrodynamics for longitudinally expanding systems},
		Acta Phys.\ Hung.\ A {\bf 21}, 53 (2004),
	arXiv:hep-ph/0203204.
		\bibitem{gubser2010-1}
  S.~S.~Gubser,
\textit{Symmetry constraints on generalizations of Bjorken flow},
  Phys.\ Rev.\ D {\bf 82}, 085027 (2010),
  arXiv:1006.0006 [hep-th].
		\bibitem{gubser2010-2}
		S.~S.~Gubser and A.~Yarom,
		\textit{Conformal hydrodynamics in Minkowski and de Sitter spacetimes},
		Nucl.\ Phys.\ B {\bf 846}, 469 (2011),
arXiv:1012.1314 [hep-th].
		\bibitem{beckenstein1978}
		J.~D.~Bekenstein and E~.Oron,
		\textit{New conservation laws in general-relativistic magnetohydrodynamics},
		Phys.\ Rev.\ D {\bf 18}, 1809 (1978).
		\bibitem{rezolla-book}
		L.~Rezzolla and O.~Zanotti, \textit{Relativistic Hydrodynamics}, Oxford University Press, Oxford, 2013.
		\bibitem{pechanski2009}
  R.~Peschanski and E.~N.~Saridakis,
  \textit{On an exact hydrodynamic solution for the elliptic flow},
  Phys.\ Rev.\ C {\bf 80}, 024907 (2009),
  arXiv:0906.0941 [nucl-th].
\par
  S.~J.~Sin, S.~Nakamura and S.~P.~Kim,
  \textit{Elliptic flow, Kasner universe and holographic dual of RHIC fireball},
  JHEP {\bf 0612}, 075 (2006),
  arXiv:hep-th/0610113.		
		\bibitem{hatta2016}
		Y.~Hatta, B.~W.~Xiao and D.~L.~Yang,
		\textit{Non-boost invariant solution of relativistic hydrodynamics in $1+3$ dimensions},
  Phys.\ Rev.\ D {\bf 93}, no. 1, 016012 (2016),
  arXiv:1512.04221 [hep-ph].
		\par
		Y.~Hatta,
		\textit{Analytic approaches to relativistic hydrodynamics},
		Nucl.\ Phys.\ A {\bf 956}, 152 (2016),
		arXiv:1601.04128 [hep-ph].		
		\bibitem{hatta2014}
		Y.~Hatta and B.~W.~Xiao,
		\textit{Building up the elliptic flow: Analytical insights},
		Phys.\ Lett.\ B {\bf 736}, 180 (2014),
		arXiv:1405.1984 [nucl-th].
		\bibitem{generalrefs}
		L.~D.~Landau and E.~M.~Lifshitz,
		\textit{Fluid mechanics}, Elsevier, Amsterdam, 1987, 2nd edition.
		\par
		A.~Zee,
		\textit{Einstein gravity in a nutshell},  Princeton University Press, Princeton, New Jersey, 2013.
		\par
		S.~Weinberg,
		\textit{Gravitation and cosmology: Principles and applications of the general theory of relativity},
		John Wiley \& Sons, Inc., New York, 1972.
		\par
		C.~Misner, K.~S~.Throne, J~.A.~Wheeler and D.~I.~Kaiser,
		\textit{Gravitation}, W. H. Freeman and Co., San Fransisco, 1973.
		\bibitem{hernandez2017}
  J.~Hernandez and P.~Kovtun, \textit{Relativistic magnetohydrodynamics},
  JHEP {\bf 1705}, 001 (2017),
arXiv:1703.08757 [hep-th].
		\bibitem{romero2005}
		R.~Romero, J.~M.~Marti, J.~A.~Pons, J.~M.~Ibanez and J.~A.~Miralles,
		\textit{The exact solution of the Riemann problem in relativistic MHD with tangential magnetic fields},
		J.\ Fluid Mech.\  {\bf 544}, 323 (2005),
		arXiv:astro-ph/0506527.
		\bibitem{dewolfe2014}
		O.~DeWolfe, S.~S.~Gubser, C.~Rosen and D.~Teaney,
		\textit{Heavy ions and string theory},
		Prog.\ Part.\ Nucl.\ Phys.\  {\bf 75}, 86 (2014),
		arXiv:1304.7794 [hep-th].
		\bibitem{gedalin1993}
		M.~Gedalin,
		\textit{Linear waves in relativistic anisotropic magnetohydrodynamics}
		Phys.\ Rev.\ E \textbf{47}, 4354 (1993).
\bibitem{kasza2018}
 T.~Cs\"org\"o, G.~Kasza, M.~Csan$\acute{\mbox{a}}$d and Z.~Jiang,
 \textit{New exact solutions of relativistic hydrodynamics for longitudinally expanding fireballs},  Universe {\bf 4}, 69 (2018),
  arXiv:1805.01427 [nucl-th].
		\bibitem{roy2015}
		V.~Roy and S.~Pu,
		\textit{Event-by-event distribution of magnetic field energy over initial fluid energy density in $\sqrt{s_{\rm NN}}$= 200 GeV Au-Au collisions},
		Phys.\ Rev.\ C {\bf 92}, 064902 (2015),
		arXiv:1508.03761 [nucl-th].
		\bibitem{roy2017}
		V.~Roy, S.~Pu, L.~Rezzolla and D.~H.~Rischke,
		\textit{Effect of intense magnetic fields on reduced-MHD evolution in $\sqrt{s_{\rm NN}}$ = 200 GeV Au+Au collisions},
		Phys.\ Rev.\ C {\bf 96}, no. 5, 054909 (2017),
		arXiv:1706.05326 [nucl-th].
		\bibitem{inghirami2018}
  G.~Inghirami, L.~Del Zanna, A.~Beraudo, M.~Haddadi Moghaddam, F.~Becattini and M.~Bleicher,
  \textit{Magnetohydrodynamic simulations of heavy ion collisions with ECHO-QGP},
  J.\ Phys.\ Conf.\ Ser.\  {\bf 1024}, no. 1, 012043 (2018).
		\bibitem{skellerud2015}
		G.~Aarts, C.~Allton, A.~Amato, P.~Giudice, S.~Hands and J.~I.~Skellerud,
		\textit{Electrical conductivity and charge diffusion in thermal QCD from the lattice},
		JHEP {\bf 1502}, 186 (2015),
		arXiv:1412.6411 [hep-lat].
		\bibitem{romatschke2018}
		H.~Bantilan, T.~Ishii and P.~Romatschke,
		\textit{Holographic heavy ion collisions: Analytic solutions with longitudinal flow, elliptic flow and vorticity},
		arXiv:1803.10774 [nucl-th].
\end{thebibliography}
\end{document}